\documentclass[final, 3p, times, number, sort&compress]{elsarticle}

\journal{J. Comput. Phys.}

\usepackage{amsmath, amssymb}
\usepackage[colorlinks,allcolors=blue]{hyperref}
\usepackage{bm}

\usepackage[english]{babel}
\usepackage{color}
\usepackage{xfrac}
\usepackage{graphicx}
\graphicspath{{./figures/}}

\DeclareMathAlphabet{\mathsfit}{\encodingdefault}{\sfdefault}{m}{sl}
\SetMathAlphabet{\mathsfit}{bold}{\encodingdefault}{\sfdefault}{bx}{sl}

\newcommand{\tens}[1]{\bm{\mathsfit{#1}}}

\newcommand{\kT}{k_{\mathrm{B}} T}

\newcommand{\ie}{\textit{i.e.}}
\newcommand{\eg}{\textit{e.g.}}
\newcommand{\etc}{\textit{etc.}}
\newcommand{\etal}{\textit{et al.}}

\newcommand{\bigO}{\ensuremath{\mathcal{O}}}
\newcommand{\grad}[1]{\ensuremath{\boldsymbol{\nabla}{#1}}}

\newcommand{\dive}[1]{\ensuremath{\boldsymbol{\nabla}\cdot{#1}}}
\newcommand{\lapl}[1]{\ensuremath{\nabla^2{#1}}}

\newcommand{\vect}[1]{\ensuremath{\boldsymbol{#1}}}

\newcommand{\bzero}{\ensuremath{\boldsymbol{0}}}

\newcommand{\bj}{\ensuremath{\boldsymbol{j}}}
\newcommand{\bff}{\ensuremath{\boldsymbol{f}}}
\newcommand{\bx}{\ensuremath{\boldsymbol{x}}}

\newcommand{\bk}{\ensuremath{\boldsymbol{k}}}
\newcommand{\br}{\ensuremath{\boldsymbol{r}}}
\newcommand{\bn}{\ensuremath{\boldsymbol{n}}}

\newcommand{\bv}{\ensuremath{\boldsymbol{v}}}
\newcommand{\bu}{\ensuremath{\boldsymbol{u}}}
\newcommand{\bp}{\ensuremath{\boldsymbol{p}}}
\newcommand{\bq}{\ensuremath{\boldsymbol{q}}}
\newcommand{\bt}{\ensuremath{\boldsymbol{t}}}
\newcommand{\bA}{\ensuremath{\boldsymbol{A}}}

\newcommand{\bF}{\ensuremath{\boldsymbol{F}}}

\newcommand{\bU}{\ensuremath{\boldsymbol{U}}}

\newcommand{\bI}{\ensuremath{\boldsymbol{I}}}

\newcommand{\bT}{\ensuremath{\boldsymbol{T}}}

\newcommand{\bR}{\ensuremath{\boldsymbol{R}}}
\newcommand{\cR}{\ensuremath{{\cal R}}}

\newcommand{\cF}{\ensuremath{{\cal F}}}
\newcommand{\cU}{\ensuremath{{\cal U}}}

\newcommand{\fM}{\ensuremath{{\frak M}}}
\newcommand{\nf}{\ensuremath{\mathrm{nf}}}
\newcommand{\ff}{\ensuremath{\mathrm{ff}}}
\newcommand{\rH}{\ensuremath{\mathrm{H}}}
\newcommand{\gamd}{\ensuremath{{\dot{\gamma}}}}

\newcommand{\tB}{\ensuremath{\tens{B}}}
\newcommand{\tE}{\ensuremath{\tens{E}}}
\newcommand{\tI}{\ensuremath{\tens{I}}}
\newcommand{\tJ}{\ensuremath{\tens{J}}}
\newcommand{\tK}{\ensuremath{\tens{K}}}

\newcommand{\tS}{\ensuremath{\tens{S}}}

\newcommand{\Rfu}{\ensuremath{\boldsymbol{R}_{{\cal FU}}}}
\newcommand{\Rfe}{\ensuremath{\boldsymbol{R}_{{\cal F}\mathrm{E}}}}
\newcommand{\Rsu}{\ensuremath{\boldsymbol{R}_{\mathrm{S}{\cal U}}}}
\newcommand{\Rse}{\ensuremath{\boldsymbol{R}_{\mathrm{SE}}}}

\newcommand{\dd}{\ensuremath{\mathrm{d}}}

\newcommand{\avg}[1]{\ensuremath{\langle{#1}\rangle}}

\newcommand{\pe}{\ensuremath{\mathrm{Pe}}}

\newcommand{\erfc}{\ensuremath{\mathrm{Erfc}}}

\newcommand{\spi}{\ensuremath{\sqrt{\pi}}}
\newcommand{\rx}{\ensuremath{r\xi}}
\newcommand{\rxp}[1]{\ensuremath{r^{#1}\xi^{#1}}}

\newcommand{\kxp}[1]{\ensuremath{k^{#1}\xi^{-#1}}}

\allowdisplaybreaks
\begin{document}








\begin{frontmatter}



\title{Spectral Ewald Acceleration of Stokesian Dynamics for polydisperse 
suspensions}


\author[caltech]{Mu Wang\corref{cor1}}
\ead{mwwang@caltech.edu}

\author[caltech]{John F. Brady}
\ead{jfbrady@caltech.edu}
\address[caltech]{Division of Chemistry and Chemical Engineering, California 
Institute of Technology, Pasadena, California 91125, USA}

\cortext[cor1]{Corresponding author.}

\begin{abstract}
In this work we develop the Spectral Ewald Accelerated Stokesian Dynamics (SEASD), a novel computational method for dynamic simulations of polydisperse colloidal suspensions with full hydrodynamic interactions.  SEASD is based on the framework of Stokesian Dynamics (SD) with extension to compressible solvents, and uses the Spectral Ewald (SE) method [Lindbo \& Tornberg, \textit{J. Comput. Phys.} \textbf{229} (2010) 8994] for the wave-space mobility computation.  To meet the performance requirement of dynamic simulations, we use Graphic Processing Units (GPU) to evaluate the suspension mobility, and achieve an order of magnitude speedup compared to a CPU implementation.  For further speedup, we develop a novel far-field block-diagonal preconditioner to reduce the far-field evaluations in the iterative solver, and \mbox{SEASD-nf}, a polydisperse extension of the mean-field Brownian approximation of Banchio \& Brady [\textit{J. Chem. Phys.} \textbf{118} (2003) 10323].  We extensively discuss implementation and parameter selection strategies in SEASD, and demonstrate the spectral accuracy in the mobility evaluation and the overall $\bigO(N\log N)$ computation scaling.  We present three computational examples to further validate SEASD and \mbox{SEASD-nf} in monodisperse and bidisperse suspensions: the short-time transport properties, the equilibrium osmotic pressure and viscoelastic moduli, and the steady shear Brownian rheology.  Our validation results show that the agreement between SEASD and \mbox{SEASD-nf} is satisfactory over a wide range of parameters, and also provide significant insight into the dynamics of polydisperse colloidal suspensions. 
\end{abstract}


\end{frontmatter}


\section{Introduction}

Colloidal suspensions are dispersions of small particles in a viscous solvent, and are found in almost every aspect of our life, ranging from dairy milk to printer ink.  They have two distinguishing features: (\textit{i}) Brownian motion of the particles due to thermal fluctuations, and (\textit{ii}) the long-range, non-pairwise-additive hydrodynamic interactions (HIs) mediated by the solvent.  As a result of these features, dispersions exhibit many surprising behaviors such as non-Newtonian rheology, glass transitions, phase transitions, \etc, and have attracted extensive scientific and engineering interests~\cite{colloidal_dispersions_1989}.  Using monodisperse colloidal suspensions as a model system, significant understanding has been achieved through theoretical, simulation, and experimental studies.

However, naturally occurring colloidal suspensions are seldom monodisperse, and particle size differences are often unavoidable.  In addition, particle size disparity introduces phenomena otherwise not observed in monodisperse suspensions.  For example, size polydispersity reduces suspension viscosity~\cite{bimodal-colloids-viscosity_chong_japs1971, exp-bimodal-latex-visc_wolfe_lang1992, exp-bimodal-visc_shikata_jor1998},  softens and even melts colloidal glasses~\cite{yield-binary-colloid-glass_egelhaaf_softmatt2013}, and promotes particle segregation in pressure driven flows~\cite{migration-binary-colloid_weeks_pof2008}.  Apparently, these behaviors can only be understood by studying dynamics of polydisperse colloidal suspensions.

In this work we develop a computational method based on the framework of Stokesian Dynamics~\cite{sd_durlofsky_jfm1987} (SD) for fast and realistic dynamic simulations of dense, polydisperse colloidal suspensions, with a focus on suspension rheology.  Presently, theoretical and computational studies on polydisperse colloidal suspensions, even for the simplest case of neutrally buoyant hard-sphere particles, are scarce, and heavily focus on the dilute or the short-time limits~\cite{sed-general_batchelor_jfm1982, bimodal-viscosity_wagner_jfm1994,tracer-diffusivity-bimodal_nagele_jcp2002, sd-bidisperse_wang_jcp2015,dg-sd-comp_wang_jcp_2014}:  the former restricts HIs to the two- or three-body level, and the latter ignores suspension dynamic evolution, particularly the influence of Brownian motion.  Beyond these limiting cases, we are only aware of the work of Ando \& Skolnick~\cite{crowding-hydro-cell_skolnick_pnas2010}, who studied particle diffusion in dense polydisperse colloidal suspensions using conventional SD in the context of biological molecular crowding.  Their implementation limits HIs to the force-torque level, and therefore is unsuitable for rheological investigations.

A difficulty in dynamic simulations of dense colloidal suspensions is the singular HIs due to the lubrication interactions between close particle pairs.  To directly resolve HIs, a computational method must capture the flow details in the small gap between particles.
For multipole expansion based methods~\cite{hydro-trans-coeff_ladd_jcp1990, fric-mob-many-spheres_cichocki_jcp1994,sd_durlofsky_jfm1987}, a large number of expansion terms are necessary to achieve convergence, and for methods based on surface or spatial discretization, such as the boundary element method~\cite{Acrivos_BEM_jfm1975, Pozrikidis1992} or direct numerical simulations~\cite{sph_lucy_astroj_1977, Liu2003, Hoogerbrugge1992, soft-matt-lb_duenweg_adv-pol-sci_2009}, very fine meshing is needed in the gap.  Directly resolving lubrication interactions drastically increases the computational cost and limits many studies to low volume fractions.  For example, the force coupling method study of Abbas~\etal~\cite{Maxey_bidisperse-fcm_pof2006} on the dynamics of non-Brownian bidisperse suspensions is limited to particle volume fractions below $20\%$.

A solution to the above difficulty is the SD framework~\cite{sd_durlofsky_jfm1987}, which exploits the local and pairwise additive nature of lubrication interactions.  In SD, the long-range, non-pairwise-additive HIs are computed from the mobility perspective using low-order multipole expansions, and for particles in close contact, lubrication corrections are added pairwise to the corresponding resistance formalism.  The corrections are based on the solutions of two-body problems with the far-field contributions removed.  In this way, SD avoids directly resolving the singular lubrication interactions.  The idea of lubrication correction in SD is general enough for incorporation to other computational methods.  For example, similar lubrication corrections has been developed for hydrodynamic multipole methods~\cite{hydro-trans-coeff_ladd_jcp1990, fric-mob-many-spheres_cichocki_jcp1994, self-diffusion-lubrication-3body_cichocki_jcp1999, mult-pole-bound_cichocki_jcp2000}, the force coupling method~\cite{Yeo2010}, the lattice Boltzmann method~\cite{Ladd_lub-corr-LB_pre_2002}, and the fictitious domain method~\cite{Gallier_fict-domain-susp_jcompp2014}.  Moreover, with an appropriate fluid solver, the lubrication corrections can be improved beyond the pairwise level~\cite{Nguyen_improve-lub_jfm2015}.  We feel that, by incorporating the lubrication corrections, many recent computational techniques can significantly extend their accessible parameter range without an increased computational burden.  This point is demonstrated in the present work, which essentially combines the lubrication corrections and the Spectral Ewald (SE) method of Lindbo \& Tornberg~\cite{Tornberg_se-stokes_jcompp2010, Tornberg_periodic-laplace_jcompp2011} for dynamic simulations of dense polydisperse suspensions.

The Spectral Ewald (SE) method is a new particle mesh technique for computing long-range electrostatic~\cite{Tornberg_periodic-laplace_jcompp2011} or hydrodynamic~\cite{Tornberg_se-stokes_jcompp2010} interactions, and has recently been incorporated into the boundary element method for soft particles~\cite{sw-stokesian_tornberg_ijnmf_2014}.  Particle mesh techniques including the Particle Mesh Ewald (PME) method~\cite{Darden_pme-algorihm_jcp1993} and the Smooth Particle Mesh Ewald (SPME) method~\cite{spme-method_essmann_jcp1995} have been extensively used for calculating HIs with $\bigO(N\log N)$ computation scaling.  Note that, although algorithms based on the fast multipole method \cite{Greengard_fmm_jcompp1987} can achieve a better computation scaling--down to $\bigO(N)$, they often have significant computation overheads, and require large system sizes to justify the complexity~\cite{sd-fmm_ichiki_jfm2002}.  Therefore, for many dynamic simulations, the particle mesh techniques remain the practical choice.  Notable examples are Accelerated Stokesian Dynamics (ASD)~\cite{asd_sierou_jfm01} which uses the PME method for the far-field mobility evaluation, and the work of Saintillan~\etal~\cite{Saintillan_spme-stokes-fiber_pof2005}, where the SPME method is employed to study fiber sedimentation.  Compared to other particle mesh techniques, the SE method is spectrally accurate, and can separate errors from mesh interpolation and the wave-space truncation.  Both features are essential for capturing the complicated HIs in polydisperse suspensions.

Another challenge in dynamic simulations of colloidal suspensions is Brownian motion, which is configuration dependent due to the fluctuation-dissipation relation.  When Euler-Maruyama time integration is used, the deterministic particle drift due to the Brownian motion must also be included~\cite{bd_ermak_jcp1978}.  As a result, computing Brownian related quantities requires the gradient and the square root of the mobility tensor.  Fortunately, these quantities can be evaluated in a matrix-free manner under the framework of ASD, making dynamic studies on hundreds of colloidal particles possible~\cite{asd-brownian_banchio_jcp2003,Swan2013}.  Moreover, the mean-field Brownian approximation, which estimates the mobility tensor based on the near-field HIs, is able to further speed up the computations~\cite{asd-brownian_banchio_jcp2003, Higdon_FLD_pre2010}.  In this work, these developments are fully incorporated for the dynamic simulation of Brownian polydisperse suspensions.  Note that a different approach to treat the Brownian motion is based on fluctuating hydrodynamics~\cite{Landau_FluidMech}, where the thermal fluctuations are directly incorporated in the governing fluid equations.  It has been applied to the lattice Boltzmann method~\cite{suspension-dws-simulation-correlation_ladd_pre1995}, the force coupling method~\cite{Keaveny_fluct-fcm_jcompp2014}, and the immersed boundary method~\cite{Donev_BEM-BD_jcp2014}.

The emergence of the General Purpose Graphic Processing Unit (GPGPU) programming often brings significant, sometimes orders of magnitude, speed improvements for many existing algorithms. Recently, Kopp \& H\"{o}fling~\cite{sd-gpu_hofling_epjst_2012} implemented the conventional SD for infinite solvent using GPGPU with direct HI summation.  Despite the $\bigO(N^2)$ scaling, they achieved impressive speedup over the CPU implementation.  However, to study the dynamics of homogeneous suspensions, further extension to periodic systems are necessary.  On the other hand, GPU acceleration of the SPME method~\cite{GPU-Fenzi_ieee2011,Harvey_spem-gpu_jctc2009} in molecular dynamics provides access to millisecond-scale dynamics on personal computers.  These acceleration techniques are applicable to particle mesh techniques in general, and inspired the present work.  In particular, we used GPGPU programming to compute the HIs with the SE method in homogeneous suspensions, and realized almost an order of magnitude speedup in dynamic simulations.

Furthermore, our computation method extends SD to compressible suspensions, allowing dynamic simulations of constant pressure rheology~\cite{boyer-granular-rheology_prl2011} without introducing geometric confinement.
This is possible because the flow disturbances due to rigid particles in a compressible solvent are incompressible and satisfy the Stokes equation~\cite{Brady2006}.  Another benefit of such extension is that the suspension normal stress, which is essential for particle migration in sheared suspensions~\cite{brady1993a, normal-stress-modelling-non-colloidal_morris_jor99, pressure-sd_morris_jor2008}, can be directly evaluated.

The remainder of the paper is arranged as follows: Sec.~\ref{sec:stok-susp} establishes the basic formalism for HIs in compressible Stokes flow. In Sec.~\ref{sec:mobility-computation}, various aspects of mobility computations with the SE method are presented.  Here, we also discuss different approaches to incorporate particle size polydispersity and the GPGPU implementation.  In Sec.~\ref{sec:stokesian-dynamics}, we present the Spectral Ewald Accelerated Stokesian Dynamics (SEASD) and its mean-field Brownian approximation, \mbox{SEASD-nf}, for dynamic simulations of Brownian polydisperse suspensions.  In Sec.~\ref{sec:accuracy-performance} we carefully discuss the accuracy and parameter selections for the SE method, and the computation scaling of various SEASD implementations.  Sec.~\ref{sec:results-discussions} presents a series of validation calculations for monodisperse and bidisperse suspensions with SEASD and \mbox{SEASD-nf}: Sec.~\ref{sec:short-time-transport} addresses the short-time transport properties, Sec.~\ref{sec:equil-susp} evaluates the equilibrium osmotic pressure and viscoelastic moduli, and Sec.~\ref{sec:rheol-bidisp-susp} presents various aspects of the steady shear rheology of Brownian suspensions.  The results also reveal the role of particle sizes in the dynamics of bidisperse suspensions.  Finally, we conclude this work with a few comments in Sec.~\ref{sec:conclusions}.

\section{Hydrodynamic interactions in (compressible) Stokes flow}
\label{sec:stok-susp}

\subsection{The mobility and resistance formalism}
\label{sec:mobil-resist-form}

We first consider a suspension of $N$ spherical rigid particles, each with radius $a_i$ and position $\br_i$, in an \emph{incompressible} solvent of viscosity $\eta_0$ and density $\rho_0$, occupying a volume $V$.  For the special case of bidisperse suspensions with particle sizes $a_1$ and $a_2$, the suspension composition is fully characterized by three dimensionless parameters, 
\begin{equation}
  \label{eq:composition}
 \lambda = a_1 / a_2\text{, }  \phi = \phi_1 + \phi_2\text{, and }y_2 = 
\phi_2/\phi,
\end{equation}
where $\lambda$ is the size ratio, $\phi$ is the total volume fraction, and $y_2$ is the volume ratio of species $2$.  The species volume fraction is $\phi_\alpha = \tfrac{4}{3}\pi a_\alpha^3 n_\alpha$, $\alpha\in\{1,2\}$, and the species number density is $n_\alpha$.  The total number density satisfies $n = n_1 + n_2$, and the species number fraction is $x_\alpha = n_\alpha / n$.  Without loss of generality, we take $a_2>a_1$.

If the particles are sufficiently small, the particle Reynolds number $\mathrm{Re}_{\mathrm{p},\alpha}=\rho_0 a_\alpha U_\alpha/\eta_0 \ll 1$, where $U_\alpha$ is the species characteristic velocity. In this limit, the velocity field $\bv(\br)$ and the pressure field $p(\br)$ of the solvent satisfy the Stokes equation,
\begin{equation}
  \label{eq:Stokes}
\grad{p} = \eta_0\lapl{} \bv,\; \dive{\bv} = 0,
\end{equation}
supplemented by no-slip boundary conditions on particle surfaces.  Due to the linearity of Eq.~\eqref{eq:Stokes}, there is a linear relation between the velocity disturbance on the surface of a particle $i$, $\bu_i'$, and the surface force density of another particle $j$, $\bff_j$,
\begin{equation}
  \label{eq:mob}
  \bu_i'(\br) =  - \int \dd \br' \sum_{j}\tens{M}_{ij}(\br, \br'; X) \cdot 
\bff_j(\br'),
\end{equation}
where $\tens{M}_{ij}(\br, \br'; X)$ is a mobility operator depending on positions $\br$ and $\br'$ and the suspension configuration $X = \{\br_1, \br_2, \ldots\}$.  The surface force density is localized on the particle surface, \ie, $\bff_j(\br) = \tens{\sigma}(\br)\cdot \bn_j\delta(\|\br\| - a_j)$, where $\tens{\sigma}$ is the stress tensor, $\bn_j$ is the surface normal of particle $j$, and $\delta(x)$ is the Dirac delta function.  The stress tensor $\tens{\sigma} =-p\tI + \eta_0[\grad{\bv} + (\grad{\bv})^\dagger]$, with $\dagger$ indicating transposition and $\tI$ is the idem tensor.  The velocity disturbance $\bu_i'(\br) = \bU_i + \vect{\Omega}_i\times (\br - \br_i) - \bv^\infty(\br) $, where $\bv^\infty(\br)$ is the ambient flow satisfying $\dive{\bv^\infty} = 0$, and $\bU_i$ and $\vect{\Omega}_i$ are respectively the linear and angular velocities of particle $i$.  By stacking the force density vectors $\bff = (\bff_1, \bff_2, \ldots)^\dagger$ and the velocity disturbance vectors $\bu' = (\bu_1', \bu_2', \ldots)^{\dagger}$ the grand mobility operator $\tens{M}$ is constructed from elements $\tens{M}_{ij}$ in Eq.~\eqref{eq:mob}, such that
\begin{equation}
  \label{eq:granmob}
  \bu'(\br) = - \int \dd \br' \tens{M}(\br, \br'; X) \cdot \bff(\br'),
\end{equation}
for the $N$ particles in the suspension.  Eqs.~\eqref{eq:mob} and \eqref{eq:granmob} are known as the mobility formalism, and the inverse relation is the resistance formalism,
\begin{equation}
  \label{eq:granres}
\bff(\br) = - \int \dd \br'\tens{R}(\br, \br'; X) \cdot \bu'(\br'),
\end{equation}
where $\tens{R}(\br, \br'; X)$ is the grand resistance operator.

The integral representations in Eqs.~\eqref{eq:granmob} and \eqref{eq:granres} can be equivalently expressed as multipole expansions of $\bff(\br)$ and $\bu'(\br)$, $\mathfrak{f}$ and $\mathfrak{u}'$ respectively, around the particle centers, \ie,
\begin{equation}
  \label{eq:multipole-expand}
  \bff(\br) \rightarrow {\mathfrak{f}} = 
  \begin{bmatrix}
    {\cal F}^\rH \\
    \tS^\rH \\
    \vdots
  \end{bmatrix}
  \text{ and }   \bu'(\br) \rightarrow {\mathfrak{u}}' = 
  \begin{bmatrix}
    {\cal U}' \\
    -\tE^\infty \\
    \vdots
  \end{bmatrix},
\end{equation}
where ${\cal F}^\rH$ is the generalized hydrodynamic force, $\tS^\rH$ is the hydrodynamic stresslet, ${\cal U}'$ is the generalized velocity disturbance, and $\tE^\infty$ is the rate of strain tensor for the ambient flow. Note that ${\cal F}^\rH = (\bF^\rH, \bT^\rH)^\dagger$, where $\bF^\rH$ and $\bT^\rH$ are respectively the particle hydrodynamic force and torque for all particles, and ${\cal U}' = (\bU - \bU^\infty,  \vect{\Omega}-\vect{\Omega}^\infty )^\dagger$, where  $\bU - \bU^\infty$ and $\vect{\Omega}-\vect{\Omega}^\infty$ are respectively the linear and angular velocity disturbances.  The hydrodynamic force, torque, and stresslet for particle $i$ are defined  as integrals of the localized surface force density $\bff_i$,
\begin{align}
  \label{eq:def-f}
  \bF_i^\rH & = -\int \dd \br \,\bff_i(\br)  ,\\
  \label{eq:def-t}
  \bT_i^\rH & = -\int \dd \br \, (\br-\br_i)\times \bff_i(\br)  , \\
  \label{eq:def-s}
  \tS_i^\rH & = - \int \dd \br \,\tfrac{1}{2}[(\br - \br_i)\bff_i + \bff_i (\br - \br_i) ].
\end{align}
In Eq.~\eqref{eq:multipole-expand} the ambient velocities are evaluated at particle centers, \ie, $\bU_i^\infty = \bv^\infty(\br_i)$, $\vect{\Omega}_i^\infty = \tfrac{1}{2}\grad{} \times \bv^\infty|_{\br_i}$, and $\tE^\infty = \tfrac{1}{2}[\grad{\bv^\infty} + (\grad{\bv^\infty})^\dagger]_{\br_i}$.  The expansions in Eqs.~\eqref{eq:granmob} and \eqref{eq:granres} lead to the following infinite dimension linear relation,
\begin{equation}
  \label{eq:mobrestensor}
  {\mathfrak{f}} = - {\mathfrak{M}}(X) \cdot {\mathfrak{u}}'  \text{ and } 
  {\mathfrak{u}}' = - {\mathfrak{R}}(X) \cdot {\mathfrak{f}}
\end{equation}
where $\mathfrak{M}(X)$ and $\mathfrak{R}(X)$ are the multipole grand mobility and resistance tensors of operators $\tens{M}(\br, \br'; X)$ and $\tens{R}(\br, \br'; X)$, respectively.  Evidently, $\mathfrak{M} =\mathfrak{R}^{-1}$, and from the Lorentz reciprocal theorem~\cite{Kim2005}, both are positive definite.

The infinite dimension vectors $\mathfrak{f}$ and $\mathfrak{u}'$ can be reduced to finite dimensions by projection.  To the stresslet level of $\mathfrak{f}$ and the strain rate level of $\mathfrak{u}'$, we introduce projection matrices $\mathcal{P}$ and $\mathcal{Q}$, such that ${\cal P}\cdot {\frak f} = ({\cal F}^\rH, \tS^\rH)^\dagger$ and ${\cal Q}\cdot {\frak u}' = ({\cal U}', -\tE^\infty)^\dagger$.  Moreover, ${\cal P}\cdot{\cal P}^{\dagger} = {\cal Q}\cdot{\cal Q}^{\dagger} = {\cal I}$, where ${\cal I}$ is an identity matrix.  The following linear relation holds:
\begin{equation}
  \label{eq:mob-res-fs}
  \begin{bmatrix}
    {\cal U}'\\
    - \tE^\infty
\end{bmatrix} = - {\cal M} \cdot 
  \begin{bmatrix}
    {\cal F}\\
    \tS
\end{bmatrix},\text{ and }
{\cal R} = {\cal M}^{-1},
\end{equation}
where ${\cal M} = {\cal Q} {\frak M} {\cal P}^\dagger$ is the (exact) grand mobility tensor and ${\cal R} = {\cal P} {\frak R} {\cal Q}^\dagger$ is the (exact) grand resistance tensor.
For convenience, the grand resistance tensor is partitioned as 
\begin{equation}
  \label{eq:res-part}
\cR = 
  \begin{bmatrix}
    \Rfu & \Rfe \\
    \Rsu & \Rse
  \end{bmatrix},
\end{equation}
where, for example, $\Rfu$ describes the coupling between the generalized force and the generalized velocity.  The linear relation in Eq.~\eqref{eq:mob-res-fs} can also be deduced from the linearity of Eq.~\eqref{eq:Stokes} without appealing to the multipole expansion, but here we establish a connection with other works, particularly the multipole methods of Cichocki and coworkers~\cite{fric-mob-many-spheres_cichocki_jcp1994, Szymczak_diag-prob_jsmte2008}.  Note that for rigid spherical particles, external flows can only affect the first two moments of $\mathfrak{f}$ and $\mathfrak{u}'$ due to symmetry and the no-slip boundary condition.

Elements of ${\frak M}$ and ${\frak R}$ can be computed from, for example, the induced force multipole~\cite{vanSaarloos_many-sphere-hydro_physa1982, hydro-multipole_ladd_jcp_1988}, eigenfunction expansions~\cite{creeping-flow_felderhof_physa1982, fric-mob-many-spheres_cichocki_jcp1994,mult-pole-bound_cichocki_jcp2000}, and multipole expansions~\cite{sd_durlofsky_jfm1987}.  To the stresslet level, ${\frak M}$ can be conveniently evaluated by combining the Fax\'{e}n formulae and the multipole expansions.  For a rigid particle $i$ in an incompressible solvent, the Fax\'{e}n formulae are~\cite{sd_durlofsky_jfm1987},
\begin{align}
\label{eq:faxen-f}
  \bU_i - \bU^\infty & =   -\frac{\bF_i^\rH}{6\pi\eta_0 a_i} + \left( 
1+\tfrac{1}{6}a_i^2 \lapl{} \right) \bv'\big|_{\br_i}\\
\label{eq:faxen-t}
  \vect{\Omega}_i - \vect{\Omega}^\infty & = -  \frac{\bT_i^\rH}{8\pi\eta_0 
a_i^3} + \tfrac{1}{2}\grad{}\times\bv'\big|_{\br_i}\\
\label{eq:faxen-s}
  -\overline{\tE^\infty}& = -\frac{\overline{\tS_i^\rH}}{\tfrac{20}{3}\pi 
\eta_0 a_i^3} +
\left(1 + \tfrac{1}{10}a_i^2 \lapl{} \right) \tfrac{1}{2}[\grad{\bv'} + 
(\grad{\bv'})^\dagger]\big|_{\br_i},
\end{align}
where the overline indicates the traceless part of the symmetric tensor, and $\bv'(\br)$ is the velocity field in the absence of particle $i$.  With the fundamental solution of  Stokes equation $\tJ(\br)$ and the force density $\bff$, the velocity field $\bv'(\br)$ can be computed as~\cite{Kim2005},
\begin{equation}
  \label{eq:funda-sol}
  \bv(\br) = - \frac{1}{8\pi\eta_0 }\int \dd \br' \tJ(\br - \br') \cdot 
\bff(\br').
\end{equation}
Expanding the force density around particle centers, we have
\begin{equation}
  \label{eq:mul-expand}
  \bv'(\br) = \frac{1}{8\pi\eta_0}\sideset{}{'}\sum_{j} 
\left(1+\tfrac{1}{6}a_j^2\lapl{}\right)\tJ\cdot\bF_j^\rH + \tens{R}\cdot 
\bT_j^\rH - \left(1+\tfrac{1}{10}a_j^2\right) \tens{K}:\overline{\tS_j^\rH} + 
\cdots,
\end{equation}
where the prime on the summation excludes the case $i=j$, and the functions $\tJ$, $\tens{R}$, and $\tK$ are evaluated at $\br-\br_j$.  In the Cartesian tensor form, $\tens{R} = R_{\alpha\beta} = \tfrac{1}{4}\epsilon_{\delta\gamma\beta} (\nabla_{\gamma} J_{\alpha\delta} - \nabla_{\delta} J_{\alpha\gamma})$ and $\tens{K} = K_{\alpha\beta\gamma} = \tfrac{1}{2}[\nabla_\gamma J_{\alpha\beta} + \nabla_\beta J_{\alpha\gamma}]$, with $\epsilon_{\alpha\beta\gamma}$ the Levi-Civita symbol.  With Eqs.~\eqref{eq:faxen-f}--\eqref{eq:faxen-s} and \eqref{eq:mul-expand}, the grand mobility tensor $\mathfrak{M}$ for incompressible solvents can be constructed in a pairwise fashion.

\subsection{The fundamental solutions}
\label{sec:fundamental-solution}

The formalism in Sec.~\ref{sec:mobil-resist-form} relies on $\tJ(\br)$, the fundamental solution of Stokes equation.  Different boundary conditions such as periodicity~\cite{stokes-fundamental-sol_hasimoto_jfm1959,Beenakker_ewald-rotne-prager_jcp86}, confinement~\cite{Swan_2walls_pof2010,mult-pole-bound_cichocki_jcp2000}, or a combination of both~\cite{confined-suspension_swan_2011}, can be incorporated to $\tJ(\br)$.  For an infinite expanse of fluid, we have the well-known Oseen tensor, 
\begin{equation}
  \label{eq:oseen}
  \tJ(\br) = \frac{1}{r}(\tI + \hat{\br} \hat{\br}),
\end{equation}
where $r = \|\br\|$ and $\hat{\br} =\br/r$.

To study dynamics of homogeneous suspensions, periodic boundary conditions are necessary to assess the HIs.  In this case, the proper fundamental solution $\tJ(\br)$ describes the fluid velocity disturbance due to an array of periodic forces $\bF\sum_{\bp} \delta(\br -\bR_{\bp})$, where $\bR_{\bp} = \sum_{d=1}^3p_d {\bold a}_d$ is the location of the periodic forcing.  Here, $\bp=(p_1, p_2, p_3)\in \mathbb{Z}^3$, $\delta(\br)$ is the 3D Dirac delta function, and ${\bold a}_1$, ${\bold a}_2$, and ${\bold a}_3$ are the Bravais lattice vectors describing the spatial periodicity.  From Fourier expansion of Stokes equation [Eq.~\eqref{eq:Stokes}], we have for the periodic $\tJ(\br)$:
\begin{equation}
  \label{eq:oseen-peridoic}
  \tJ(\br) = -\frac{8\pi}{V} (\tI \lapl{} - \grad{}\grad{}) \sum_{\bk\neq 
0}\frac{1}{k^4} \exp(-\imath \bk\cdot\br),
\end{equation}
where $\imath = \sqrt{-1}$, the unit cell volume $V={\bold a}_1\cdot ({\bold a}_2\times  {\bold a}_3)$, the wave vector $\bk = \sum_{d=1}^3 j_d {\bold b}_d$ is defined by the reciprocal vectors ${\bold b}_1$, ${\bold b}_2$, and ${\bold b}_3$, $\bj=(j_1, j_2, j_3)\in \mathbb{Z}^3$, and $k^2 = \bk\cdot\bk$.  Writing the lattice and the reciprocal vectors as column vectors and defining matrices $\tens{A} = [{\bold a}_1 {\bold a}_2  {\bold a}_3]$ and $\tens{B} = [{\bold b}_1 {\bold b}_2  {\bold b}_3]$,  we have $ \tens{B}^\dagger = 2\pi \tens{A}^{-1}$ and $\exp(\imath \bk\cdot\bR_{\bp}) = 1$.  By requiring $\bk \neq 0$ in Eq.~\eqref{eq:oseen-peridoic}, the external forces are balanced by the pressure gradient~\cite{stokes-fundamental-sol_hasimoto_jfm1959}, a necessary condition for convergent HIs~\cite{brady-sd-ew_jfm_88}.

A difficulty associated with HIs is the long range nature of $\tJ(\br)$, \ie, Eq.~\eqref{eq:oseen} decays as $r^{-1}$ in the real space and Eq.~\eqref{eq:oseen-peridoic} as $k^{-2}$ in the wave space.  For periodic systems, however, the conditionally converging sum in Eq.~\eqref{eq:oseen-peridoic} can be split into two exponentially fast converging series, \ie,
\begin{equation}
  \label{eq:oseen-split}
  \tJ(\br) = \tJ_R(\br) + \tJ_W(\br),
\end{equation}
where $\tJ_R(\br)$ is the real-space sum, $\tJ_W(\br)$ is the wave-space sum.  Although the splitting in Eq.~\eqref{eq:oseen-split} is not unique~\cite{Tornberg_se-stokes_jcompp2010}, a particularly efficient scheme by Hasimoto~\cite{stokes-fundamental-sol_hasimoto_jfm1959} utilizes the integral
\begin{equation}
  \label{eq:ewald-theta}
  \frac{1}{k^4} = \pi^2 \int^{\infty}_0 \beta \exp(-\pi k^2 \beta)\dd \beta, 
(k\neq 0),
\end{equation}
and the Poisson summation formula.  The result is
\begin{align}
\label{eq:oseen-ewald-rs}
\tJ_R (\br)& =\sum_{\bp\neq 0} (\tI \lapl - \grad{}\grad{}) \left[ r\erfc(\rx) 
-\frac{1}{\xi\spi}e^{-\rxp{2}}
\right], \\
\label{eq:oseen-ewald-ws}
\tJ_W (\br) & =  
\frac{8\pi}{V} \sum_{\bk \neq 0} (\tI \lapl - \grad{}\grad{}) 
\left(-1-\frac{k^2}{4\xi^2}\right)  \frac{1}{k^4} e^{-\tfrac{1}{4}k^2 
\xi^{-2}}e^{-\imath \bk\cdot \br},
\end{align}
where $\xi$ is the splitting parameter and $\erfc(x)$ is the complementary error function. The real-space sum $\tJ_R$ only covers the \emph{neighboring} periodic cells.  The parameter $\xi$ is consistent with the convention of Beenakker\cite{Beenakker_ewald-rotne-prager_jcp86} and satisfies $4\pi\alpha \xi^2 =1 $, where $\alpha$ is the splitting parameter introduced by Hasimoto~\cite{stokes-fundamental-sol_hasimoto_jfm1959}.

\subsection{Extension to compressible fluid}
\label{sec:extens-compr-fluid}

The formalism in Sec.~\ref{sec:mobil-resist-form} is limited to an incompressible fluid, \ie, the imposed flow must satisfy $\dive{\bv^\infty} = 0$.  This requirement is relaxed by imposing a \emph{uniform} rate of expansion everywhere in the fluid, such that $\dive{\bv^\infty} = E^\infty$, and the fluid is assumed compressible with a bulk viscosity $\kappa_0$.  The rigid particles, unable to expand with the compressible fluid, generate a velocity disturbance that satisfies the incompressible Stokes equation~\cite{Brady2006}.  From the linearity of Stokes flow, this velocity disturbance can be superimposed with other flows in the suspension, extending the existing formalism to compressible fluids.

For a rigid particle of radius $a_i$ located at $\br_i = 0$, the velocity disturbance $\bv_{s}$ due to a compressible flow with an expansion rate $E^\infty$ is 
\begin{equation}
  \label{eq:compress-disturb}
  \bv_{s}(\br) = -\tfrac{1}{3}a_i^3 E^\infty\frac{\br}{r^3}.
\end{equation}
This isotropic flow disturbance generates an isotropic stress contribution.  Introducing the pressure moment as the trace of the stresslet in Eq.~\eqref{eq:def-s}, \ie,
\begin{equation}
  \label{eq:press-moment}
  S_i^\rH = - \int \dd \br \, (\br - \br_i) \cdot \bff_i(\br) ,
\end{equation}
we have $S_i^\rH =-\tfrac{16}{3} \pi\eta_0 a_i^3 E^\infty$ from Eq.~\eqref{eq:compress-disturb}.  Therefore, the velocity disturbance due to a pressure moment $S_i^\rH$ at the origin is 
\begin{equation}
  \label{eq:pres-vel-disturb}
  \bv_{s}(\br) =  \frac{1}{16\pi\eta_0} \frac{\br}{r^3} S_i^\rH 
= \vect{Q}(\br) S_i^\rH.
\end{equation}
Adding the compressible velocity disturbances $\bv_s(\br)$ from other particles to the incompressible velocity disturbance $\bv'(\br)$ in Eq.~\eqref{eq:mul-expand}, the general velocity disturbance in a compressible suspension is
\begin{equation}
  \label{eq:compress-vel-disturb}
  \bv_c'(\br) = \bv'(\br) + \sideset{}{'}\sum_j\vect{Q}(\br-\br_j) S_j^\rH.
\end{equation}
When applying the Fax\'{e}n formulae [Eqs.~\eqref{eq:faxen-f}--\eqref{eq:faxen-s}] in compressible suspensions, the velocity disturbance $\bv_c'$, instead of $\bv'$, is used.

In addition to Eqs.~\eqref{eq:faxen-f}--\eqref{eq:faxen-s}, the Fax\'{e}n relation for the pressure moment in a compressible fluid is~\cite{pres-moment_jeffrey_pof1993, compres-res_khair_pof2006}
\begin{equation}
  \label{eq:faxen-p}
S_i^\rH = -\tfrac{16}{3}\pi\eta_0 a_i^3 E^\infty + 4\pi a_i^3 p'(\br_i),
\end{equation}
where $p'$ is the pressure disturbance without the particle at $\br_i$.  The pressure disturbance can be obtained from the pressure fundamental solution of Stokes equation,
\begin{equation}
  \label{eq:pres-funda}
  \vect{P}(\br) = \frac{\br}{r^3},
\end{equation}
such that the pressure distribution due to a force density is
\begin{equation}
  \label{eq:pres-distro}
  p(\br) = -\frac{1}{4\pi} \int \dd \br' \vect{P}(\br - \br')\cdot \bff(\br').
\end{equation}
For the pressure disturbance $p'$ in Eq.~\eqref{eq:faxen-p}, expanding the surface force densities leads to
\begin{equation}
  \label{eq:pres-multi-expans}
   p'(\br) = \frac{1}{4\pi}\sideset{}{'}\sum_{j} \vect{P}(\br-\br_j) \cdot 
\bF_j^\rH - \grad{}\vect{P} : \tens{S}_j^\rH|_{(\br-\br_j)} + \cdots.
\end{equation}
Eq.~\eqref{eq:faxen-p} is different from the Fax\'{e}n formulae in Eqs.~\eqref{eq:faxen-f}--\eqref{eq:faxen-s} as it presents the pressure moment or trace of the stresslet on the left hand side.  This subtle difference highlights a distinct feature of the compressible flow disturbances: in a compressible fluid, the pressure moment can cause particle movement satisfying the incompressible Stokes equation, but the incompressible force moments cannot generate compressible disturbances.  As a result, the interaction part of the pressure moment can only be evaluated after $\bF_i^\rH$, $\bT_i^\rH$, and $\overline{\tens{S}^\rH_i}$ are known.  Otherwise, the resulting hydrodynamic interactions contain spurious contributions due to the unphysical coupling between the incompressible force moments and the compressible flow disturbances.

To extend the above results for $\bv_s$ and $S_i^\rH$ to periodic boundary conditions,  we note that the divergence of $\vect{Q}$ in Eq.~\eqref{eq:pres-vel-disturb} satisfies
\begin{equation}
  \label{eq:fluid-source}
  \dive{\vect{Q}} = \frac{1}{4\eta_0} \delta(\br),
\end{equation}
since $\lapl{r^{-1}} = -4\pi \delta(\br)$.  This means that, for uniform expansion in compressible suspensions, the particles act as fluid sources, each with a strength proportional to its pressure moment.  In a periodic system, the velocity disturbance corresponds to an array of sources are obtained by replacing the delta function in Eq.~\eqref{eq:fluid-source} with $\sum_{\bp}\delta(\br-\bR_{\bp})$.  From Fourier transform, the solution is
\begin{equation}
  \label{eq:wave-dist-S}
  \vect{Q}(\br) = \frac{1}{4\eta_0 V} \sum_{\bk\neq 0}  \grad{}  
\frac{1}{k^{2}} e^{-\imath \bk\cdot \br}.
\end{equation}
The above wave-space sum can be split to two exponentially converging series~\cite{stokes-fundamental-sol_hasimoto_jfm1959, Tornberg_periodic-laplace_jcompp2011}
\begin{equation}
  \label{eq:source-split}
  \sum_{\bk\neq 0}  \frac{1}{k^{2}} e^{-\imath \bk\cdot \br}
= \frac{V}{4\pi} \sum_{\bp\neq 0} \frac{1}{r}\erfc(r\xi) 
+ \sum_{\bk\neq 0}\frac{1}{k^2} e^{-\frac{1}{4}k^2\xi^{-2}}e^{-\imath 
\bk\cdot\br}.
\end{equation}
Similar to $\vect{Q}(\br)$, the pressure fundamental solution $\vect{P}(\br)$ in Eq.~\eqref{eq:pres-funda} can also be extended to periodic systems.

\section{The mobility computation}
\label{sec:mobility-computation}

The mobility problem seeks the action of the grand mobility tensor $\fM$ on the force moments such as $\cF^\rH$ and $\tens{S}^\rH$.  It can be constructed in a pairwise fashion using the formalism in Sec.~\ref{sec:stok-susp} for compressible suspensions.  Na\"{i}vely, this is an $\bigO(N^2)$ operation for an $N$-particle system since the long-range HIs necessitate considerations of all particle pairs.  However, with the Ewald summation that splits the fundamental solutions $\tJ(\br)$, $\vect{Q}(\br)$, and $\vect{P}(\br)$ into exponentially fast converging wave-space and real-space series, the particle mesh techniques can improve the computation scaling to $\bigO(N\log N)$.  In the following, our implementation of the mobility computation is discussed.

\subsection{Wave-space computation: the Spectral Ewald (SE) method}
\label{sec:spectr-ewald-meth}

The wave-space computation concerns the part of grand mobility tensor associated with $\tJ_W(\br)$ of Eq.~\eqref{eq:oseen-ewald-ws} and the wave-space sum of Eq.~\eqref{eq:source-split} in $\vect{P}(\br)$ and $\vect{Q}(\br)$.  Using the Fast Fourier Transform (FFT) algorithm, the computation cost can be reduced to $\bigO(N\log N)$.  To illustrate this, let us consider the wave-space velocity disturbance $\bU_i^W$ on particle $i$ at the Rotne-Prager level, obtained by combining Eqs.~\eqref{eq:faxen-f}, \eqref{eq:mul-expand}, and \eqref{eq:oseen-ewald-ws}, \ie,
\begin{equation}
  \label{eq:rp-wave}
  \bU_i^W =  \frac{1}{\eta_0 V}\sum_{\bk\neq 0} e^{- \imath\bk\cdot \br_i} 
\left(1-\tfrac{1}{6}a_i^2k^2\right) \tens{g}_1(\bk) \cdot \sum_j 
\left(1-\tfrac{1}{6}a_j^2k^2\right) e^{\imath \bk\cdot\br_j} \bF_j^\rH,
\end{equation}
and the wave-space kernel
\begin{equation}
  \label{eq:low-pass-filter}
  \tens{g}_1(\bk) = \left(1+\tfrac{1}{4}\kxp{2}
\right) k^{-4} e^{-\tfrac{1}{4}k^2 \xi^{-2}} (\tI k^2 - \bk\bk).
\end{equation}
Different from Eq.~\eqref{eq:mul-expand}, the summation over particle $j$ in Eq.~\eqref{eq:rp-wave} is unrestricted and includes the case of $i=j$.  Therefore, the self interaction term for $i=j$, which is
\begin{equation}
  \label{eq:uf-self}
  \frac{\xi(9-10a_i^2\xi^2+7a_i^4\xi^4)}{18\eta_0\pi^{3/2}}\bF_i^\rH,
\end{equation}
should be removed later.  Eq.~\eqref{eq:rp-wave} exposes the basic idea behind many particle mesh techniques including the PME method and the SPME method.  From an inverse Fourier transform, the real-space force distribution corresponding to the summation over $j$ in Eq.~\eqref{eq:rp-wave} is
\begin{equation}
  \label{eq:real-force-distro}
  \sum_j (1+\tfrac{1}{6}a_j^2 \lapl{}) \bF_j^\rH \delta(\br-\br_j).
\end{equation}
The force distribution in Eq.~\eqref{eq:real-force-distro} is assigned to a regular spatial grid by approximating the delta functions by Lagrangian polynomials in the PME method~\cite{Petersen_pme-accuracy_jcp1995} or Cardinal B-splines in the SPME method~\cite{spme-method_essmann_jcp1995}.  The interpolated forces are then transformed to the wave space by FFT and the wave-space computation in Eq.~\eqref{eq:rp-wave} is performed.  The wave-space results is then brought back to the real space by inverse FFTs.  Subsequently, the velocity on each particle, $\bU^W_i$, is interpolated back from the grid, preferably using the same interpolation scheme for the force assignment~\cite{Deserno_mesh-up-ewald-pt1_jcp1998}.
Here, the action of the mobility tensor on the force $\bF^\rH$, rather than the tensor itself, is computed.  The kernel $\tens{g}_1(\bk)$ in Eq.~\eqref{eq:low-pass-filter} is effectively a low-pass filter that cuts off the spatial signals at high $k$.  Computationally, for $M^3$ grid points the FFT scales as $\bigO(M^3\log M^3)$.  To ensure reasonable accuracy, $M^3\propto N$, and the wave-space computation scales as $\bigO(N\log N)$.

There are two sources of error affecting the accuracy of particle mesh techniques.  The first is associated with the truncation of the wave-space sum ($k$-summation) in Eq.~\eqref{eq:rp-wave}.  This is only affected by the number of grid points $M$ in the simulation box.  The second error is the interpolation error, and arises from polynomial approximation of the $\delta$-functions in Eq.~\eqref{eq:real-force-distro}.  For a simulation box of size $L$, this error scales as $(L/M)^p$, where $p$ is the polynomial order of the approximation scheme.  Since both errors are associated with $M$, we cannot separate the two error sources. Consequently, to maintain a satisfactory overall accuracy, a large $M$ is often used in the wave-space computations to keep the interpolation error small, resulting in unnecessary FFT computations.

In addition, for polydisperse suspensions, different particle sizes introduce additional complications to traditional particle mesh techniques. If the Laplacian in Eq.~\eqref{eq:real-force-distro} is computed in the real space in the SPME method, the interpolation error increases to $(L/M)^{p-2}$, which further increases the $M$ requirement.  For the PME method, real-space differentiation is unsuitable due to the discontinuity of Lagrangian polynomials, and all the computations have to be carried out in the wave space.  This significantly increases the total number of FFTs.  In addition, different particle sizes increase the complexity in the algorithm implementation.  Therefore, a simple method with flexible error control is crucial for accurate and efficient wave-space computation in polydisperse systems.

To address these concerns, we use a new particle mesh technique, the Spectral Ewald (SE) method~\cite{Tornberg_se-stokes_jcompp2010,Tornberg_periodic-laplace_jcompp2011,sw-stokesian_tornberg_ijnmf_2014} for the wave-space mobility computation.  The SE method decouples the $k$-space truncation and interpolation errors, and is accurate, efficient, and flexible for polydisperse systems.  To show this, we use Eq.~\eqref{eq:rp-wave} again as an example and consider the general case of non-orthogonal lattice vectors.  We first introduce the fractional coordinate $\bt = (t_1, t_2, t_3)^\dagger \in [0,1)^3$.  For each point $\br$ in the simulation box, $\br = t_1 {\bold a}_1 +t_2 {\bold a}_2 + t_3 {\bold a}_3 = \tens{A} \cdot \bt$.  Accordingly, defining $\bq=(q_1, q_2, q_3)^\dagger$ such that $\bk = q_1 {\bold b}_1 +q_2 {\bold b}_2 + q_3 {\bold b}_3 = \tens{B}\cdot \bq$, $\exp(\imath \bk\cdot \br) = \exp(2\pi \imath\bq \cdot \bt)$, and $k^2 = \bq^\dagger\cdot\tens{B}^\dagger\cdot \tens{B} \cdot \bq$. Eq.~\eqref{eq:rp-wave} is rewritten in $\bt$ and $\bq$ as
\begin{align}
  \label{eq:rp-wave-tq}
    \bU_i^W = &
  \frac{1}{\eta_0 V}\sum_{\bq\neq 0} e^{- 2\pi \imath\bq\cdot 
\bt_i-\frac{1}{8}\theta q^2 \xi^{-2}}  \left(1-\tfrac{1}{6}a_i^2 
\bq^\dagger\cdot\tens{B}^\dagger\cdot \tens{B}  \cdot \bq  \right)  
e^{\frac{1}{4}\theta q^2 \xi^{-2}} \tens{g}_1(\tens{B} \cdot \bq)  \nonumber \\ 
& \cdot\sum_j \left(1-\tfrac{1}{6}a_j^2 \bq^\dagger\cdot\tens{B}^\dagger\cdot 
\tens{B} \cdot \bq \right) e^{2\pi \imath \bq \cdot\bt_j-\frac{1}{8}\theta q^2 
\xi^{-2}} \bF_j^\rH,
\end{align}
with two $e^{-\frac{1}{8}\theta q^2 \xi^{-2}}$ multiplied after particle positions and one $e^{\frac{1}{4}\theta q^2 \xi^{-2}}$ before $\tens{g}_1$, and $\theta$ is a parameter.  Introducing the Fourier transform pair
\begin{equation}
  \label{eq:fourier-transform}
  \hat{f}_{\bq} = \int \dd \bt f(\bt) e^{2\pi\imath \bq \cdot \bt } \text{ and 
}   f(\bt) = \int \dd \bq \hat{f}_{\bq} e^{-2\pi\imath \bq \cdot \bt },
\end{equation}
the basic idea of SE is to note that 
\begin{equation}
  \label{eq:shape}
h(\bt) =  \int \dd \bq e^{- 2\pi \imath \bq \cdot \bt-\frac{1}{8}\theta q^2 
\xi^{-2}}  = \left(\frac{8\pi\xi^2}{\theta}\right)^{\frac{3}{2}} 
\exp\left(-\frac{8\pi^2\xi^2}{\theta} \|\bt\|_*^2\right),
\end{equation}
\ie, the the Fourier transform of a Gaussian remains a Gaussian, and the shape of the Gaussian is controlled by $\theta$.  Here, $\|\cdot\|_*$ indicates distance computation using the minimum image convention for periodic systems.  The inverse Fourier transform of the second line of Eq.~\eqref{eq:rp-wave-tq} with respect to $\bq$ is
\begin{equation}
  \label{eq:grid-space}
  \vect{H}(\bt) = \sum_j \left(
1+\tfrac{1}{24} a_j^2 \pi^{-2}  \grad{}_t^\dagger \cdot \tens{B}^\dagger\cdot 
\tens{B} \cdot \grad{}_t
\right) h\big|_{(\bt-\bt_j)} \bF_j^\rH,
\end{equation}
where $\grad{}_t = (\partial/\partial t_1, \partial/\partial t_2, \partial/\partial t_3)^{\dagger}$.  Eq.~\eqref{eq:grid-space} facilitates interpolation of a discrete force distribution onto a uniform grid of coordinate $\bt$ via the Gaussian shape function $h(\bt)$ in Eq.~\eqref{eq:shape}.  The effect of particle size is automatically incorporated in the grid assignment scheme in the real space.  After converting the real-space $\vect{H}(\bt)$ to the wave-space $\hat{\vect{H}}_{\bq}$ using FFTs, the wave-space computation produces
\begin{equation}
  \label{eq:wave-result}
  \hat{\vect{G}}_{\bq} = 
  \begin{cases}
e^{\frac{1}{4}\theta q^2 \xi^{-2}} \tens{g}_1(\tens{B}\cdot \bq) \cdot 
\hat{\vect{H}}_{\bq}, &  \bq\neq 0 \\
    0 & \text{otherwise.}
  \end{cases}
\end{equation}
From Parseval's theorem,
\begin{equation}
  \label{eq:parseval}
  \int_\mathrm{T} \dd \bt f(\bt) g^*(\bt) = \sum_{\bq} \hat{f}_{\bq} 
\hat{g}^*_{\bq},
\end{equation}
where  $\mathrm{T}$ is a periodic lattice and ${(\cdot)}^*$ indicates complex conjugation, Eq.~\eqref{eq:rp-wave-tq} becomes a convolution integral with the Gaussian shape function,
\begin{equation}
  \label{eq:parseval-sum}
\bU^W_i = \frac{1}{\eta_0 V}\int_{\mathrm{T}}\dd\bt
\vect{G}(\bt) 
\left(1+ \tfrac{1}{24} a_i^2 \pi^{-2}
 \grad{}_t^\dagger \cdot \tens{B}^\dagger\cdot \tens{B} \cdot \grad{}_t
\right) h\big|_{(\bt-\bt_i)},
\end{equation}
where $\vect{G}(\bt)$ is the inverse Fourier transform of $\hat{\vect{G}}_{\bq}$.  Extending the SE method to couplings beyond Rotne-Prager level is straightforward, with adjusted $\vect{H}(\bt)$ and $\vect{G}(\bt)$ based on the Fax\'{e}n laws and multipole expansions in Sec.~\ref{sec:stok-susp}.  In this work, we have implemented the mobility computation to the stresslet and the strain rate level.

Unlike other particle mesh techniques, the SE formulation in Eqs.~\eqref{eq:rp-wave-tq}--\eqref{eq:parseval-sum} is exact and therefore the errors are entirely from the numerical implementations.  Since the FFT algorithm is accurate to machine precision, the sources of error include the discretization and truncation of the shape function [Eq.~\eqref{eq:shape}], and the numerical integration in Eq.~\eqref{eq:parseval-sum}.  Practically, the evaluation of each shape function is limited to $P^3$ points ($P\leq M$) around the particle.  Due to the exponential decay of $h(\bt)$, the truncation error decreases exponentially with increasing $P$.  Meanwhile, the integral in Eq.~\eqref{eq:parseval-sum} is evaluated using trapezoidal quadrature~\cite{Tornberg_se-stokes_jcompp2010, Tornberg_periodic-laplace_jcompp2011}, which also exhibits exponential error decay with increasing $P$.  Therefore, the interpolation error in SE method depends exclusively on $P$ for sufficiently large $M$, and can be separately controlled from the $k$-space truncation error.  The rapid, exponential error decay is known as spectral accuracy~\cite{Tornberg_se-stokes_jcompp2010, Tornberg_periodic-laplace_jcompp2011}, and this is the namesake of the SE method.

The computation cost of the SE method also becomes apparent with the truncation of $h(\bt)$.  The grid assignment in Eq.~\eqref{eq:grid-space} and the convolution Eq.~\eqref{eq:parseval-sum} are $\bigO(NP^3)$ for an $N$-particle system, and the FFTs to and from the wave space are $\bigO[M^3\log (M^3)]$.  With $M^3\propto N$, the time limiting step is the FFT, and the SE method also scales as $\bigO(N\log N)$ as other particle mesh techniques.

The Gaussian shape in $h(\bt)$ of Eq.~\eqref{eq:shape} is controlled by $\theta$, which is parameterized as
\begin{equation}
  \label{eq:theta-param}
  \theta = \left( \frac{2\pi P \xi}{M m}\right)^2,
\end{equation}
on a regular grid of $M^3$ points with $P^3$ points for each shape function evaluation.  The shape parameter $m$ in Eq.~\eqref{eq:theta-param} ensures that at the edge of $h(\bt)$ evaluation, \ie, $t^2 = P^2/(2M)^2$, $h\propto e^{-m^2/2}$.  Therefore, with fixed $M$ and $P$, $m$ describes the truncation of $h(\bt)$ on the discretized grid and is consistent with the original SE method of Lindbo \& Tornberg~\cite{Tornberg_se-stokes_jcompp2010, Tornberg_periodic-laplace_jcompp2011}.

The computation efficiency of the SE method relies on rapidly computing the $\bigO(NP^3)$ different Gaussian shape functions $h(\bt)$, which involves expensive exponential evaluations.  To reduce these expensive operations, Lindbo \& Tornberg~\cite{Tornberg_se-stokes_jcompp2010,Tornberg_periodic-laplace_jcompp2011} introduced the fast Gaussian gridding (FGG) technique~\cite{Greengard_accel-fft_SIAMRev2004} to the SE method.  In essence, the FGG technique evaluates the exponential function on a regular grid as 
\begin{equation}
  \label{eq:fgg-exp}
  e^{-\alpha (\delta t + i\Delta t)^2} = e^{-\alpha (\delta t)^2 } \times 
\left( e^{-2\alpha \delta t\Delta t}  \right)^i \times \left[e^{-\alpha (\Delta 
t)^2}\right]^{i^2},
\end{equation}
where $\alpha$ is a constant, $\delta t$ is the off-grid value, $\Delta t$ is the spacing of the regular grid, and $i$ is an integer within the range $[-P/2, P/2]$.  It reduces the $P$ exponential evaluations in each direction in the SE method to $3$ exponential computations and at most $2P$ multiplications.  In addition, the last term of Eq.~\eqref{eq:fgg-exp} is independent of $\delta t$, and therefore only needs to be computed once.

\subsection{Wave-space computation: the particle size effect}
\label{sec:wave-space-comp}

In Sec.~\ref{sec:spectr-ewald-meth} the terms associated with finite particle sizes in the Fax\'{e}n laws and the multipole expansions are incorporated in the real-space derivatives of the shape function $h(\bt)$.  For example, in a simple shear flow with lattice vectors ${\bold a}_1 = (L,0,0)$, ${\bold a}_2 = (\gamma L,L,0)$, and ${\bold a}_3 = (0,0,L)$, where $\gamma$ is the strain, the relevant term in Eqs.~\eqref{eq:grid-space} and~\eqref{eq:parseval-sum} is
\begin{align}
\label{eq:non-ortho-shape}
& \left(\tfrac{1}{24} a_i^2 \pi^{-2}
 \grad{}_t^\dagger \cdot \tens{B}^\dagger\cdot \tens{B} \cdot \grad{}_t
\right)
 h(\bt) \nonumber =\\  
& \tfrac{8}{3}\left(\frac{\pi\xi a_i}{\theta L}\right)^2\left\{
-\theta(3+\gamma^2) + 16\pi^2\xi^2[(1+\gamma^2)t_1^2 + t_2^2 + t_3^2 - 2 \gamma 
t_1 t_2]
\right\} h(\bt).
\end{align}
The finite particle sizes introduce additional features to the shape function, and for non-orthogonal simulation boxes, non-trivial anisotropy.  As a result, compared to the case of point forces, more points $P$ are needed to resolve the details in Eq.~\eqref{eq:non-ortho-shape}.  On the other hand, the benefit of evaluating the particle size effects in the real space is that fewer FFTs are involved.  To compute the mobility problem of compressible suspensions to the stresslet and the strain rate levels, only four pairs of FFTs are necessary: three are associated with $\tJ_W$ in Eq.~\eqref{eq:oseen-ewald-ws}, and one associated with the $\vect{Q}$ in Eq.~\eqref{eq:pres-vel-disturb}.

Alternatively, the particle size effect can be completely accounted in the wave space.  This requires, for each particle $j$, $\bF_j^\rH$, $\bT_j^\rH$, and $\tens{S}_j^\rH$, as well as $a_j^2 \bF_j^\rH$ and $a_j^2\tens{S}_j^\rH$, to be separately interpolated to the grid via $h(\bt)$ and brought to the wave space for computation.  The derivatives associated with the Fax\'{e}n laws and multipole expansions in Sec.~\ref{sec:stok-susp} are carried out in the wave space as multiplication of wave vectors.  The final results are then combined from different convolutions and weighted by the particle sizes.  To demonstrate this, we again take the wave-space Rotne-Prager velocity, Eq.~\eqref{eq:rp-wave-tq}, as an example.  In this approach, the grid assignment is split into two parts,
\begin{equation}
  \label{eq:wave-only-f-assign}
  \vect{H}'(\bt) = \sum_j h(\bt-\bt_j)\bF_j^\rH \text{ and }   \vect{H}''(t) = 
\sum_j h(\bt-\bt_j) a_j^2 \bF_j^\rH.
\end{equation}
The wave-space computation for $\bq \neq 0$ is also split as
\begin{align}
  \label{eq:wave-only-compute-1}
    \hat{\vect{G}}'_{\bq} = & e^{\frac{1}{4}\theta q^2 \xi^{-2}} 
\tens{g}_1(\tens{B}\cdot \bq) \cdot 
\left[\hat{\vect{H}}'_{\bq} - (\tfrac{1}{6} \bq^\dagger \cdot \tens{B}^\dagger 
\cdot \tens{B}\cdot \bq) \hat{\vect{H}}''_{\bq}\right], \\
  \label{eq:wave-only-compute-2}
    \hat{\vect{G}}''_{\bq} = &  (-\tfrac{1}{6} \bq^\dagger \cdot 
\tens{B}^\dagger \cdot \tens{B}\cdot \bq)
 e^{\frac{1}{4}\theta q^2 \xi^{-2}} \tens{g}_1(\tens{B}\cdot \bq) \cdot
\left[\hat{\vect{H}}'_{\bq} - (\tfrac{1}{6} \bq^\dagger \cdot \tens{B}^\dagger 
\cdot \tens{B}\cdot \bq) \hat{\vect{H}}''_{\bq}\right],
\end{align}
and $\hat{\vect{G}}'_{\bq} =\hat{\vect{G}}''_{\bq} =0$ when $\bq = 0$.  The wave-space velocity disturbance is a sum of two convolutions,
\begin{equation}
  \label{eq:parseval-sum-2}
\bU^W_i = \frac{1}{\eta_0 V}\int_{\mathrm{T}}\dd\bt
\vect{G}'(\bt) h(\bt-\bt_i) + \frac{a_i^2}{\eta_0 V}\int_{\mathrm{T}}\dd\bt
\vect{G}''(\bt) h(\bt-\bt_i).
\end{equation}
Note that the convolution associated with $\vect{G}''(\bt)$ is weighted by the particle size $a_i$.  Compared to the other approach, the wave-space computation is rather straightforward for the force interpolation and convolution.  With the same $P$, the accuracy is expected to be higher as the derivatives are calculated in the wave space~\cite{Deserno_mesh-up-ewald-pt1_jcp1998}.  However, the computation burden is shifted to the FFTs: for the mobility problem to the $\tS$ and $\tE$ level, a total of 20 pairs of FFTs are necessary: 12 for $\bF_j^\rH$, $\bT_j^\rH$, and $\tS_j^\rH$, three for $a_j^2\bF_j^\rH$, and five for the traceless part of $a_j^2\tens{S}_j^\rH$.

A third approach, a hybridization between the wave- and the real-space approaches above, aims to reduce the errors associated with the high order derivatives of $h(\bt)$ in the real space.  It retains the real-space derivatives in the force interpolation step, but when evaluating the Fax\'{e}n laws, the second order derivatives are computed in the wave space for improved accuracy.  The first order derivatives are computed in the real space to keep the total number of FFTs low.  As a result, this hybrid approach requires $12$ FFTs: four to the wave space and eight from the wave space.  Taking Eq.~\eqref{eq:rp-wave-tq} again for example, the most significant error in Sec.~\ref{sec:spectr-ewald-meth} is due to applying the operator $(\grad{}_t^\dagger \cdot \tB^\dagger \cdot \tB \cdot \grad{}_t)$ twice to $h(\bt)$, once during the force interpolation, and another time during the convolution.  The hybrid approach retains the real-space grid assignment using $\vect{H}(\bt)$ in Eq.~\eqref{eq:grid-space}, but evaluates the convolution using Eq.~\eqref{eq:parseval-sum-2} with modified $\hat{\vect{G}}'(\bt)$ and $\hat{\vect{G}}''(\bt)$:  in the wave-space computations, the content in the square bracket on the right hand side of Eqs.~\eqref{eq:wave-only-compute-1} and \eqref{eq:wave-only-compute-2} is replaced with $\hat{\vect{H}}_{\bq}$ in Eq.~\eqref{eq:grid-space}.  We adopted this hybrid approach in this work to compute the HIs, and discuss the accuracy of various approaches in Sec.~\ref{sec:impl-accur}.

\subsection{Real-space computation}
\label{sec:real-space-comp}

The real-space contributions to the grand mobility tensor $\fM$ are computed pairwise using the formalism in Sec.~\ref{sec:stok-susp}.  Since $\tJ_R(\br)$ [Eq.~\eqref{eq:oseen-ewald-rs}] decays exponentially fast with distance, when the parameter $\xi$ is sufficiently large, only particle pairs within a cutoff distance $r_c$ need to be evaluated.  Introducing the cutoff radius $r_c$ for pair evaluation allows fast neighbor searching algorithms such as the linked list~\cite{CompSimLiq} or the chaining mesh~\cite{simulation_particles} method to be used.  These methods divide the simulation box into cells of size slightly larger than $r_c$, and sort the particles into the cells.  To find the neighbors of a particle, only particles in the residing cell and its 26 neighboring cells need to be searched.  This effectively improves the operation count to $\bigO(N\log N)$ for the real-space computations.

To accommodate the iterative scheme for HI computations in Sec.~\ref{sec:stokesian-dynamics}, the real-space grand mobility tensor is constructed as a sparse matrix at each time step.  After the matrix construction, the action of the real-space contributions to $\fM$ is simply a matrix-vector multiplication.  Otherwise, neighbor searching and pair HI evaluations need to be carried out at every iteration.  Note that we also include the self-contributions from the wave-space computations, \eg, Eq.~\eqref{eq:uf-self}, and the self-part of the pressure Fax\'{e}n law [Eq.~\eqref{eq:faxen-p}], in the real-space grand mobility tensor.

\subsection{GPGPU acceleration of the mobility computation}
\label{sec:gpu-accel-mobil}

The mobility computation with the SE method was first implemented on CPU and the performance was unsatisfactory for dynamic simulations.  The bottlenecks are the force interpolation step and the convolution step.  These are common speed limiting steps in particle mesh techniques due to ineffective memory caching between the particle and the grid data.  For polydisperse systems in this work, the situation is aggravated as more interpolation points $P$ are needed for satisfactory HI resolution.  After a few optimization iterations on CPU, we realized that the key to the performance is the memory bandwidths.  Since modern GPUs typically have significantly higher memory bandwidths compared to CPUs, in this work the entire mobility computation is carried out on GPU using CUDA C, a popular GPGPU programming model with a relatively mature environment for scientific computations.

The GPU mobility computations are carried out in Single Precision (SP) for the highest GPU performance.  The cost of the performance in SP computation is the accuracy, as the SP arithmetics can be severely limited by the number of significant digits compared to the Double Precision (DP).  However, this is not a problem in this work for at least three reasons: (\textit{i}) For dynamic simulations with iterative solvers, the SP accuracy is often sufficient; (\textit{ii}) The SE method is able to reach the round-off error of the SP arithmetics with proper parameter selection due to its spectral accuracy; and (\textit{iii}) The far-field HIs captured by the mobility computations are smooth compared to the lubrication interactions, which are evaluated in DP on CPUs.  The split of the near- and far-field HIs in SD allows a natural mixed precision HI computation that captures the most significant contributions from each part.

The GPGPU computations exploit the massively parallel structure of modern GPUs by simultaneously executing a large number of similar tasks, or threads, on the data.  To maintain performance, data dependencies and communications between threads should be minimized.  This makes the GPU implementation of the SE method different from its CPU counterpart.  Inspired by earlier GPU implementations of particle mesh techniques, this work combines the grid-based method of Ganesan~\etal~\cite{GPU-Fenzi_ieee2011} for force interpolation and the particle-based approach of Harvey \& De~Fabritiis~\cite{Harvey_spem-gpu_jctc2009} for convolution.  The grid-based force interpolation keeps a list of contributing particles for each grid point, and the list is updated when the particle configurations are changed.  The grid values are computed in parallel using $M^3$ threads: with the particle list, each thread sums the force, torque, and stresslet contributions independently for each grid point.  On the other hand, the particle-based convolution is a weighted summation on $P^3$ grid points for each particle.  To maximize parallelization, the summation for each particle is performed by a group of $P$ threads cooperatively.  Each thread in the group first sums $P^2$ grid points on the transverse plane, and for the final result, the first thread in the group adds up the values from other threads using the shared memory of the GPU.  Moreover, on the GPU we use the \texttt{cufft} package for the FFTs and the \texttt{cusparse} package for the sparse matrix-vector multiplication.

\section{Dynamic simulation with Stokesian Dynamics}
\label{sec:stokesian-dynamics}

The framework of SD~\cite{sd_durlofsky_jfm1987,brady-sd-ew_jfm_88} approximates the projected grand resistance tensor~$\cR$ in Eq.~\eqref{eq:res-part} as
\begin{equation}
  \label{eq:sd-orig}
  \cR = \fM^{-1} + \cR^{\nf},
\end{equation}
where $\fM$ is the multipole grand mobility tensor, and $\cR^{\nf} $ is the pairwise additive lubrication correction without the far-field contributions.  Recall that the inversion of $\fM$ captures the many-body aspect of HIs, and the short-range correction $\cR^{\nf}$ captures the lubrication effects.  The SD recovers the exact result for two-body problems and agrees well with the exact solutions of three-body problems~\cite{three-spheres_wilson_jcp2013}.  It can provide significant insights to the HIs of dense suspensions~\cite{sd-brownian-susp_brady_jfm2000, rheo-non-colloidal-suspension_brady_jor02}.

\subsection{Iterative computation of hydrodynamic interactions}
\label{sec:iter-comp}

We incorporate the SE mobility computation into the framework of SD using the iterative scheme of Swan \& Brady~\cite{confined-suspension_swan_2011}, and call the resulting method the Spectral Ewald Accelerated Stokesian Dynamics (SEASD).  Here, a matrix-free iterative scheme is necessary as the grand mobility tensor $\fM$ is not explicitly constructed.  The iterative scheme splits the overall hydrodynamic force,
\begin{equation}
  \label{eq:fh-expr}
  \cF^{\mathrm{H}} = -\Rfu\cdot \cU^{\mathrm{H}} + \Rfe \cdot \tE^\infty,
\end{equation}
where $\cU^{\mathrm{H}}$ is the velocity disturbances due to HIs, into a near-field part and a far-field part.  The near-field part satisfies 
\begin{equation}
  \label{eq:HI-nf}
    0 = -\Rfu^{\nf}\cdot \cU^{\mathrm{H}} +  \cF^{\mathrm{H},\ff} + 
\widetilde{\cF}^{\mathrm{P}},
\end{equation}
where $\Rfu^{\nf}$ is the $\cF\cU$ coupling in $\cR^\nf$ and is stored as a sparse matrix, $\widetilde{\cF}^{\mathrm{P}}=\cF^{\mathrm{P}}  +\Rfe^\nf\cdot \tE^\infty$ contains the interparticle force $\cF^{\mathrm{P}}$ and the near-field contributions from $\tE^\infty$. The far-field hydrodynamic force $\cF^{\mathrm{H},\ff}$ satisfies
\begin{equation}
  \label{eq:HI-ff}
  \begin{bmatrix}
    \cU^{\mathrm{H}}\\
    -\tE^\infty
  \end{bmatrix}
= -\fM \cdot 
  \begin{bmatrix}
    \cF^{\mathrm{H},\ff}\\
    \tS^{\mathrm{H},\ff}
  \end{bmatrix},
\end{equation}
where $\tS^{\mathrm{H},\ff}$ is the far-field stresslet from HIs.  Solving Eqs.~\eqref{eq:HI-nf} and \eqref{eq:HI-ff}, the far-field hydrodynamic forces and stresslets are
\begin{equation}
  \label{eq:HI-solve}
  \begin{bmatrix}
    \widetilde{\cF}^{\mathrm{H},\ff}\\
    \tS^{\mathrm{H},\ff}
  \end{bmatrix}
 = 
\widetilde{\fM}^{-1} 
\cdot
\left(
(\lambda \fM -{\cal I}) \cdot
    \begin{bmatrix}
      (\widetilde{\bR}_{\cF\cU}^{\nf})^{-1}\cdot \widetilde{\cF}^{\mathrm{P}} \\
      \bzero
    \end{bmatrix}
    +
    \begin{bmatrix}
      \bzero \\
      \tE^\infty
    \end{bmatrix}
\right),
\end{equation}
where
\begin{equation}
  \label{eq:fM-comp}
\widetilde{\fM}= \left(
    ({\cal I} - \lambda \fM)\cdot
    \begin{bmatrix}
      (\widetilde{\bR}_{\cF\cU}^{\nf})^{-1} & \bzero \\
      \bzero & \bzero
    \end{bmatrix}
 + \fM
\right).
\end{equation}
To ensure invertibility, a diagonal matrix $\lambda \bI$, with $\lambda$ a parameter, is added to $\Rfu^{\nf}$, \ie, $\widetilde{\bR}_{\cF\cU}^{\nf} = \Rfu^{\nf} + \lambda \bI$, and accordingly $\widetilde{\cF}^{\mathrm{H},\ff} = \cF^{\mathrm{H},\ff} + \lambda\cU^{\mathrm{H}}$. A convenient choice for $\lambda$ is $6\pi\eta_0 a$, where $a$ is the reference particle radius~\cite{confined-suspension_swan_2011}.

Solving Eq.~\eqref{eq:HI-solve} requires nested iteration as each evaluation of $\widetilde{\fM}$ contains the solution of the near-field problem with $\widetilde{\bR}_{\cF\cU}^{\nf}$.  The near-field problem is efficiently solved by the Generalized Minimum Residual (GMRES) method with an Incomplete Cholesky preconditioner with zero fill-in (IC0)~\cite{iterative-book}.  To reduce the IC0 breakdown, prior to applying the preconditioner particles are reordered using the reverse Cuthill-McKee algorithm.  For isotropic suspensions, the near-field problem typically converges to an error of $10^{-4}$ within $10$ iterations~\cite{asd_sierou_jfm01}.  For suspensions with strong structural anisotropy, however, the convergence becomes more difficult and the IC0 preconditioner breaks down even with the reordering.  This is resolved by increasing $\lambda$ in $\widetilde{\bR}_{\cF\cU}^{\nf}$, or introducing a threshold value $\lambda_{\mathrm{IC}}$ in during the IC0 preconditioner computation~\cite{iterative-book}.  Increasing $\lambda$ in $\widetilde{\bR}_{\cF\cU}^{\nf}$ does not change the convergence of the near-field problem, but increases the number of expensive $\widetilde{\fM}$ iterations.  On the other hand, increasing $\lambda_{\mathrm{IC}}$ deteriorates the quality of the IC0 preconditioner and increases the iterations required for the near-field problem, but has little effect on the far-field evaluations.  In dynamic simulations, both $\lambda$ and $\lambda_{\mathrm{IC}}$ are adjusted for optimal computation efficiency.

The pressure moment computation in SEASD also follows the near- and far-field splitting scheme in Eqs.~\eqref{eq:HI-nf} and~\eqref{eq:HI-ff}.  Due to the special coupling between the pressure moments and other force moments in compressible suspensions (Sec.~\ref{sec:extens-compr-fluid}), the interaction contribution to the far-field pressure moment is evaluated after ${\bF}^{\mathrm{H},\ff}$ and the traceless part of ${\tS^{\mathrm{H},\ff}}$ are solved in Eq.~\eqref{eq:HI-solve}.  On the other hand, the near-field part of the pressure moment is evaluated along with other parts of the stresslets using the near-field resistance functions.

The near-field pairwise lubrication corrections $\cR^{\mathrm{nf}}$ are based on the exact solutions of two-body problems in series form~\cite{resist-func_jeffrey_jfm1984, resist-func_jeffrey_pof1992, pres-moment_jeffrey_pof1993, compres-res_khair_pof2006} up to $s^{-300}$, where $s = 2r/(a_i+a_j)$, with  $a_i$ and $a_j$ the radii of the pair, is the scaled particle center-center distance.  In the simulations, the lubrication corrections are activated when $s<4$: for $s>2.1$ the interpolation of tabulated data and for $s\leq 2.1$ the analytical expressions are used.  Note that $\cR^{\mathrm{nf}}$ constructed from two-body problems contains both the relative and the collective motions of the particle pair and,  as pointed out by Cichocki~\etal~\cite{self-diffusion-lubrication-3body_cichocki_jcp1999}, the lubrication corrections corresponding to the collective motion can destroy the far-field asymptotics beyond the pair level.  However, for dense suspensions, this only leads to a minor quantitative difference on the suspension static properties~\cite{sd-bidisperse_wang_jcp2015} in conventional SD.  Therefore, we retain the full lubrication correction here for consistency with the existing SD framework.  The SD implementations of Ando \& Skolnick~\cite{crowding-hydro-cell_skolnick_pnas2010} removed the pair collective motion in the lubrication corrections.

\subsection{Far-field preconditioner}
\label{sec:far-field-prec}

\begin{figure}
  \centering
  \includegraphics[width=3.5in]{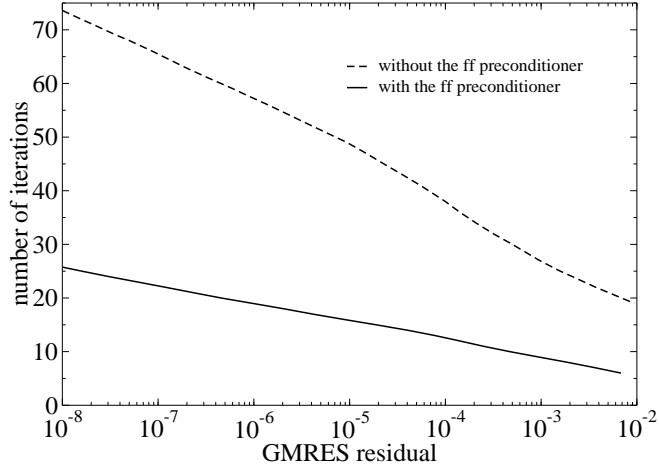}
  \caption{The number of far-field iterations, \ie, the number of the grand mobility tensor $\widetilde{\fM}$ evaluations, as a function of the GMRES residual with (solid line) and without (dashed line) the far-field preconditioner for a bidisperse suspension of $N=200$, $\lambda=2$, $x_2 = 0.3$, and $\phi=0.2$.
}
  \label{fig:pc_res}
\end{figure}

Here we introduce a preconditioner for $\widetilde{\fM}$ to reduce the number of expensive far-field mobility evaluations when solving Eq.~\eqref{eq:HI-solve}.  Since $\widetilde{\fM}$ is not explicitly constructed, the preconditioner needs to be built from a suitable approximation.  For mobility problems without the lubrication corrections, Saintillan~\etal~\cite{Saintillan_spme-stokes-fiber_pof2005} and Keaveny~\cite{Keaveny_fluct-fcm_jcompp2014} found substantial iteration improvement even with the diagonal mobility approximation.  Unfortunately, the approximation of $\widetilde{\fM}$ is more involved due to the presence of $(\widetilde{\bR}_{\cF\cU}^{\nf})^{-1}$.  In this work, a block diagonal approximation of $\widetilde{\fM}$ for the far-field preconditioner is adopted.  First, the near-field resistance tensor $\widetilde{\bR}_{\cF\cU}^{\nf}$ is approximated by $N$ blocks of $6\times 6$ submatrices along its diagonal.  Using the direct sum notation, this is $\bigoplus_{i=1}^N(\widetilde{\bR}_{\cF\cU}^{\nf})_{ii}$, where $\bigoplus$ is the direct sum, and $(\widetilde{\bR}_{\cF\cU}^{\nf})_{ij}$ is the block submatrix between particles $i$ and $j$ in $\widetilde{\bR}_{\cF\cU}^{\nf}$.  To approximate $\widetilde{\fM}$,  we use 
\begin{equation}
  \label{eq:rfu-approx}
  (\widetilde{\bR}_{\cF\cU}^{\nf})^{-1} 
\approx \bigoplus_{i=1}^N[(\widetilde{\bR}_{\cF\cU}^{\nf})_{ii}]^{-1},
\end{equation}
which only involves $N$ inversion of $6\times 6$ matrices.  The mobility tensor $\fM$ is approximated by its block-diagonal components using direct Ewald summation, \ie, for each particle, the approximation only considers the interactions with its periodic images.  To obtain the preconditioner, we apply the Incomplete LU decomposition with zero fill-in (ILU0)~\cite{iterative-book} on the approximated $\widetilde{\fM}$, which is constructed following Eq.~\eqref{eq:fM-comp} with the approximated $(\widetilde{\bR}_{\cF\cU}^{\nf})^{-1}$ and $\fM$. Unlike Saintillan~\etal~\cite{Saintillan_spme-stokes-fiber_pof2005}, including close pair interactions has an adverse effect on the preconditioner due to the diagonal approximation of $\widetilde{\bR}_{\cF\cU}^{\nf}$.

The effectiveness of this preconditioner on the far-field iteration is demonstrated in Fig.~\ref{fig:pc_res}.  In this case, the HIs corresponding to random forces and strain rates are solved for a random bidisperse suspension of $200$ particles with $\lambda=2 $, $x_2 = 0.3$, and $\phi = 0.2$.  The far-field preconditioner substantially reduces the number of GMRES iterations.  Evidently, its usage is justified when the required GMRES residual is small, since constructing the approximate $\widetilde{\fM}$ and the ILU0 decomposition also take time.  In dynamic simulations, further time saving can be achieved by updating the preconditioner every few time steps.  In addition, the exact break-even time also depends on the far-field mobility computation parameters, including $M$, $P$, and $r_c$ that indirectly affect the iterative solver.  Finally, since the preconditioner construction is an $\bigO(N)$ operation and the $\widetilde{\fM}$ evaluation scales as $\bigO(N\log N)$, preconditioning is almost always justified for large systems.

\subsection{Dynamics simulation of Brownian suspensions}
\label{sec:brownian-suspensions}

Particle dynamics in a suspension are described by the generalized $N$-body Langevin equation,
\begin{equation}
  \label{eq:langevin}
  \tens{m}\cdot \frac{\dd \cU}{\dd t} = \cF^\mathrm{H} + \cF^\mathrm{P} + 
\cF^\mathrm{B}
\end{equation}
where $\tens{m}$ is the generalized mass/moment of inertial matrix, $\cU$ is the generalized particle velocity and $\cF^\mathrm{H}$, $ \cF^\mathrm{P}$, and $\cF^\mathrm{B}$ are the forces on particles.  The hydrodynamic force $\cF^{\rH}$ arises from the HIs and can be computed from Eq.~\eqref{eq:fh-expr}.  The interparticle force $ \cF^\mathrm{P}$  originates from the interparticle potentials.  The Brownian force $\cF^\mathrm{B}$ is due to thermal fluctuations in the solvent, and from the fluctuation-dissipation theorem~\cite{evans_morriss}, $\cF^\mathrm{B}$ satisfies
\begin{equation}
  \label{eq:br-force}
\overline{ \cF^\mathrm{B}(t)} = 0 \text{ and }
\overline{ \cF^\mathrm{B}(0)\cF^\mathrm{B}(t)} = 2\kT \delta(t)\Rfu ,
\end{equation}
where the overline denotes an average over the solvent fluctuations and $\kT$ is the thermal energy scale.

The configuration evolution is obtained by integrating Eq.~\eqref{eq:langevin} twice over an appropriate time scale $\Delta t$, and the result is~\cite{bd_ermak_jcp1978, Brady_diffusion-susp_jcp1987} 
\begin{equation}
  \label{eq:conf-change}
  \Delta X = \left[\cU^\infty + \Rfu^{-1}\cdot\left(\Rfe\cdot \tE^\infty + 
\cF^\mathrm{P}\right)\right] \Delta t + \kT\dive{\Rfu^{-1}} \Delta t +  \Delta 
X^\mathrm{B},
\end{equation}
where $\Delta X$ is the suspension configuration change over time $\Delta t$, $\cU^\infty$ is the generalized velocity from the imposed flow, and $\Delta X^\mathrm{B}$ is the Brownian displacement which satisfies
\begin{equation}
  \label{eq:xb-expre}
  \overline{\Delta X^\mathrm{B}} = 0 \text{ and }\overline{\Delta X^\mathrm{B} 
\Delta X^\mathrm{B}} = 2\kT \Delta t \Rfu^{-1}.
\end{equation}
The second term on the right hand side of Eq.~\eqref{eq:conf-change} is the deterministic drift due to the configuration dependent Brownian force $\cF^\mathrm{B}$, and the divergence operator is acting on the last index of $\Rfu^{-1}$.  The divergence can be numerically evaluated following Banchio \& Brady~\cite{asd-brownian_banchio_jcp2003}.

The suspension bulk stress is obtained by spatially averaging the Cauchy stress~\cite{brady1993a, Brady2006}, \ie,
\begin{equation}
  \label{eq:bulk-stress-st}
\avg{\tens{\Sigma}} =  -\langle p\rangle_\mathrm{f} \tI + 2\eta_0 
\avg{\overline{\tens{E}^\infty}} + (\kappa_0 - \tfrac{2}{3}\eta_0)E^\infty\tI 
-n\kT\tI+ n (\langle\tS^\mathrm{E} \rangle + \langle\tS^\mathrm{P} \rangle 
+\langle\tS^\mathrm{B} \rangle) ,  
\end{equation}
where $\langle p\rangle_\mathrm{f}$ is the average solvent pressure, $\langle \cdot \rangle$ is the volume average over the entire suspension, $\kappa_0$ is the fluid bulk viscosity, and $n$ is the particle number density.  The particle stresslets $\tS^\mathrm{H}$ are broken down as $\tS^\mathrm{H}=\tS^\mathrm{E}+\tS^\mathrm{P}+\tS^\mathrm{B}$, where $\tS^\mathrm{E}$ is the contributions from the the imposed flow, $\tS^\mathrm{P}$ from the interparticle potential, and $\tS^\mathrm{B}$ from the Brownian motion.  Their suspension averages are expressed in resistance tensors
\begin{align}
\label{eq:se}
\langle\tS^\mathrm{E}\rangle = & -\langle \Rsu\cdot \Rfu^{-1}\cdot \Rfe - \Rse 
\rangle,\\
\label{eq:sp}
\langle\tS^\mathrm{P}\rangle = & -\langle (\Rsu\cdot \Rfu^{-1} + \br\tI)\cdot 
\bF^\mathrm{P} \rangle,\\
\label{eq:sb}
\langle\tS^\mathrm{B}\rangle = & -\kT \langle\grad\cdot(\Rsu\cdot \Rfu^{-1}) 
\rangle,
\end{align}
where the divergence in Eq.~\eqref{eq:sb} is applied to the last index in the bracket. For hard-sphere suspensions, $\langle\tS^\mathrm{P}\rangle = 0$ as the HI and the interparticle force contributions exactly cancel each other~\cite{brady1993a}.  The Brownian stresslet $\langle\tS^\mathrm{B}\rangle$ can also be computed using the modified mid-point scheme~\cite{asd-brownian_banchio_jcp2003}.

In dynamic simulations, the Brownian displacement $\Delta X^\mathrm{B}$ is evaluated from the Brownian force $\cF^\mathrm{B}$ in Eq.~\eqref{eq:br-force} as
\begin{equation}
  \label{eq:xb-comp1}
  \Delta X^\mathrm{B} = \Rfu^{-1}\cdot \cF^\mathrm{B} \Delta t.
\end{equation}
Following Banchio \& Brady~\cite{asd-brownian_banchio_jcp2003}, the Brownian force can be split into a near-field part and a far-field part,
\begin{equation}
  \label{eq:br-f-split}
  \cF^\mathrm{B} = \cF^{\mathrm{B},\mathrm{nf}} + \cF^{\mathrm{B},\mathrm{ff}}.
\end{equation}
Both $\cF^{\mathrm{B},\mathrm{nf}}$ and $\cF^{\mathrm{B},\mathrm{ff}}$ have zero mean and satisfy
\begin{align}
  \label{eq:br-cond}
  \overline{\cF^{\mathrm{B},\mathrm{nf}}\cF^{\mathrm{B},\mathrm{nf}}} =&  
\frac{2\kT}{\Delta t} \Rfu^\mathrm{nf}, \\
\overline{\cF^{\mathrm{B},\mathrm{ff}}\cF^{\mathrm{B},\mathrm{ff}}} =& 
\frac{2\kT}{\Delta t} (\fM^{-1})_{\cF\cU}, \\
\overline{\cF^{\mathrm{B},\mathrm{ff}}\cF^{\mathrm{B},\mathrm{nf}}} = & 0,
\end{align}
where $(\fM^{-1})_{\cF\cU}$ is the $\cF\cU$ block of the inverted far-field grand mobility tensor.  The pairwise-additive lubrication corrections allow pairwise evaluation of the near-field Brownian force $\cF^{\mathrm{B},\mathrm{nf}}$~\cite{asd-brownian_banchio_jcp2003}.  Since $\fM$ is not explicitly constructed, to compute $\cF^{\mathrm{B},\mathrm{ff}}$, it is necessary to solve
\begin{equation}
  \label{eq:br-f-ff}
  \begin{bmatrix}
    \cF^{\mathrm{B},\mathrm{ff}} \\
    \Delta\tS^\mathrm{B}
  \end{bmatrix}
=\frac{2\kT}{\Delta t} (\fM^{-1/2} )\cdot \vect{\Psi},
\end{equation}
where $\vect{\Psi}$ is a Gaussian noise of zero mean and unit variance, and $\Delta\tS^\mathrm{B}$ is the fluctuation part of the Brownian stress in Eq.~\eqref{eq:sb}.  The inverse square root of the grand mobility tensor $\fM^{-1/2}$ in Eq.~\eqref{eq:br-f-ff} can be approximated using Chebychev polynomials with eigenvalue estimations~\cite{asd-brownian_banchio_jcp2003, Jendrejack_cheby-brown_jcp2000}, or solved as an Initial Value Problem (IVP)~\cite{Swan2013,Higham_FuncMatrix2008}, which was first used by Swan \& Brady~\cite{Swan2013} in ASD.  The solution of the following IVP~\cite{Boyd_MatInverSq_la-app_2000} with matrix $\bA$,
\begin{equation}
  \label{eq:ivp-1}
  \frac{\dd {\bx}}{ \dd \tau} = 
-\tfrac{1}{2}\left[\tau\bI +(1-\tau) \bA  \right]^{-1}\cdot (\bA - \bI) \cdot 
\bx, \; \bx(0) = \vect{c},
\end{equation}
at $\tau=1$ satisfies $\bx(1) = \bA^{-1/2}\cdot \vect{c}$.  Swan \& Brady~\cite{Swan2013} devised a numerical scheme to solve  Eq.~\eqref{eq:ivp-1} in ASD:  at each time step with step size $\Delta \tau$, Eq.~\eqref{eq:ivp-1} is marched first with a Euler forward half-step then a Euler backward half-step, \ie,
\begin{align}
  \label{eq:ivp-forward}
  \frac{\bx_{i+\frac{1}{2}} - \bx_{i}}{\Delta\tau /2 } = & 
-\tfrac{1}{2}\left[\tau_i\bI +(1-\tau_i) \bA  \right]^{-1}\cdot (\bA - \bI) 
\cdot \bx_i, \\
  \label{eq:ivp-backward}
  \frac{\bx_{i+1} - \bx_{i+\frac{1}{2}}}{\Delta\tau /2 } = & 
-\tfrac{1}{2}\left[\tau_{i+1}\bI +(1-\tau_{i+1}) \bA  \right]^{-1}\cdot (\bA - 
\bI) \cdot \bx_{i+1}.
\end{align}
With $\bA=\fM$ and $\vect{c}  = (2\kT/\Delta t)\vect{\Psi}$, Eq.~\eqref{eq:br-f-ff} is solved at $\tau = 1$.  In SEASD, both Eqs.~\eqref{eq:ivp-forward} and~\eqref{eq:ivp-backward} are solved iteratively, usually with a smaller tolerance compared to $\Delta \tau$.  The results with $\Delta \tau = 0.1$ are often satisfactory.

For dynamic simulation of Brownian suspensions under a simple shear flow with strain rate $\gamd$, the ratio of the convective transport rate $\gamd$ and the diffusive transport rate $\kT/(6\pi\eta_0 a_\mathrm{p}^3)$ defines the P\'{e}clet number,
\begin{equation}
  \label{eq:pe-num}
  \pe = \frac{6\pi\eta_0 a_\mathrm{p}^3 \gamd}{\kT}.
\end{equation}
Small $\pe$ indicates Brownian motion dominance, and large values suggest negligible Brownian influences.  For bidisperse suspensions, we define $\pe$ based on the size of the small particles to capture the dynamics of the most rapid changes, \ie, $a_\mathrm{p} = a_1$.  In dynamic simulations, the time in Eq.~\eqref{eq:conf-change} is scaled according to the P\'{e}clet number: when $\pe\le 1$, it is scaled with the diffusive time scale of the small particles, $6\pi\eta_0 a_1^3/(\kT)$, and when $\pe > 1$, the convective time scale $\gamd^{-1}$.

\subsection{The mean-field Brownian approximation}
\label{sec:mean-field-brownian}

The most time-consuming step in dynamic simulations of Brownian suspensions is computing $\cF^{\mathrm{B},\ff}$ from Eq.~\eqref{eq:br-f-ff} due to the large number of $\fM$ evaluations, although the IVP approach in Sec.~\ref{sec:brownian-suspensions} is expected to be faster than the Chebychev approximation~\cite{Swan2013}.  Further speed improvement is possible by introducing a mean-field approximation of the Brownian-related quantities~\cite{asd-brownian_banchio_jcp2003}.  In this approach, the far-field grand mobility tensor $\fM$ is approximated as a diagonal matrix for all Brownian related computations, and the full HI computations are retained for the flow-related quantities such as $\tS^\mathrm{E}$.  As a result, this method retains the $\bigO(N\log N)$ scaling, but with an order of magnitude smaller prefactor for monodisperse suspensions~\cite{asd-brownian_banchio_jcp2003}.  The diagonal approximation of $\fM$ uses the single particle result for the $\mathrm{ES}$ coupling, and the far-field translational and rotational short-time self-diffusivities for the $\cU\cF$ coupling.  These far-field values are from Monte-Carlo computations of equilibrium configurations at the same volume fraction \emph{without} the lubrication corrections.  Extending this approach to polydisperse suspensions is trivial: the suspension far-field diffusivities in the diagonal elements are replaced by the far-field diffusivities for each species.  The mean-field Brownian approximation is especially suitable for studying dense suspension rheology, where the HIs are dominated by the near-field lubrication interactions.  Following Brady \& Banchio~\cite{asd-brownian_banchio_jcp2003}, we designate this approximation scheme \mbox{SEASD-nf}.

\section{Accuracy and performance}
\label{sec:accuracy-performance}

\subsection{Mobility computation accuracy}
\label{sec:impl-accur}

The accuracy of the mobility computation is characterized by the relative $\infty$-norm of the strain rate, \ie, 
\begin{equation}
  \label{eq:err-E}
  e_{\infty,r}(E) = \max_{i\in\{1,\ldots, N\}}\frac{\|\tens{E}^{\mathrm{SE}}_i 
- \tens{E}^*_i \|}{\| \tens{E}^*_i \|},
\end{equation}
where the $\tens{E}^{\mathrm{SE}}_i$ is the particle strain rate from the SE method and $\tens{E}^*_i$ is a well-converged value from direct Ewald summation.  Other error measurements can be similarly defined. For example, $e_{\infty,r}(U)$ for the linear velocity was used by Lindbo \& Tornberg~\cite{Tornberg_se-stokes_jcompp2010} to characterize the accuracy of the SE method for point forces.  For the stresslet-strain rate level mobility computation here, we found $e_{\infty,r}(E)$ the most stringent error criteria, possibly because more derivatives are involved in Eq.~\eqref{eq:faxen-s}.

To facilitate quantitative discussions, in this section we focus on a random bidisperse hard-sphere system of $N=50$,  $\phi=0.05$, $\lambda = 2$, and $x_2 = 0.3$.  The imposed force, torque, and stresslet on each particle are randomly drawn from a normal distribution, and rescaled to ensure $\|\bF_i\|=1$, $\|\bT_i\|=1$, and $\| \tS_i \| = 1$.  The the simulation box lattice vectors are ${\bold a}_1 = (L,0,0)$, ${\bold a}_2 = (\gamma L,L,0)$, and ${\bold a}_3 = (0,0,L)$, with $\gamma$ the strain.  The computations are carried out in DP accuracy on CPU.

\subsubsection{Wave-space accuracy}
\label{sec:wave-space-accuracy}

\begin{figure}
  \centering
  \includegraphics[width=5.2in]{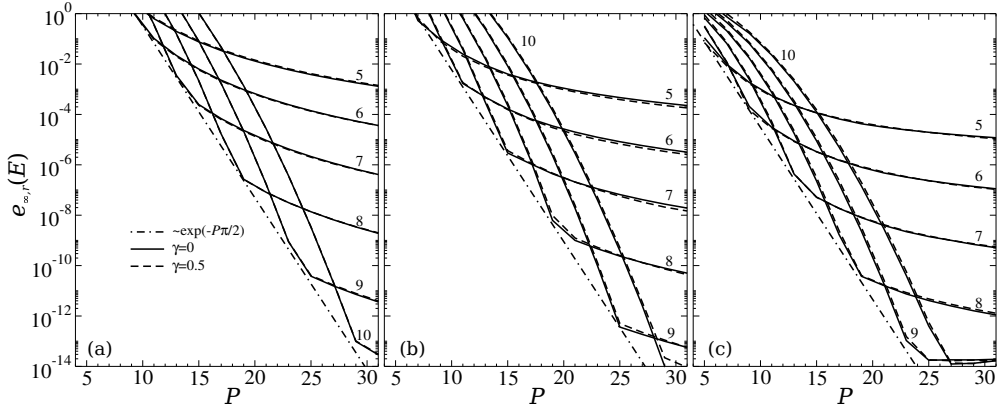}
  \caption{
The wave-space accuracy measured by $e_{\infty,r}(E)$ [Eq.~\eqref{eq:err-E}] as a function of the interpolation point $P$ with various shape parameter $m$ at $M=64$ and $\xi a_1 = 0.1$.  The particle size effects are incorporated using (a): the real-space, (b): the hybrid, and (c): the wave-space approaches in Sec.~\ref{sec:wave-space-comp}.  The values of $m$ are annotated in each figure.  The solid and dashed lines represent the case of $\gamma = 0$ and $0.5$, respectively.  The dashed dotted lines show the exponential minimum error decay, $e_{\infty,r}(E)\sim \exp({-P \pi/2})$.
}
  \label{fig:wave_err}
\end{figure}

Fig.~\ref{fig:wave_err} presents the accuracy of wave-space computation using different SE implementations with orthogonal ($\gamma = 0$) and sheared ($\gamma = 0.5$) simulation boxes in solid and dashed lines, respectively.  The error $e_{\infty,r}(E)$ is shown as a function of the interpolation point $P$ with various shape parameter $m$ at $M = 64$ and $\xi a_1 = 0.1$.  Different particle size incorporation approaches discussed in Sec.~\ref{sec:wave-space-comp} are presented: in Fig.~\ref{fig:wave_err}a the real-space approach, in Fig.~\ref{fig:wave_err}b the hybrid approach, and in Fig.~\ref{fig:wave_err}c the wave-space approach.

There are several key observations in Fig.~\ref{fig:wave_err}.  First of all, the errors associated with orthogonal and sheared simulation boxes are almost identical.  This validates the general formalism for non-orthogonal simulation boxes in Sec.~\ref{sec:spectr-ewald-meth}.  Secondly, the SE method is sensitive to $P$ and $m$, which respectively correspond to the discretization and truncation of the shape function $h(\bt)$.  At a given $m$, $e_{\infty, r}(E)$ first decreases exponentially, followed by a much slower reduction with increasing $P$.  The two-stage reduction of $e_{\infty, r}(E)$ is well understood for point forces~\cite{Tornberg_se-stokes_jcompp2010}: the exponential decrease is due to the improved resolution of the shape function, and the slower reduction is associated with the Gaussian truncation from the shape parameter $m$.  Therefore, at large $P$ and $m$ the result is  expected to be accurate; indeed, in Fig.~\ref{fig:wave_err} the minimum errors are all close to the machine precision.  Such accuracy is inaccessible using the PME or the SPME method at this grid number ($M=64$) due to the inherent coupling between the interpolation and the wave-space truncation errors.  Moreover, for a given $P$, $e_{\infty, r}(E)$ first decreases to a minimum and then increases with increasing $m$.  At the minimum, $e_{\infty, r}(E)$ is transitioning from exponential to slower decay, and the errors from the shape resolution is about the same as the errors from the Gaussian truncation.  From the error estimation of Lindbo \& Tornberg~\cite{Tornberg_se-stokes_jcompp2010, Tornberg_periodic-laplace_jcompp2011}, at a given $P$, the minimum wave-space error $e_{\infty, r}(E)$ and the corresponding shape parameter $m$ are
\begin{equation}
  \label{eq:error-scale}
e_{\infty, r}(E)\sim \exp(-P\pi/2) \text{  and  } m\sim \sqrt{\pi P},
\end{equation}
respectively.  The asymptotic exponential decay of the minimum $e_{\infty, r}(E)$ is also shown as dash-dotted lines in Fig.~\ref{fig:wave_err}.  The exponential decay of the minimum error with respect to $P$ to the round-off precision at large $P$ and $m$ clearly demonstrate the spectral accuracy~\cite{Trefethen_SpecMethod2000} of the SE method.

In Fig.~\ref{fig:wave_err} different particle size incorporation approaches exhibit similar qualitative behaviors with quantitative differences.  For example, to achieve an accuracy of $e_{\infty, r}(E) \sim 10^{-4}$ at the optimal $m$, in Fig.~\ref{fig:wave_err}a, \ref{fig:wave_err}b, and \ref{fig:wave_err}c the required $P$ are respectively $15$, $13$, and $9$, corresponding to the real-space, hybrid, and wave-space approaches discussed in Sec.~\ref{sec:wave-space-comp}.  The latter two approaches reduce the $h(\bt)$ evaluations by $35\%$ and $78\%$ compared to the real-space approach at a cost of the number of required FFTs.  Therefore, there is a subtle balance between the number of interpolation points $P$ and the number of FFTs in the SE method implementation.  The hybrid approach in Fig.~\ref{fig:wave_err}b achieves a good balance between accuracy and computation efficiency, and therefore is adopted in SEASD.

Finally, Fig.~\ref{fig:wave_err} shows that, in addition to the spectral accuracy and the ease of implementation, the SE method also allows flexible error control by adjusting $P$ and $m$ without changing the grid points $M$.  As a result, the errors from the wave-space summation and the interpolation can be separated, and this permits more flexible error control when computing HIs in polydisperse systems.  On the other hand, such error separation is not possible in other particle mesh techniques such as the PME and the SPME methods.

\subsubsection{Overall mobility accuracy}
\label{sec:total-accuracy}

\begin{figure}
  \centering
  \includegraphics[width=5in]{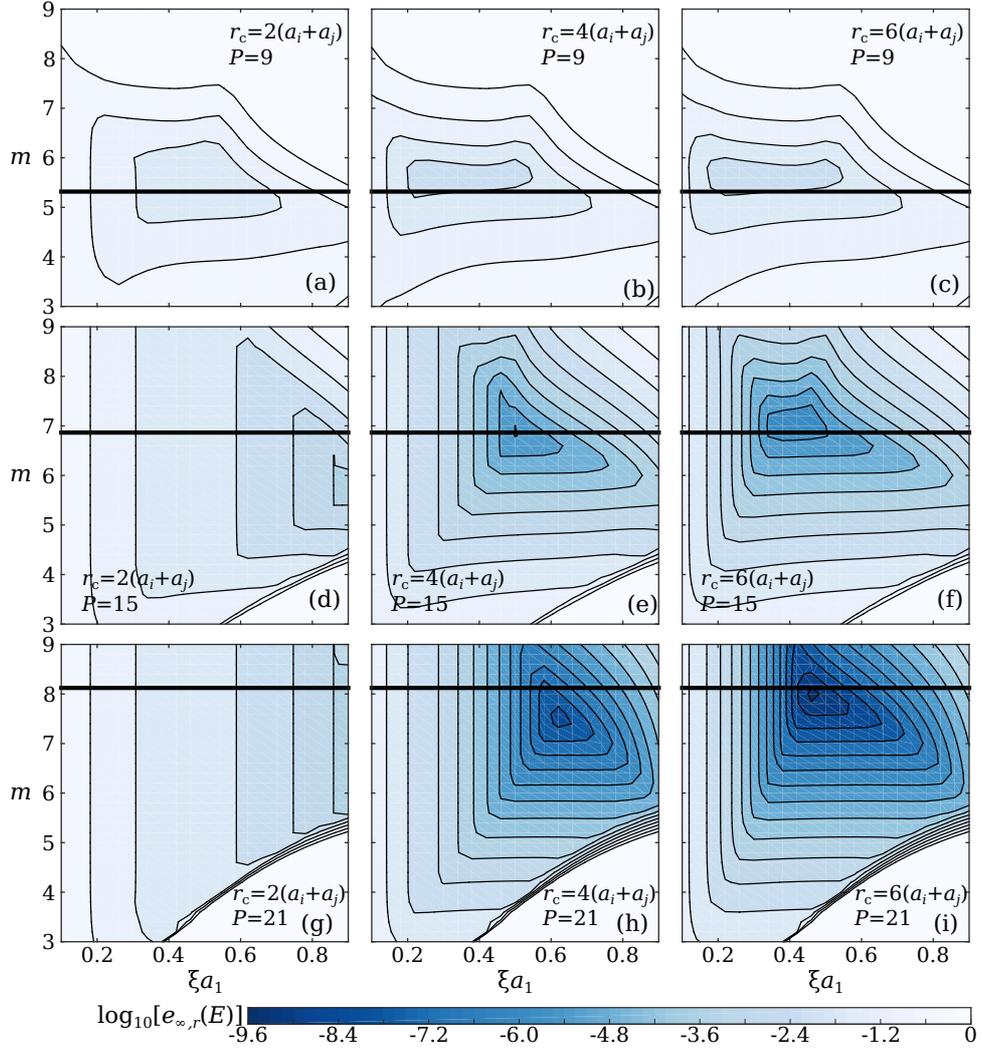}
  \caption{(Color online) The overall accuracy measured in $e_{\infty, r}(E)$ as a function of the splitting parameter $\xi a_1$ and the shape parameter $m$ at $M = 64$ for a real-space cutoff radius $r_c = 2 (a_i+a_j)$  (left column), $ 4 (a_i+a_j) $ (middle column), and $ 6 (a_i+a_j)$ (right column), and the interpolation point $P = 9$ (top row), $15$ (middle row), and $21$ (bottom row). The thick black lines represent $m=\sqrt{\pi P}$.  The simulation cell is orthogonal ($\gamma$ = 0), and the particle size effects are accounted using the hybrid approach.}
  \label{fig:M64_err}
\end{figure}

\begin{figure}
  \centering
  \includegraphics[width=5in]{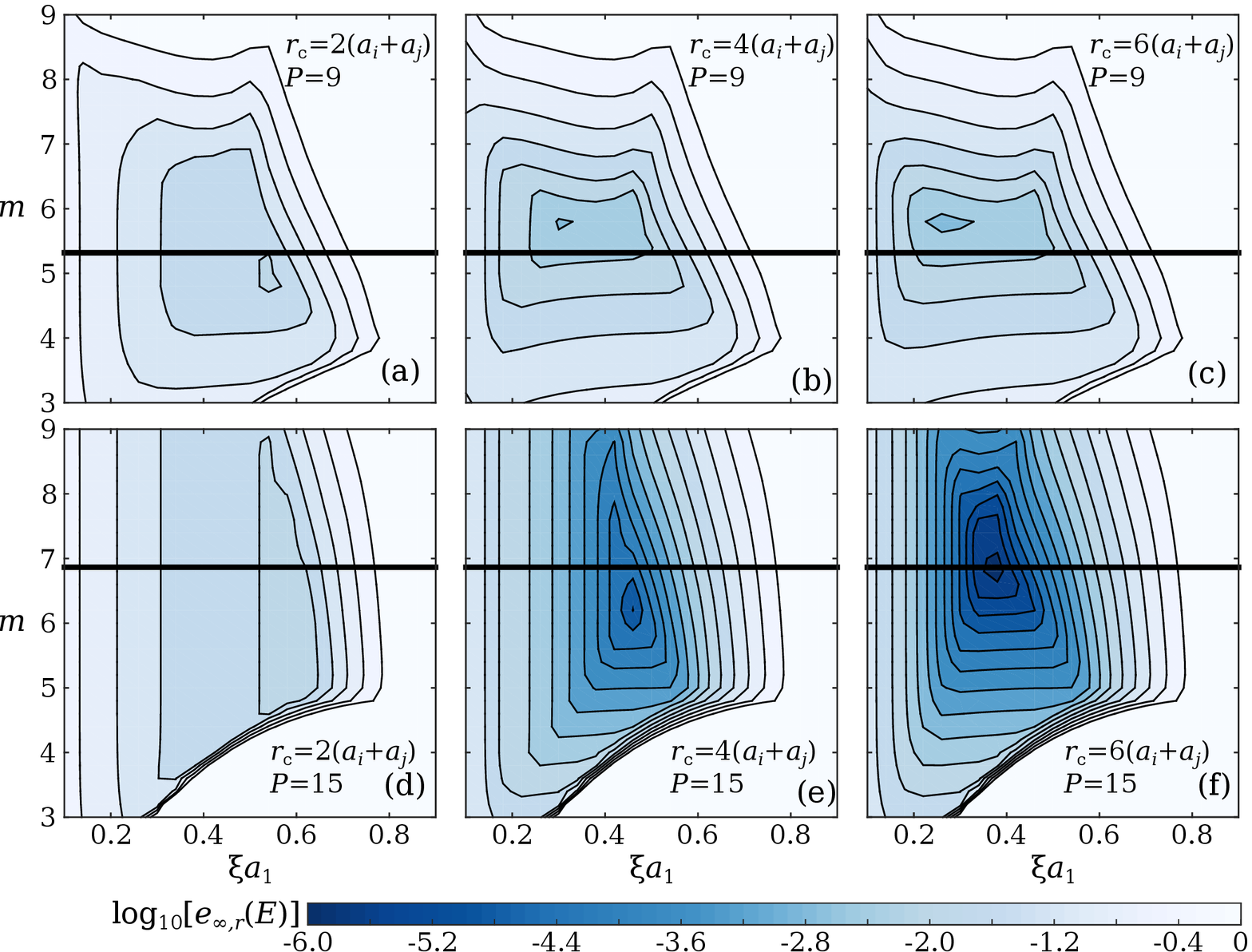}
  \caption{(Color online) The overall accuracy measured in $e_{\infty, r}(E)$ as a function of the splitting parameter $\xi a_1$ and the shape parameter $m$ with $M = 32$ for a real-space cutoff radius $r_c = 2 (a_i+a_j)$  (left column), $ 4 (a_i+ a_j) $ (middle column), and $ 6( a_i+a_j)$ (right column), and the interpolation point $P = 9$ (top row) and $15$ (bottom row). The thick black lines represent $m=\sqrt{\pi P}$.  The simulation cell is orthogonal ($\gamma$ = 0), and the particle size effects are accounted using the hybrid approach.}
  \label{fig:M32_err}
\end{figure}

Both the wave-space and the real-space computations affect the overall mobility accuracy, and the controlling parameters are the grid point $M$, the interpolation point $P$, the Gaussian shape parameter $m$, the real-space cutoff radius $r_c$, and the splitting parameter $\xi$.  Out of the five parameters, only changes in $\xi$ and $m$ do not affect the computational cost since adjusting $M$ affects the FFT size, changing $r_c$ influences the neighbor search, \etc~With fixed computation cost, \ie, fixed $M$, $P$, and $r_c$, it is desirable to find the combination of $m$ and $\xi$ that minimizes the overall error.

Fig.~\ref{fig:M64_err} and~\ref{fig:M32_err} present the effects of $m$ and $\xi$ on the overall mobility accuracy with various $P$ and $r_c$ for $M=64$ and $32$, respectively.  The wave-space computation uses the hybrid approach in Sec.~\ref{sec:wave-space-comp}, and the simulation box is orthogonal ($\gamma = 0$).  The thick black lines in these figures indicate the theoretical optimal shape parameter $m=\sqrt{\pi P}$~\cite{Tornberg_periodic-laplace_jcompp2011,Tornberg_se-stokes_jcompp2010}.  Note that in our implementation, the cutoff radius $r_c$ depends on the radius $a_i$ and $a_j$ in a particle pair.

Fig.~\ref{fig:M64_err}i with $M=64$, $P=21$, and $r_c=6(a_i+a_j)$ best illustrates the influences of $m$ and $\xi$.  Here, the mobility computation can reach $e_{\infty,r}(E)<10^{-9}$ at $(\xi a_1, m) = (0.46, 8)$.  With fixed $m$, $e_{\infty, r}(E)$ exhibits a minimum with increasing  $\xi a_1$, and when $m \leq 8$, the minimum degenerates to a plateau due to the wave-space Gaussian truncation, which is also illustrated in Fig.~\ref{fig:wave_err} at low $m$.  At low $\xi$, the overall error is dominated by the real-space error, which decreases with increasing $\xi$.  At high $\xi$, the overall error is mainly from the wave space, and increases with increasing $\xi$.
With fixed $\xi$ on the other hand, $e_{\infty, r}(E)$ also shows a minimum with increasing $m$.  When $\xi a_1 \leq 0.46$, the $e_{\infty, r}(E)$ minimum becomes a plateau since the real-space error is independent of $m$. Here, the reduction of $e_{\infty, r}(E)$ with increasing $m$ at small $m$ comes almost entirely from the reduced Gaussian truncation.  When $\xi a_1 > 0.46$, the minimum plateau disappears as in this region the wave-space error is sensitive to $m$, a point also illustrated in Fig.~\ref{fig:wave_err}.
Furthermore, in Fig.~\ref{fig:M64_err}i there is a region of $e_{\infty, r}(E)>1$ at high $\xi$ and low $m$ due to large wave-space errors.

Comparison across rows and columns in Fig.~\ref{fig:M64_err} and \ref{fig:M32_err} reveals the influences of $r_c$ and $P$ on the overall accuracy, respectively.  For both cases, reducing $r_c$ or $P$ increases the minimum value of $e_{\infty, r}(E)$ and changes the corresponding $\xi a_1$ and $m$.  Comparing Fig.~\ref{fig:M64_err}g, \ref{fig:M64_err}h, and \ref{fig:M64_err}i shows that reducing $r_c$ increases the real-space error and shifting the minimum of $e_{\infty, r}(E)$ towards larger $\xi a_1$.  The decrease of $e_{\infty, r}(E)$ with respect to increasing $\xi$ at small $\xi a_1$ also becomes slower.  In Fig.~\ref{fig:M64_err}g, the $e_{\infty, r}(E)$ minimum is at $\xi a_1 > 1$.  Comparing Fig.~\ref{fig:M64_err}i, \ref{fig:M64_err}f, and \ref{fig:M64_err}c reveals the effects of reducing the interpolation point $P$.  With diminishing $P$, the wave-space error increases due to poor Gaussian resolution, and the $e_{\infty, r}(E)$ minimum is shifted towards lower $m$.  In addition, the overall accuracy decreases significantly for large $m$ at small $P$, \eg, in Fig.~\ref{fig:M64_err}c, $e_{\infty, r} (E)>1$ when $m>8$.

Comparing Fig.~\ref{fig:M64_err} and \ref{fig:M32_err} shows the effect of grid point $M$ on the mobility accuracy.  Note that the color scales in Fig.~\ref{fig:M64_err} and \ref{fig:M32_err} are different, and the minimum $e_{\infty, r} (E)$ in Fig.~\ref{fig:M64_err}f and \ref{fig:M32_err}f is approximately the same.  The most apparent effect of reducing $M$ is the shrinkage of the parameter space corresponding to $e_{\infty, r }(E)<1$ due to the truncation of the wave-space sum.  As a result, at $M=32$, the mobility evaluation is more sensitive to $\xi a_1$ compared to the case of $M=64$.  Otherwise, the qualitative aspects of Fig.~\ref{fig:M32_err} are similar to Fig.~\ref{fig:M64_err}.  Moreover, the thick black lines representing the theoretical optimal shape parameter $m=\sqrt{\pi P}$ is almost always in the vicinity of the regions of the highest accuracy in both Fig.~\ref{fig:M64_err} and \ref{fig:M32_err}.  This substantially simplifies the search for the optimal $\xi$.

\begin{figure}
  \centering
  \includegraphics[width=5.2in]{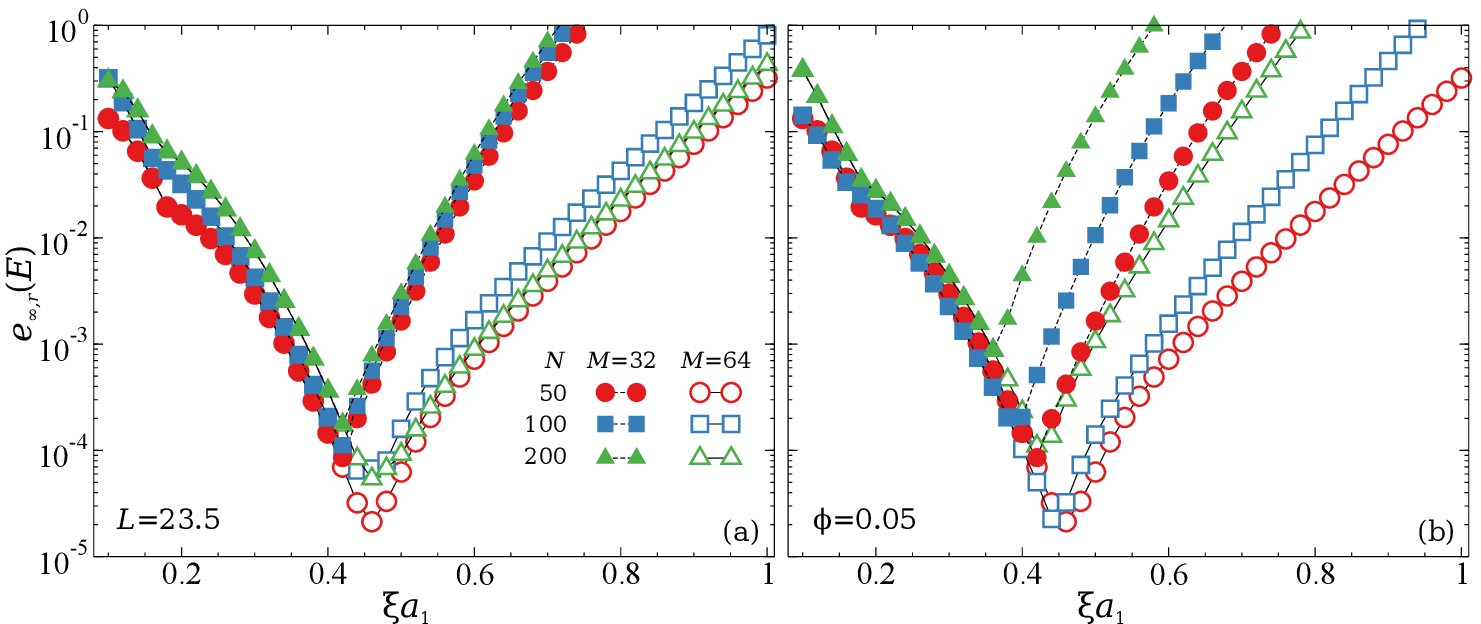}
  \caption{(Color online) The overall mobility accuracy measured in $e_{\infty,r}(E)$ as a function of the splitting parameter $\xi$ with $N=50$, $100$, and $200$, and $M=32$ (filled symbols) and $64$ (open symbols) for (a): constant box size $L/a_1=23.5$ and (b): constant volume fraction $\phi = 0.05$.  Changes are based on the baseline case in Sec.~\ref{sec:impl-accur}.  Other parameters are $P=13$, $m = 6.7$, and $r_c = 4(a_i+a_j)$.
}
  \label{fig:err_size}
\end{figure}

The influences of the particle number $N$ on the overall mobility accuracy is presented in Fig.~\ref{fig:err_size} for $M=32$ and $64$.  The simulation box size is fixed at $L/a_1 = 23.5$ in Fig.~\ref{fig:err_size}a, and the suspension volume fraction is fixed at $\phi = 0.05$ in Fig.~\ref{fig:err_size}b.  Other parameters remain unchanged from the baseline case, and the mobility computation parameters are $P=13$, $m = 6.7$, and $r_c = 4(a_i+a_j)$.  The mobility accuracy is more sensitive to changes in $L$ than changes in $\phi$.  In Fig.~\ref{fig:err_size}a, $e_{\infty,r}(E)$ changes little, but in Fig.~\ref{fig:err_size}b, the $e_{\infty,r}(E)$ minimum increases drastically with different $N$.  The almost identical decrease in $e_{\infty,r}(E)$ at small $\xi a_1$ suggests the real-space error are not significantly changed by $N$ in either case.  The diverging $e_{\infty,r}(E)$ at higher $\xi a_1$ in Fig.~\ref{fig:err_size}b suggests the wave-space computation is sensitive to the box size at fixed $P$ and $m$.  This is well-known for particle mesh techniques in general~\cite{Deserno_mesh-up-ewald-pt1_jcp1998,Tornberg_se-stokes_jcompp2010}.  Therefore, to retain the computational accuracy with larger systems at the same volume fraction, it is necessary to increase the grid point $M$ or the interpolation point $P$.  Finally, we note in passing that the same qualitative error behaviors are found in the pressure moment computations.

\subsection{Accuracy of the GPGPU implementation}
\label{sec:gpgpu-mobil-accur}

\begin{figure}
  \centering
  \includegraphics[width=6in]{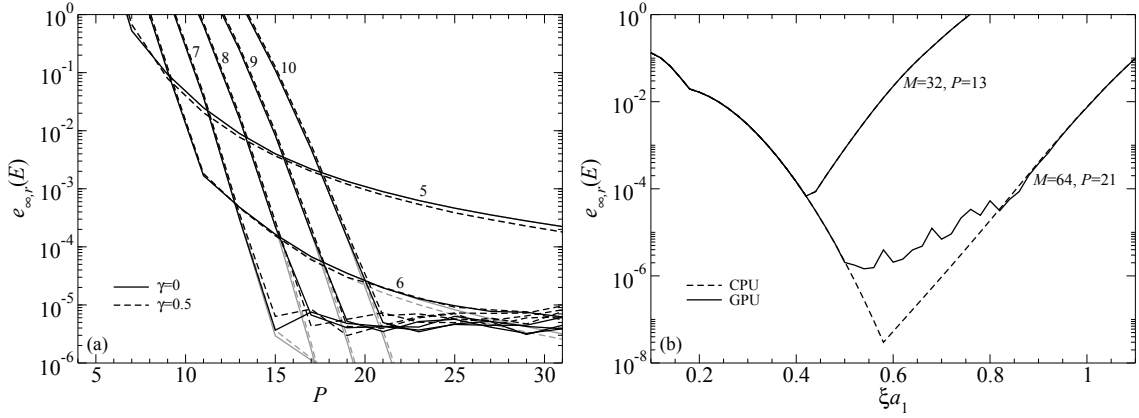}
  \caption{The accuracy of GPGPU mobility computation measured in $e_{\infty,r}(E)$. (a): the wave-space accuracy as a function of $P$ for various $m$ with the same parameters in Fig.~\ref{fig:wave_err}b.  The GPU results are shown in black lines, and the CPU results in Fig.~\ref{fig:wave_err}b are reproduced in gray lines.  The values of $m$ are annotated in the figure.  The solid and dashed lines represent the case of $\gamma=0$ and $0.5$, respectively. (b): The overall mobility accuracy from the GPU (solid lines) and the CPU (dashed lines) computations as a function $\xi a_1$ with $r_c=4(a_i+a_j)$ and $m=\sqrt{\pi P}$.  The corresponding $M$ and $P$ are annotated in the figure.
}
  \label{fig:accu-GPU}
\end{figure}

The accuracy of mobility computation using GPGPU programming discussed in Sec.~\ref{sec:gpu-accel-mobil} is presented in Fig.~\ref{fig:accu-GPU}.  Clearly, the GPU computations provide sufficient accuracy for dynamic simulations.  Fig.~\ref{fig:accu-GPU}a shows the GPU wave-space accuracy as a function of the interpolation point $P$ for various shape parameters $m$ for orthogonal ($\gamma = 0$) and sheared ($\gamma=0.5$) simulation boxes.  Here, the particle size effects are incorporated using the hybrid approach in Sec.~\ref{sec:wave-space-comp}, and the SE method parameters are identical to those of Fig.~\ref{fig:wave_err}b.  Moreover, for comparison the data in Fig.~\ref{fig:wave_err}b are reproduced in gray.  In Fig.~\ref{fig:accu-GPU}a, the GPU results in black lines are indistinguishable from the CPU results in gray lines when $e_{\infty, r}(E)> 10^{-5}$ for all $m$ and $\gamma$, indicating that the GPU computations are only limited by the SP arithmetics.  When the error $e_{\infty, r}(E)$ reaches $10^{-5}$, increasing the interpolation point $P$ does not improve the computation accuracy on GPUs, while the error in the CPU computations using DP arithmetics continue to decrease until $e_{\infty, r}(E)\sim 10^{-14}$.  In addition, the wave-space error remain $e_{\infty, r}(E)\sim 10^{-5}$ after reaching the SP limit even with further increase in $P$, \ie, increasing $P$ does not adversely affect the wave-space accuracy.

The overall GPU mobility accuracy as a function of $\xi a_1$ is presented in Fig.~\ref{fig:accu-GPU}b for two $M$ and $P$ combinations with $m=\sqrt{\pi P}$ and $r_c = 4(a_i+a_j)$ in orthogonal simulation boxes.  The errors $e_{\infty, r}(E)$ are computed using the baseline case of Sec.~\ref{sec:impl-accur}.  The GPU results are shown in solid lines and the CPU results in dashed lines.  When the overall error $e_{\infty, r}(E)>10^{-5}$, \ie, the case of $(M,P)=(32,13)$ in Fig.~\ref{fig:accu-GPU}b, the GPU and the CPU results are indistinguishable from each other.  However, the differences are evident for the case of $(M,P)=(64,21)$.  When $0.5<\xi a_1<0.85$, the GPU computations deviate from the CPU results with larger errors due to the SP arithmetics.  Beyond this range, the CPU and the GPU results overlap again.  In both cases, the accuracy achieved by the GPU mobility computation is sufficient for dynamic simulations, where the error tolerance is typically set at $10^{-3}$.  The results in Fig.~\ref{fig:accu-GPU} dispel any concerns over the SP accuracy in the GPU mobility computations for dynamic simulations.

\subsection{Overall performance}
\label{sec:performance}

\begin{figure}
  \centering
  \includegraphics[width=3.5in]{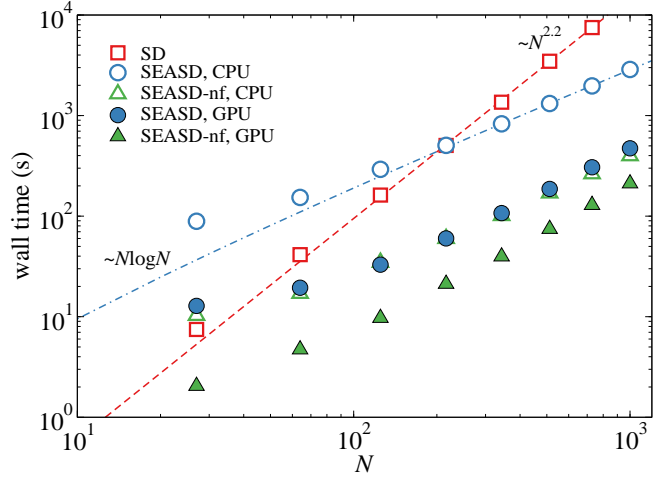}
  \caption{(Color online) The wall time (in second) of $100$ time steps in dynamic simulations at $\pe = 1$ as a function of the particle number $N$ using the conventional SD, SEASD, and \mbox{SEASD-nf}.  The open symbols represent the CPU mobility computation and the filled symbols the GPU mobility computation.  The dashed line show the $\bigO(N^{2.2})$ scaling, and the dash-dotted line show the $\bigO(N\log N)$ scaling.  The suspension is bidisperse with $\lambda = 2$, $y_2 = 0.5$, and $\phi=0.45$ starting from equilibrium configurations.
}
  \label{fig:timing}
\end{figure}

Fig.~\ref{fig:timing} presents the overall performance of various implementations of the SEASD and the conventional SD as a function of the system size $N$.  The program performance is characterized by the wall time, \ie, the actual time of program execution, to march $100$ steps in a dynamic simulation of Brownian suspensions at $\pe = 1$ starting from an equilibrium configuration.  The suspension composition is $\lambda = 2$, $y_2 = 0.5$, and $\phi = 0.45$.  The SEASD mobility computation parameters are fixed at $M = 32$, $P = 11$, $r_c = 4 (a_i+a_j) $ with appropriate $\xi$ and $m$ as they provide sufficient accuracy. The tolerance of the iterative solvers is set at $10^{-3}$.  For SEASD the far-field Brownian forces are calculated using Eqs.~\eqref{eq:ivp-forward} and \eqref{eq:ivp-backward} with $\Delta \tau = 0.2$, and for \mbox{SEASD-nf} the far-field diffusivities are from Table~\ref{tab:ff-coeff}.  The conventional SD result is from an efficient polydisperse implementation~\cite{sd-bidisperse_wang_jcp2015,dg-sd-comp_wang_jcp_2014, Wang_sbm-rerevist_POF2015}.  All the timing results are collected from a workstation with Intel i7-3770K CPU and NVIDIA GeForce GTX 680 GPU.

Fig.~\ref{fig:timing} demonstrates the expected $\bigO(N\log N)$ asymptotic scaling of various SEASD implementations, highlighted by the dash-dotted line.  The implementations with the CPU mobility computation are shown in open symbols and the GPU mobility computation in filled symbols.  The GPU SEASD has almost the same time scaling as the CPU \mbox{SEASD-nf} at all the system size $N$.  Both are almost an order of magnitude faster than the CPU SEASD at a typical system size $N\approx 200$.  This clearly demonstrates the power and promise of GPGPU programming in the dynamic simulation of colloidal suspensions.  More significant speedup is achieved by combining the mean-field Brownian approximation and the GPU mobility computation.  In this case, the speedup of GPU \mbox{SEASD-nf} computation relative to the CPU SEASD ranges between $40$ times for small systems and $15$ times for large systems.  We believe further speedup is still possible by optimizing the GPU implementation.  With the speedup shown in Fig.~\ref{fig:timing}, we are able to study dynamics of larger systems at longer times.  In addition, compared to the conventional SD, all the SEASD implementations are faster at large enough $N$ due to their favorable scaling.  Here, the conventional SD scales as $\bigO(N^{2.2})$, highlighted by the dashed line in Fig.~\ref{fig:timing}.  This peculiar scaling is a combined effect of the pairwise grand mobility tensor construction and explicit matrix inversion.  At $N\gg 1000$, the scaling should recover $ \bigO(N^3)$.  In Fig.~\ref{fig:timing}, the break-even between the CPU SEASD and SD is $N = 216$, and for GPU SEASD at $N\approx 40$.  At all the system sizes studied here, the GPU \mbox{SEASD-nf} is always faster than the conventional SD.

\section{Static and dynamic simulation results}
\label{sec:results-discussions}

\subsection{Short-time transport properties}
\label{sec:short-time-transport}

In this section we present static SEASD simulation results on the short-time transport properties of monodisperse and bidisperse hard-sphere suspensions.  With the iterative computation scheme in Sec.~\ref{sec:iter-comp}, the short-time translational and rotational self-diffusivities, instantaneous sedimentation velocities, and high-frequency dynamic shear and bulk viscosities can be straightforwardly evaluated.  Other transport properties can also be calculated with an appropriate computation scheme.

The suspension short-time limit refers to a time scale $t$ satisfying $\tau_I\ll t \ll \tau_D$, where $\tau_I$ is the inertial time and $\tau_D$ is the diffusion time.  The inertia time $\tau_I = \tfrac{2}{9}\rho_\mathrm{p} a_\mathrm{p}^2/\eta_0 $, where $\rho_\mathrm{p}$ and $a_\mathrm{p}$ are the characteristic particle density and radius, describes the time required for the \emph{particle} momentum to dissipate by interacting with the solvent.  When $\tau_I\ll t$, the particle momentum dissipates almost instantaneously and the particle dynamics are completely overdamped.  The diffusion time $\tau_D = 6\pi \eta_0 a_\mathrm{p}^3 / \kT$ characterizes the time scale of suspension configuration change and $t \ll \tau_D$ ensures that the transport properties entirely arise from the (instantaneous) HIs.  Therefore, they are only determined by the configuration $X$, and can be calculated by sampling independent but equivalent configurations.  In this work we use the Monte-Carlo procedure of Wang \& Brady~\cite{sd-bidisperse_wang_jcp2015}: the hard-sphere configurations are first generated by an event-driven Lubachesky-Stillinger algorithm~\cite{Lubachevsky1990, packing-gen-code_torquato_pre2006}, followed by a short equilibration.  The transport properties are then computed statically.  Here we compare the results from the SEASD with CPU mobility computation with our recent conventional SD results~\cite{sd-bidisperse_wang_jcp2015}.  Although SEASD and SD are based on the same formalism, the grand mobility tensor $\fM$ constructed from SD includes an additional mean-field quadrupole term~\cite{brady-sd-ew_jfm_88}, which can have quantitative consequences.  For bidisperse hard-sphere suspensions, we focus on the composition with $\lambda=2$ and $y_2 = 0.5$.  In the SEASD computations, the system size is $N=800$, and the results are averaged over $500$ independent configurations.  Note that for simple cubic array of monodisperse particles, SEASD  produces identical results as those of Sierou \& Brady~\cite{asd_sierou_jfm01}.

\subsubsection{Short-time translational and rotational self-diffusivities}
\label{sec:short-time-transl}

The microscopic definition of the short-time translational and rotational self-diffusivities, $d^t_{s,\alpha}$ and $d^r_{s,\alpha}$ respectively, for homogeneous suspensions are,
\begin{equation}
  \label{eq:dstr-def}
  d^t_{s,\alpha} = \frac{\kT}{N_\alpha}\Big \langle \sum_{i\in \alpha} 
\hat{\bq} \cdot \boldsymbol{\mu}_{ii}^{tt}\cdot \hat{\bq}  \Big \rangle \text{, 
and }   d^r_{s,\alpha} = \frac{\kT}{N_\alpha}\Big\langle \sum_{i\in \alpha} 
\hat{\bq} \cdot \boldsymbol{\mu}_{ii}^{rr}\cdot \hat{\bq}  \Big\rangle,
\end{equation}
where $\hat{\bq}$ is a vector of unit length for the averaging process and $\boldsymbol{\mu}_{ii}^{tt}$ and $\boldsymbol{\mu}_{ii}^{rr}$ are respectively the diagonal blocks of the force-linear velocity and torque-angular velocity couplings in $\Rfu^{-1}$.  Note that $i\in \alpha$ in Eq.~\eqref{eq:dstr-def} suggests the summation is restricted to particles of species $\alpha$.  The diffusivities are computed using the matrix-free approach of Sierou \& Brady~\cite{asd_sierou_jfm01}: the velocity disturbance $\cU^\mathrm{R}$ corresponding to a stochastic external force $\cF^\mathrm{R}$ satisfying  $\langle{\cF^\mathrm{R}}\rangle = 0$ and $\langle{\cF^\mathrm{R}  \cF^\mathrm{R}}\rangle = \mathcal{I}$ is evaluated.  It is straightforward to show that the ensemble average $\avg{\cU^\mathrm{R} \cF^\mathrm{R}} =  \mathrm{diag} (\Rfu^{-1}) $, allowing extraction of the diffusivities in Eq.~\eqref{eq:dstr-def}.

The computed short-time translational self-diffusivities $d^t_{s,\alpha}$  exhibit a strong  $ N^{-1/3}$ size dependence due to the periodic boundary conditions.  The size dependence from an $N$-particle system can be eliminated by adding the following quantity to the results,
\begin{equation}
  \label{eq:dst-corr}
  \Delta_N d_{s,\alpha}^t = \frac{1.76 
d_{0,1}^t}{(x_1+x_2\lambda^3)^{\frac{1}{3}}} \frac{\eta_0}{\eta_s}
\left( \frac{\phi}{N} \right)^{\frac{1}{3}},
\end{equation}
where $d^t_{0,1} = \kT/(6\pi\eta_0 a_1)$ is  Stokes-Einstein-Sutherland diffusivity for species $1$, and $\eta_s$ is the high-frequency dynamic shear viscosity from the same configurations.  The shear viscosity exhibits little size dependence, and can be directly used.  The effectiveness of Eq.~\eqref{eq:dst-corr} has been demonstrated by Wang \& Brady \cite{sd-bidisperse_wang_jcp2015} in the wave-number-dependent hydrodynamic functions.  The results here always contain this finite size $N$ correction.

\begin{figure}
  \centering
  \includegraphics[width=6in]{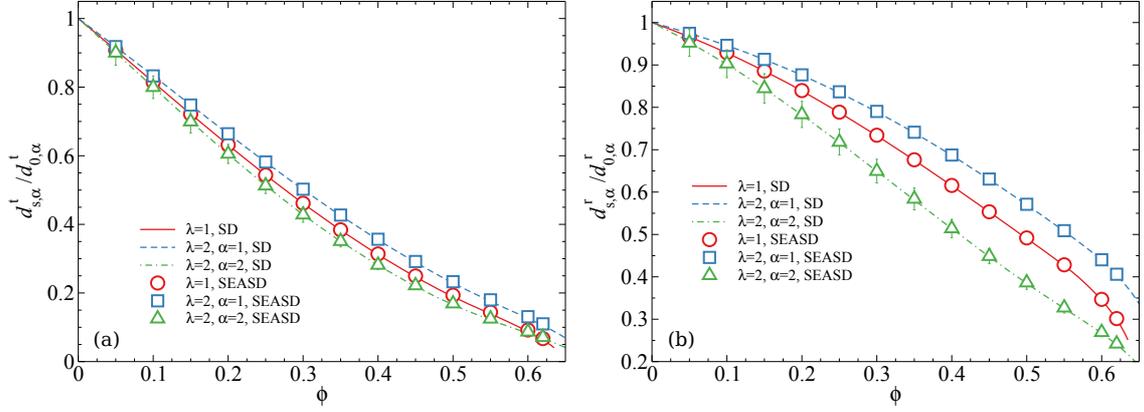}
  \caption{(Color online) The species short-time (a): translational and (b): rotational self-diffusivities, $d_{s,\alpha}^t$ and $d_{s,\alpha}^r$ respectively, as a function of the total volume fraction $\phi$ for monodisperse and bidisperse hard-sphere suspensions with $\lambda = 2$, $y_2 = 0.5$.  The results are scaled with the single particle translation and rotational diffusivity, $d^t_{0,\alpha}$ and $d^r_{0,\alpha}$, respectively.  The SEASD results are shown in symbols and the conventional SD results from Wang \& Brady~\cite{sd-bidisperse_wang_jcp2015} are shown as lines.
}
  \label{fig:dstr}
\end{figure}

Fig.~\ref{fig:dstr}a and Fig.~\ref{fig:dstr}b respectively present $d^t_{s,\alpha}/d^t_{0,\alpha}$ and $d^r_{s,\alpha}/d^r_{0,\alpha}$ of monodisperse and bidisperse suspensions, where the single particle translational and rotational self-diffusivities are $d^t_{0,\alpha} = \kT/(6\pi\eta_0 a_\alpha)$ and $d^r_{0,\alpha} = \kT/(8\pi\eta_0 a_\alpha^3)$.  The SEASD results, shown in symbols, agree well with the conventional SD results shown in lines.  As expected, both $d_{s,\alpha}^t$ and $d_{s,\alpha}^t$ decrease with increasing volume fraction $\phi$, and for bidisperse suspensions, the small particles show diffusivity enhancement while the large particles exhibit diffusivity supression.  Compared to $d^t_{s,\alpha}$, $d^r_{s,\alpha}$ are less sensitive to the volume fractions $\phi$, but more sensitive to the particle sizes $\lambda$.  The SEASD results for large particles show larger error bars compared to the SD results~\cite{sd-bidisperse_wang_jcp2015}, most likely due to the stochastic computation procedure.

\begin{table}[tp]
  \caption{The polynomial coefficient fitted from the far-field diffusivities in Fig.~\ref{fig:dinf-l2}.  The data is for polydisperse suspensions with $\lambda = 2$ and $y_2 = 0.5$.  The far-field self-diffusivity $d_s^\ff$ can be expressed as $d_s^\ff/d_0 = 1+ c_1\phi + c_2 \phi^2 + c_3 \phi^3 $, where $d_0$ is the single particle diffusivity.
}
  \label{tab:ff-coeff}
  \centering
  \begin{tabular}{c|c|c|c|c}
    \hline     \hline
    &      $d_{s,1}^{t,\ff}$  &  $d_{s,2}^{t,\ff}$  &   $d_{s,1}^{r,\ff}$ & 
$d_{s,2}^{r,\ff}$ \\
\hline
    $c_1$ & -1.27  & -1.70 & -0.207 & -0.538 \\
    $c_2$ & 0.536  & 1.005 & -0.131 & -0.312 \\
    $c_3$ & -0.018 & -0.12 & -0.091 &  0.19  \\
    \hline\hline
  \end{tabular}
\end{table}

\begin{figure}
  \centering
  \includegraphics[width=3.5in]{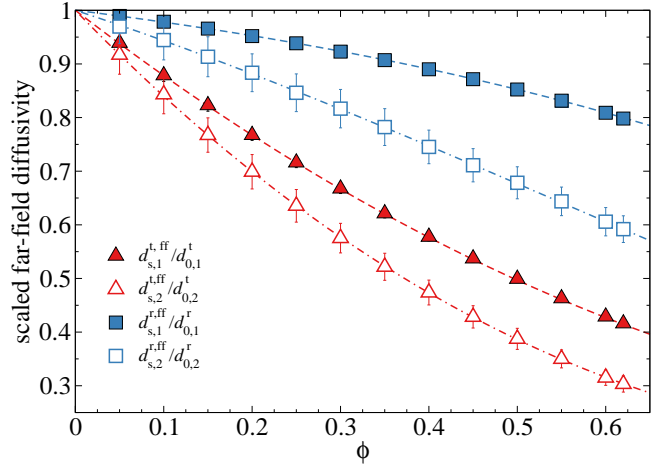}
  \caption{(Color online) The species far-field short-time translational and rotational self-diffusivities, $d_{s,\alpha}^{t,\ff}$ and $d_{s,\alpha}^{r,\ff}$, respectively, as a function of the total volume fraction $\phi$ for bidisperse hard-sphere suspensions with $\lambda = 2$ and $y_2 = 0.5$. The results scaled with the single particle translation and rotational diffusivity, $d^t_{0,\alpha}$ and $d^r_{0,\alpha}$, respectively.  The symbols are the computation results, and the dashed and the dash-dotted lines are polynomial fittings for the small and the large particles, respectively.
}
  \label{fig:dinf-l2}
\end{figure}

We also calculated the far-field short-time translational and rotational self-diffusivities $d_{s,\alpha}^{t, \ff}$ and $d_{s,\alpha}^{r, \ff}$, where ``$\ff$'' suggests only the far-field HIs without the lubrication corrections are considered.  They are the input for subsequent \mbox{SEASD-nf} computations in Sec.~\ref{sec:equil-susp} and~\ref{sec:rheol-bidisp-susp}.  The $N^{-1/3}$ size dependency in the far-field translational diffusivity $d_{s,\alpha}^{t, \ff}$ is corrected using Eq.~\eqref{eq:dst-corr} with the corresponding far-field viscosity.  Fig.~\ref{fig:dinf-l2} shows $d_{s,\alpha}^{t, \ff}$ and $d_{s,\alpha}^{r, \ff}$ for bidisperse suspensions up to $\phi = 0.62$.  Compared to Fig.~\ref{fig:dstr}, the far-field diffusivities exhibit weaker volume fraction dependence, and they do not have sharp reductions at high volume fractions.  Consistent with Fig.~\ref{fig:dstr}, $d_{s,\alpha}^{r, \ff}$ also exhibits stronger particle size dependence compared to its translational counterpart.  In general, the $\phi$ dependence of any scaled far-field diffusivity $d_s^\ff/d_0$, with $d_0$ the corresponding single-particle data, can be adequately captured by a cubic polynomial $d_s^\ff/d_0 = 1+ c_1\phi + c_2 \phi^2 + c_3 \phi^3$, where the coefficients $c_i$, $i\in\{1,2,3\}$, only depend on the suspension composition.  The fitting coefficients for bidisperse suspensions with $\lambda = 2$ and $y_2=0.5$ are presented Table \ref{tab:ff-coeff}.  The polynomial fittings, also shown in Fig.~\ref{fig:dinf-l2} in dashed and dash-dotted lines for the small and the large particles, respectively, indeed describe the computation data.  Not shown in Fig.~\ref{fig:dinf-l2} are the SEASD far-field diffusivities for monodisperse suspensions, which are identical to those of Banchio \& Brady~\cite{asd-brownian_banchio_jcp2003}.

\subsubsection{Instantaneous sedimentation velocity}
\label{sec:inst-sedim-veloc}

\begin{figure}
  \centering
  \includegraphics[width=3.5in]{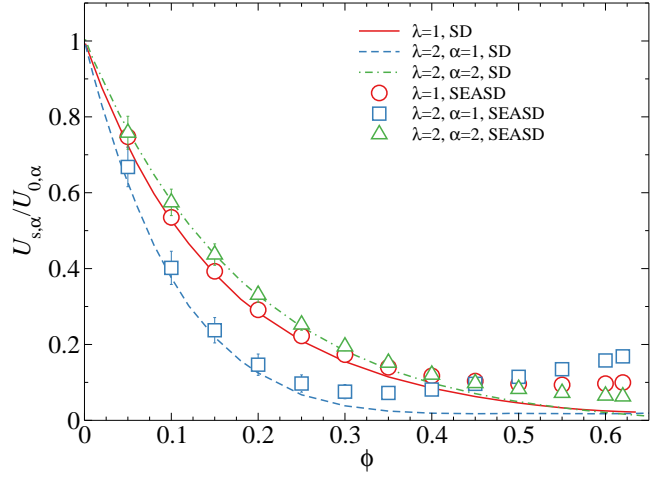}
  \caption{(Color online) 
The scaled species instantaneous sedimentation velocity, $U_{s,\alpha}/U_{0,\alpha}$, as a function of the total volume fraction $\phi$ for monodisperse and bidisperse hard-sphere suspensions with $\lambda = 2$ and $y_2 = 0.5$.  The single particle sedimentation velocity is $U_{0,\alpha}$.  The SEASD results are shown in symbols and the conventional SD results from Wang \& Brady~\cite{sd-bidisperse_wang_jcp2015} are shown as lines.
}
  \label{fig:used}
\end{figure}

The species instantaneous sedimentation velocities $U_{s,\alpha}$ are computed by applying a uniform external force $F_\alpha$ to each species.  For bidisperse suspensions, the sedimentation velocity $U_{s,\alpha}$ also depends on the species density ratio~\cite{sed-general_batchelor_jfm1982}, $\gamma = \Delta \rho_2/\Delta \rho_1$, with $\Delta\rho_\alpha = \rho_\alpha - \rho_0$ the density difference of species $\alpha$.  The species force ratio satisfies $F_2 / F_1 = \gamma \lambda^3$, and here we set $\gamma=1$ to facilitate comparison with earlier results.  To eliminate the $N^{-1/3}$ size dependence, the following corrections are added to the results:
\begin{align}
  \label{eq:used-1-corr}
  \Delta_N U_{s,1} & = \frac{1.76 U_{0,1} }{(x_1+x_2\lambda^3)^{\frac{1}{3}}} 
\frac{\eta_0}{\eta_s}
\left( \frac{\phi}{N} \right)^{\frac{1}{3}} \left[ S_{11}(0) + \lambda^3\gamma 
\sqrt{\frac{x_2}{x_1}} S_{12}(0) \right], \\
  \label{eq:used-2-corr}
  \Delta_N U_{s,2} & =   \frac{1.76 U_{0,1} }{(x_1+x_2\lambda^3)^{\frac{1}{3}}} 
\frac{\eta_0}{\eta_s}
\left( \frac{\phi}{N} \right)^{\frac{1}{3}} \left[ \sqrt{\frac{x_1}{x_2}} 
S_{21}(0)  + \lambda^3\gamma  S_{22}(0)  \right],
\end{align}
where $U_{0,\alpha} = F_{\alpha}/(6\pi\eta_0 a_\alpha)$ is the single particle sedimentation velocity and $S_{\alpha\beta}(0)$ is the partial static structural factors in the zero wave number limit.  Eqs.~\eqref{eq:used-1-corr} and \eqref{eq:used-2-corr} are based on the finite-size correction for partial hydrodynamic functions~\cite{sd-bidisperse_wang_jcp2015}.  Here, the partial static structural factors are computed from the polydisperse Percus-Yevic integral equations~\cite{Percus1958,py-hs-mixture_lebowitz_physrev1964, sk-binary-mix_ashcroft_pr1967,sk-binary-mix-errata_ashcroft_pr1968}.

Fig.~\ref{fig:used} presents the SEASD $U_{s,\alpha}/U_{0,\alpha}$ in symbols, which are not the identical to the conventional SD results shown in lines.  The difference is especially pronounced at high volume fractions.  For monodisperse suspensions, the SEASD and the conventional SD agree with each other satisfactorily up to $\phi\approx 0.3$, and at higher $\phi$, the SEASD results become significantly higher.  This difference is from the mean-field quadrupole term, which is absent in SEASD.  Despite the quantitative differences, the  SEASD monodisperse sedimentation velocity remain positive and physical.  A similar overestimation of the sedimentation velocity is also found when comparing ASD results~\cite{asd_sierou_jfm01} and the conventional SD results~\cite{brady-sd-ew_jfm_88} for simple cubic arrays.

The differences between the SEASD and the conventional SD results are more significant for bidisperse suspensions.  For $U_{s,1}$ of the small particles, the differences are not evident until $\phi = 0.3$, and for $U_{s,2}$ of the large particles, the differences are obvious even at $\phi \approx 0.2$.  Moreover, $U_{s,2}$ exhibits a minimum and increases with $\phi$ at higher volume fraction, leading to a crossing of $U_{s,1}$ and $U_{s,2}$ at $\phi = 0.45$.  These unphysical behaviors are caused by inaccurate HI computations at the stresslet-strain rate level.  Apparently, the HIs of the large particles, which are surrounded by many small particles, are more complicated than those of the small particles and more difficult to capture accurately.  Note that for sedimentation the lubrication interactions are not important and one must rely on the far-field mobility for all HIs.

Fig.~\ref{fig:used} also illustrates that sedimentation problems in dense bidisperse suspensions, even at $\lambda = 2$, is challenging for SEASD.  Incorporating the mean-field quadrupole term~\cite{brady-sd-ew_jfm_88}, $(1-\tfrac{1}{5}\phi)$, in the grand mobility tensor can significantly improve the results~\cite{sd-bidisperse_wang_jcp2015}.  However, such incorporation is not carried out in this work.

\subsubsection{High-frequency dynamic shear and bulk viscosities}
\label{sec:high-freq-dynam}

\begin{figure}
  \centering
  \includegraphics[width=6in]{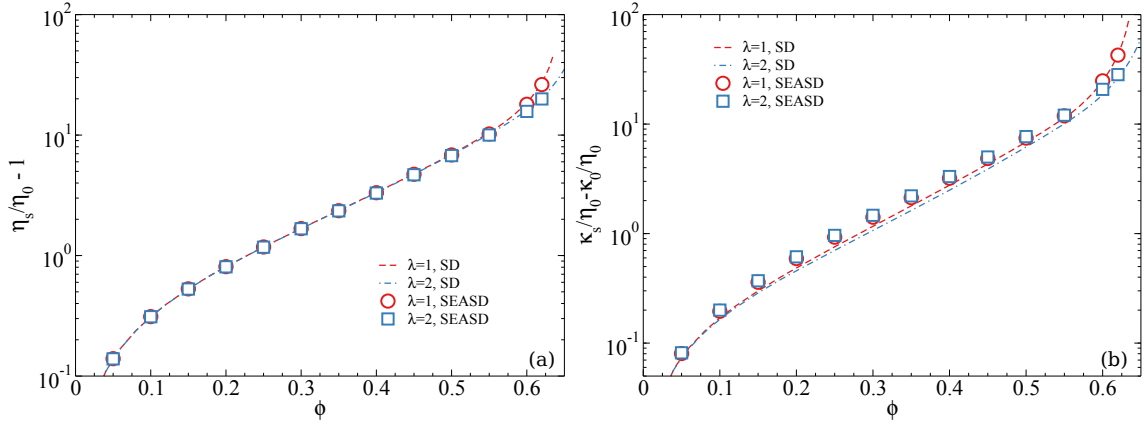}
  \caption{(Color online) 
The high-frequency dynamic (a): shear viscosity $\eta_s$ and (b): bulk viscosity $\kappa_s$ as functions of the total volume fraction $\phi$ for monodisperse and bidisperse hard-sphere suspensions with $\lambda = 2$ and $y_2 = 0.5$.  The results are scaled with the solvent viscosity $\eta_0$, and only the particle contributions, $\eta_s/\eta_0 -1$ and $(\kappa_s-\kappa_0)/\eta_0$ are presented. The SEASD results are shown as symbols and the conventional SD results~\cite{sd-bidisperse_wang_jcp2015} are shown as lines.
}
  \label{fig:etakappa}
\end{figure}

The high-frequency dynamic shear and bulk viscosities, $\eta_s$ and $\kappa_s$, are respectively defined as,
\begin{equation}
\label{eq:eta-s}
  \eta_s = \eta_0 + n\langle \tS^\mathrm{E} \rangle_{xy}/\dot{\gamma}\text{, 
and }\kappa_s = \kappa_0 + \tfrac{1}{3}n\langle \tS^\mathrm{E} 
\rangle:\tI/\dot{e},  
\end{equation}
where $\gamd$ is the imposed strain rate, $\dot{e}$ is the imposed uniform expansion rate, $\tS^\mathrm{E}$ is the hydrodynamic stresslet in Eq.~\eqref{eq:se}, and the subscript $xy$ denotes the velocity-velocity gradient component.  They are directly computed from SEASD and exhibit little size dependencies.  Experimentally, $\eta_s$ and $\kappa_s$ are measured by imposing high-frequency, low-amplitude deformations, such that the suspension microstructures are only slightly perturbed, and the Brownian stress contributions are out of phase with the applied deformations~\cite{sd-ew-transport-coeff-pt1_phillips_pof1988}.

Fig.~\ref{fig:etakappa}a and \ref{fig:etakappa}b present the volume fraction $\phi$ dependency of the particle contributions to the high-frequency dynamic shear and bulk viscosities, $\eta_s/\eta_0 -1$ and $(\kappa_s-\kappa_0)/\eta_0$, respectively.  The SEASD calculations are shown in symbols, and the corresponding conventional SD results are shown in lines.  For $\eta_s$, the SEASD and the conventional SD results agree well over the entire $\phi$ range.  The results for monodisperse and bidisperse suspensions with $\lambda = 2$ are almost identical when $\phi<0.55$.  At higher volume fractions, the monodisperse $\eta_s$ are more sensitive to $\phi$ compared to the bidisperse results, as introducing particles of difference sizes significantly alters the suspension hydrodynamic environment in this limit.  Unlike sedimentation, for the shear viscosity lubrication interactions are important and dominate the behavior at high $\phi$.

For the high-frequency dynamics bulk viscosity $\kappa_s$ in Fig.~\ref{fig:etakappa}b, the SEASD and conventional SD results show qualitative agreement with noticeable quantitative differences at moderate $\phi$: the SEASD results are higher and less sensitive to the particle size ratio $\lambda$.  The differences are caused by different pressure moment computation procedures.  Recall that the far-field grand mobility tensor $\fM$ is not symmetric by construction, and the symmetry of $\fM^{-1}$ must be restored for subsequent calculations.  This is done in conventional SD by explicit copy of matrix elements after the matrix inversion~\cite{swaroopthesis2010}.  This is not applicable for the matrix-free computation of $\fM$ in SEASD.  Here, the pressure moment is computed from the far-field forces and stresses.  Fig.~\ref{fig:etakappa}b shows that the two conceptually equivalent approaches do lead to small quantitative differences.  Moreover, for dense suspensions, such differences are masked by the dominance of lubrication interactions.  Therefore, the SEASD and the conventional SD results agree well at low and high $\phi$.  Near the close packing limit, $\kappa_s$ for bidisperse suspensions is significantly lower than that of the monodisperse case, since the particle size polydispersity improves the particle packing.

\subsection{Equilibrium suspensions}
\label{sec:equil-susp}

Here we present the dynamic simulation results with SEASD and \mbox{SEASD-nf} for monodisperse and bidisperse Brownian suspensions at zero P\'{e}clet number.  In particular, we are interested in the following equilibrium properties: the osmotic pressure $\Pi$, the high-frequency dynamic bulk modulus $K'_\infty$, and high-frequency dynamic shear modulus, $G'_\infty$.  The dynamic simulations are carried out with $100$ particles over $200$ diffusive time units with a time step $\Delta t d_{0,1}^t / a_1^2 =10^{-3} $.  The mobility computation in SEASD is performed on GPUs with $M=32$, $P=11$, and $r_c=4(a_i+a_j)$, and the far-field Brownian force is calculated using the IVP method in Sec.~\ref{sec:brownian-suspensions} with $\Delta\tau = 0.1$.  The tolerance for the iterative solver is $10^{-3}$ and the tolerance for matrix inversion in Eqs.~\eqref{eq:ivp-forward} and \eqref{eq:ivp-backward} is $0.02$.  The composition of bidisperse suspensions are $\lambda=2$ and $y_2=0.5$.  Therefore, for the \mbox{SEASD-nf} computations the coefficients in Table~\ref{tab:ff-coeff} are used.  Note that with $\pe=0$, \mbox{SEASD-nf} computations do not contain far-field mobility evaluations.

\subsubsection{Osmotic pressure}
\label{sec:osmotic-pressure}

\begin{figure}
  \centering
  \includegraphics[width=3.5in]{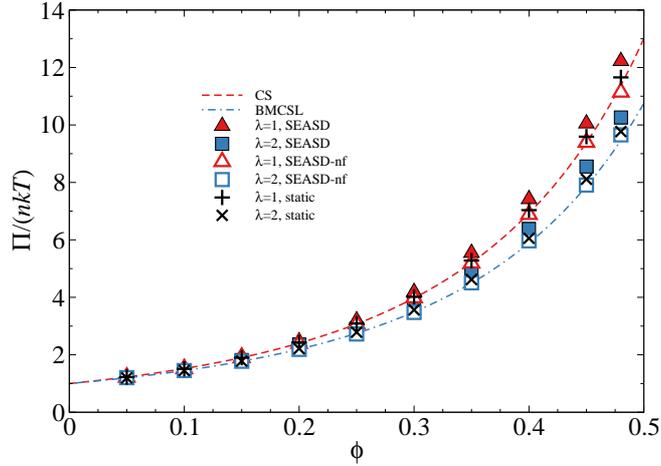}
  \caption{(Color online) The equilibrium osmotic pressure $\Pi/(n\kT)$ of monodisperse and bidisperse Brownian suspensions with $\lambda=2$ and $y_2 = 0.5$, as a function of volume fraction $\phi$. The dashed line represents the CS equation of state, Eq.~\eqref{eq:eos-cs}, and the dash-dotted line represents the BMCSL equation of state, Eq.~\eqref{eq:eos-bmcsl}.}
  \label{fig:pres}
\end{figure}

The osmotic pressure of an equilibrium suspension is defined as
\begin{equation}
  \label{eq:osm-pres-def}
  \Pi = n\kT -\tfrac{1}{3} n\langle \tS^\mathrm{B} \rangle : \tI,
\end{equation}
where $\langle \tS^\mathrm{B} \rangle$ is the Brownian stresslet in Eq.~\eqref{eq:sb}.  For rigid particles with no-slip boundary conditions, Brady~\cite{brady1993a} showed that the osmotic pressure is purely hydrodynamic in origin, and is identical to that of a hard-sphere fluid.  The osmotic pressure of monodisperse suspensions is well described by the Carnahan-Starling (CS) equation up to the fluid-solid transition,
\begin{equation}
  \label{eq:eos-cs}
  \frac{\Pi}{n\kT} = \frac{1+\phi + \phi^2 - \phi^3}{(1-\phi)^3}.
\end{equation}
The CS equation of state is extended to polydisperse suspensions as the Boublik-Mansoori-Carnahan-Starling-Leland (BMCSL) equation~\cite{Mansoori_mix-eos_jcp1971}:
\begin{equation}
  \label{eq:eos-bmcsl}
\frac{\Pi}{n\kT} = \frac{1+\phi + \phi^2 - 3\phi(z_1+z_2\phi)- z_3 \phi^3 
}{(1-\phi)^3}, 
\end{equation}
where $z_1 = \Delta_{12} (1+\lambda)/\sqrt{\lambda} $,  $z_2 = \Delta_{12}(y_1\lambda + y_2) /\sqrt{\lambda}$, and $z_3 = [({y_1^2x_1})^{1/3} + ({y_2^2x_2})^{1/3}]^3 $ with $\Delta_{12} = \sqrt{y_1y_2}\sqrt{x_1x_2} (\lambda-1)^2/\lambda$.

Fig.~\ref{fig:pres} presents the equilibrium osmotic pressure of monodisperse and bidisperse suspensions with $\lambda = 2$ and $y_2 = 0.5$ as functions of $\phi$ using SEASD and \mbox{SEASD-nf} computations.  The CS [Eq.~\eqref{eq:eos-cs}] and the BMCSL [Eq.~\eqref{eq:eos-bmcsl}] equations of state at the corresponding bidisperse compositions are respectively shown in dashed and dash-dotted lines.  Also shown in Fig.~\ref{fig:pres} are the static computation results with $N=200$, denoted ``static''.  The static computations do not consider particle dynamics, and calculate the osmotic pressure by taking a full Brownian step from independent particle configurations in a Monte-Carlo fashion.  In Fig.~\ref{fig:pres}, at each volume fraction $500$ independent configurations are used in the static computations.

The osmotic pressures from the SEASD, the \mbox{SEASD-nf}, and the static computations agree with the CS and BMCSL predictions in Fig.~\ref{fig:pres}.  The static computations show the best agreement over the entire $\phi$ range, and this directly validates the Brownian stress computation method in Sec.~\ref{sec:brownian-suspensions}.  The dynamic SEASD results are slightly higher than the theoretical predictions because the configuration evolution is affected by the finite $\Delta \tau$ in the far-field Brownian force computation.  The slight difference does not invalidate this approach as it is well within the discretization errors of Eqs.~\eqref{eq:ivp-forward} and \eqref{eq:ivp-backward}.  Note that, as long as the tolerances for the iterative solution of Eqs.~\eqref{eq:ivp-forward} and \eqref{eq:ivp-backward} are smaller than the discretization step size $\Delta \tau$, the principal source of error is the time discretization.  We have verified that reducing the iterative solver tolerance with fixed $\Delta \tau$ does not improve the results.  Finally, the agreement in the bidisperse osmotic pressures from \mbox{SEASD-nf} and the BMCSL equation validates the extension of the mean-field Brownian approximation to polydisperse systems.  The \mbox{SEASD-nf} results are only slightly lower than the theoretical predictions, which is acceptable considering the substantial speedup offered by this approach.

\subsubsection{High-frequency dynamic moduli}
\label{sec:high-freq-moduli}

\begin{figure}
  \centering
  \includegraphics[width=6in]{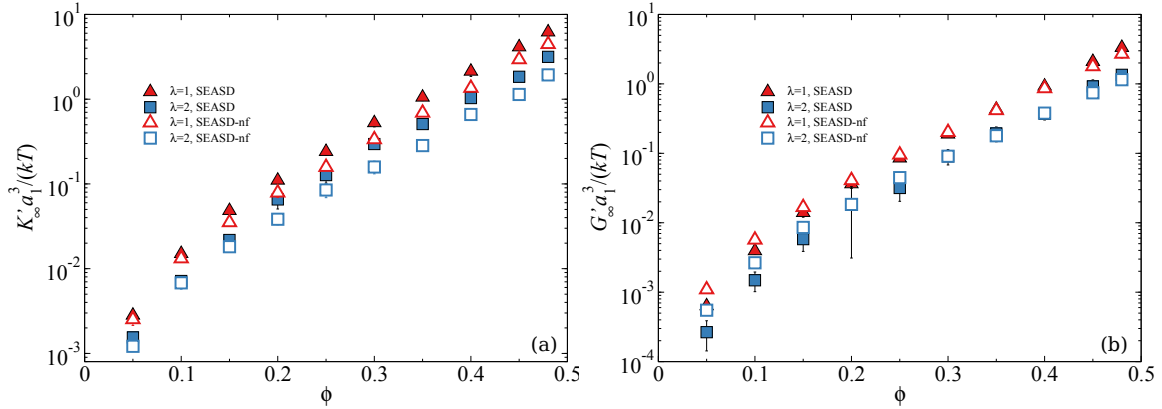}
  \caption{(Color online) The high-frequency dynamic moduli: (a) the bulk modulus $K'_\infty a_1^3/(\kT)$, and (b) the shear modulus $G'_\infty a_1^3/(\kT)$, as functions of volume fraction $\phi$ for equilibrium monodisperse and bidisperse Brownian suspensions with $\lambda=2$ and $y_2 = 0.5$. The results are computed from SEASD (filled symbols) and \mbox{SEASD-nf} (open symbols).
}
  \label{fig:bgmod}
\end{figure}

The suspension high-frequency dynamic bulk and shear moduli, $K'_\infty$ and $G'_\infty$ respectively, can be computed from the short-time limit of the pressure-pressure and stress-stress autocorrelation functions~\cite{brady1993b,Nagele_mixture-viscoelast_jcp1998,swaroopthesis2010}, \ie,
\begin{equation}
  \label{eq:def-cscp}
K'_\infty = \lim_{t\rightarrow 0} \frac{V}{\kT} \avg{\delta\Pi(t)\delta\Pi(0)} 
\text{, and }  
G'_\infty = \lim_{t\rightarrow 0} \frac{V}{\kT} \avg{\sigma(t)\sigma(0)},
\end{equation}
where $\delta \Pi$ is the osmotic pressure fluctuations and $\sigma$ is the off-diagonal components of the bulk stress $\langle \tens{\Sigma}\rangle$ in Eq.~\eqref{eq:bulk-stress-st}.  Note that the viscoelasticity of colloidal suspensions is entirely of hydrodynamic origin, and without HIs, \eg, in hard-sphere fluids, these moduli are infinite.

Fig.~\ref{fig:bgmod}a and \ref{fig:bgmod}b respectively present $K'_\infty$ and $G'_\infty$ of monodisperse and bidisperse suspensions as functions of $\phi$ from the same SEASD and \mbox{SEASD-nf} dynamic simulations of Fig.~\ref{fig:pres}.  Both $K'_\infty$ and $G'_\infty$ grow rapidly with $\phi$, and at the same volume fraction, the monodisperse moduli are always higher.  In Fig.~\ref{fig:bgmod}a, the bulk modulus $K'_\infty$ computed from SEASD and \mbox{SEASD-nf} share the same qualitative behavior.  However, the SEASD results are almost always higher than the \mbox{SEASD-nf} results except at small $\phi$, and their differences grow with increasing $\phi$.  This is consistent with the growing differences in $\Pi$ with increasing $\phi$ in Fig.~\ref{fig:pres}.  On the other hand, in Fig.~\ref{fig:bgmod}b the differences in the shear modulus $G'_\infty$ between the SEASD and the \mbox{SEASD-nf} results decrease with increasing $\phi$, with the \mbox{SEASD-nf} data higher at low volume fractions.  Note that the bidisperse SEASD results show large fluctuations when $\phi=0.2\sim 0.25$, most likely due to the small number of large particles at $N=100$ and the particular particle spacing at this volume fraction. Finally, small differences in fluctuation quantities such as $K'_\infty$ and $G'_\infty$ are expected for SEASD and \mbox{SEASD-nf} because the Brownian stresses are computed differently.  However, more importantly, the same qualitative behaviors are followed in both methods.

\subsection{Rheology of bidisperse suspensions}
\label{sec:rheol-bidisp-susp}

The final validation of SEASD and \mbox{SEASD-nf} is the steady shear rheology of Brownian suspensions at constant strain rate.  Both monodisperse and bidisperse hard-spehre suspensions are considered: the volume fractions are fixed at $\phi=0.45$ in both cases, and the bidisperse composition is $\lambda = 2$ and $y_2=0.5$.  The results are extracted from SEASD and \mbox{SEASD-nf} simulations with GPU mobility computation over a wide range of P\'{e}clet number $\pe=6\pi\eta_0a_1^3\gamd/(\kT)$.  Moreover, we introduce a small excluded volume on each particle to emulate the effects of surface asperities or polymer coating and to prevent particle overlap.  It is characterized by,
\begin{equation}
  \label{eq:excl-vol}
  \delta=1-a_i/b_i,
\end{equation}
where $b_i$ is the excluded volume radius for each particle.  The SEASD and \mbox{SEASD-nf} simulations are carried out at $\delta = 5\times 10^{-4}$ with $N=200$ over $150$ dimensionless time units with a step size $10^{-3}$.  Other simulation parameters are similar to those in Sec.~\ref{sec:equil-susp}.  The data are averaged in segments after the steady state is reached, usually after $20$ dimensionless time units.  As is customary, the $x$-direction is the velocity direction, the $y$-direction is the velocity gradient direction, and the $z$-direction is the vorticity direction.

\subsubsection{Shear viscosity}
\label{sec:shear-viscosity}

\begin{figure}
  \centering
  \includegraphics[width=6in]{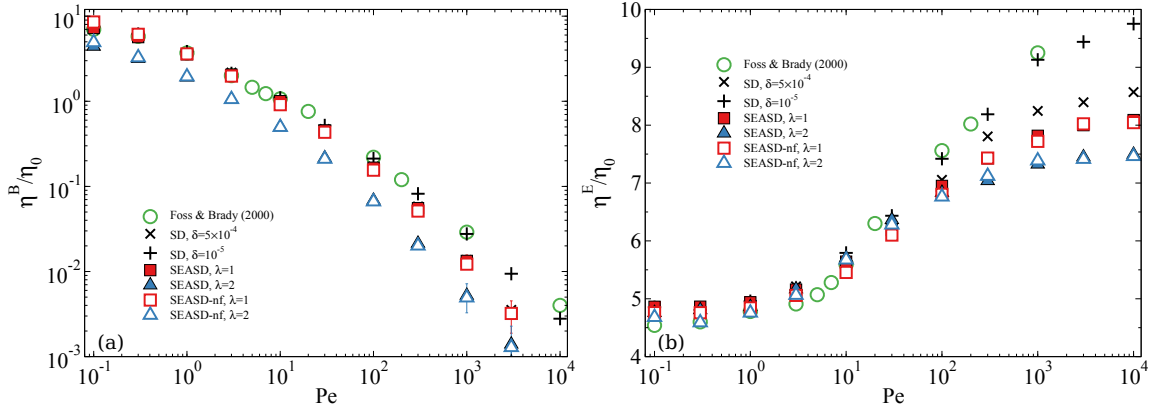}
  \caption{(Color online) Different viscosity contributions to the rheology of monodisperse and bidisperse hard-sphere suspensions: (a) the Brownian viscosity $\eta^B/\eta_0$ and (b) the flow viscosity $\eta^E/\eta_0$, as functions of $\pe$.  The volume fraction $\phi = 0.45$ in both cases, and the bidisperse composition is $\lambda = 2$ and $y_2 = 0.5$.
}
  \label{fig:etaBH}
\end{figure}

Fig.~\ref{fig:etaBH}a and \ref{fig:etaBH}b respectively present the Brownian viscosity $\eta^B$ and the flow viscosity $\eta^E$ as functions of the P\'{e}clet number. These viscosities are defined as 
\begin{equation}
  \label{eq:etahb-def}
  \eta^B = n\langle \tS^\mathrm{B}\rangle_{xy} / \gamd \text{ and }
  \eta^E = n\langle \tS^\mathrm{E}\rangle_{xy} / \gamd,
\end{equation}
with $\langle \tS^\mathrm{B}\rangle$ in Eq.~\eqref{eq:sb} and $\langle \tS^\mathrm{E}\rangle$ in Eq.~\eqref{eq:se}.  
In this figure, the monodisperse data are shown in squares and the bidisperse data in triangles, with the SEASD results in filled symbols and the \mbox{SEASD-nf} results in open symbols.  For comparison, the SD results of Foss \& Brady~\cite{sd-brownian-susp_brady_jfm2000} for monodisperse suspensions are presented in open circles.  To clarify the effects of the excluded volume parameter $\delta$ on viscosities, another set of monodisperse SD simulations with $N=30$ are performed at $\delta = 5\times 10^{-4}$ and $10^{-5}$, and the results are shown as crosses and pluses respectively.  In all cases, the stress contributions from inter-particle forces are negligible, and therefore are not presented.

In Fig.~\ref{fig:etaBH} both the Brownian viscosity $\eta^B$ and the flow viscosity $\eta^E$ exhibit the expected behaviors: with increasing $\pe$, $\eta^B$ decreases (shear-thinning) and $\eta^E$ grows (shear-thickening).  In addition, there are several important observations.  First of all, the excluded volume parameter $\delta$ introduces quantitative effects on the suspension rheology, especially at high $\pe$.  Comparing the SD results with $\delta = 5\times 10^{-4}$ and $10^{-5}$, increasing $\delta$ enhances the shear-thinning of $\eta^B$ and weakens the shear-thickening of $\eta^E$, especially at high $\pe$.  At low $\pe$, the effect of $\delta$ is almost unnoticeable.  The SD results at $\delta = 10^{-5}$ agree well with those of Foss \& Brady~\cite{sd-brownian-susp_brady_jfm2000}, and the results at $\delta = 5\times 10^{-4}$ are consistent with the monodisperse SEASD and \mbox{SEASD-nf} results, with larger differences shown in $\eta^E$.  This difference is most likely due to the number of particles in the computations.  Next, the bidisperse Brownian viscosity $\eta^B$ is always lower than the monodisperse value at all $\pe$, and for the flow viscosity $\eta^E$, their difference is most apparent at high $\pe$.  The large difference in $\eta^E$ at high $\pe$ suggests distinct HIs and structures between the monodisperse and the bidisperse suspensions, since Fig.~\ref{fig:etakappa}a suggests $\eta^E$ is insensitive to equilibrium suspension structures at $\phi=0.45$.  Finally, the SEASD and \mbox{SEASD-nf} results in Fig.~\ref{fig:etaBH} almost always overlap each other, showing that the mean-field Brownian approximation is valid over the entire P\'{e}clet number range.  At high $\pe$, the Brownian viscosity $\eta^B$ from SEASD shows larger fluctuations compared to the \mbox{SEASD-nf} results as the Brownian stresses are difficult to compute with highly anisotropic structures.  However, these fluctuations do not affect the overall viscosity since the Brownian contribution at high $\pe$ is insignificant.

\subsubsection{Non-equilibrium osmotic pressures}

\begin{figure}
  \centering
  \includegraphics[width=6in]{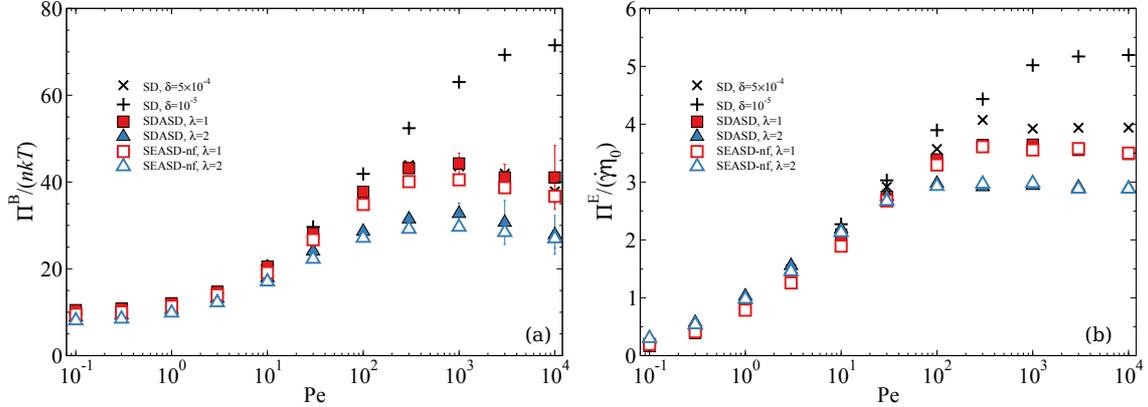}
  \caption{(Color online) Different contributions to the osmotic pressures of monodisperse and bidisperse hard-sphere suspensions: (a) the Brownian contribution scaled with $n\kT$, $\Pi^B/(n\kT)$, and (b) the flow contribution scaled with $\eta_0\gamd$, $\Pi^H/(\gamd \eta_0)$, as functions of $\pe$.  The volume fraction is $\phi = 0.45$ in both cases, and the bidisperse composition is $\lambda = 2$ and $y_2 = 0.5$.
}
  \label{fig:piBH}
\end{figure}

Fig.~\ref{fig:piBH}a and \ref{fig:piBH}b present the Brownian and the flow contributions to the suspension osmotic pressure, 
\begin{equation}
  \label{eq:nstress}
  \Pi^B = n\kT -\tfrac{1}{3}n\langle\tS^\mathrm{B}\rangle : \tI \text{ and }
  \Pi^E=-\tfrac{1}{3}n\langle\tS^\mathrm{E}\rangle : \tI,
\end{equation}
respectively, as functions of P\'{e}clet number $\pe$.  In these figures, the scaling for the Brownian contribution is $n\kT$ and the scaling for the flow contribution $\Pi^E$ is $\eta_0\gamd$.  Similar to Fig.~\ref{fig:piBH}, the monodisperse data are presented in squares and the bidisperse data in triangles, with the SEASD results in filled symbols and \mbox{SEASD-nf} results in open symbols.  Fig.~\ref{fig:piBH} also presents the $N=30$ monodisperse SD results with $\delta = 5\times 10^{-4}$ and $10^{-5}$ in crosses and pluses, respectively.  Similarly to the shear stresses, the inter-particle contribution to the osmotic pressures is also negligible compared to the contributions from HIs.

In Fig.~\ref{fig:piBH}, both $\Pi^B/(n\kT)$ and $\Pi^E/(\gamd\eta_0)$ grow with increasing $\pe$ when $\pe<100$.  The Brownian contribution $\Pi^B/(n\kT)$ asymptotes the equilibrium value as $\pe\rightarrow 0$.  At higher $\pe$, the influence of the excluded volume parameter $\delta$ becomes apparent.  For the Brownian osmotic pressure contribution $\Pi^B/(n\kT)$, the SD results at $\delta = 10^{-5}$ continuously grow with $\pe$ up to $\pe = 10^4$, the highest value in our study, while with $\delta = 5\times 10^{-4}$, a maximum in $\Pi^B/(n\kT)$ around $\pe = 10^3$ is apparent. After the maximum, $\Pi^B/(n\kT)$ decreases slowly with growing $\pe$.  In this case, the parameter $\delta$ not only brings quantitative, but also qualitative differences.  On the other hand, the flow osmotic pressure contribution $\Pi^E/(\gamd \eta_0)$ increases and reaches a plateau at high $\pe$.  Comparing the SD results with $\delta = 5\times 10^{-4}$ and $10^{-5}$, increasing $\delta$ reduces the final plateau value of $\Pi^E/(\gamd\eta_0)$ at a smaller $\pe$.  Apparently, the high $\pe$ osmotic pressure is very sensitive to the excluded volume parameter $\delta$.  In terms of the normal viscosity, \ie, $\Pi/\gamd$ with $\Pi = \Pi^B+\Pi^E$, increasing $\delta$ weakens the shear thickening of the normal viscosity.  Furthermore, the SD results at $\delta = 10^{-5}$ agree qualitatively with the results of Yurkovetsky \& Morris~\cite{pressure-sd_morris_jor2008}, with slight quantitative difference due to different osmotic pressure computations.
At $\delta=5\times 10^{-4}$, the Brownian osmotic pressures $\Pi^B$ from SD and SEASD almost overlap each other in Fig.~\ref{fig:piBH}a, and $\Pi^E$ from SEASD is lower than the SD results in Fig.~\ref{fig:piBH}b.  Similarly to Fig.~\ref{fig:etaBH}b, the difference is most likely due to the small system sizes in the SD computations.  Moreover, the SEASD $\Pi^B$ also exhibits larger error bars at high $\pe$ due to the Brownian stress computation,  but such errors are of little consequences on the suspension total osmotic pressures.

For the bidisperse results shown in triangles in Fig.~\ref{fig:piBH}, the Brownian osmotic pressure $\Pi^B$ is always lower than its monodisperse counterpart, and the bidisperse $\Pi^E$ is first slightly higher than the monodisperse results at low $\pe$ and then lower at high $\pe$.  The crossing of the monodisperse and bidisperse $\Pi^E$ demonstrates the complex interplay between HIs and structures in polydisperse systems.

The \mbox{SEASD-nf} results in Fig.~\ref{fig:piBH} agree qualitatively with the SEASD computations. However, for $\Pi^B$, there are quantitative differences at both $\lambda = 1$ and $\lambda =2$, with the \mbox{SEASD-nf} results systematically lower.  This difference is inherently associated with the far-field Brownian force computations in Sec.~\ref{sec:brownian-suspensions} and the mean-field Brownian approximations, and is also encountered in Fig.~\ref{fig:pres}.  However, the quantitative discrepancies in $\Pi^B$ are still within the discretization errors of $\Delta \tau$ in Eqs.~\eqref{eq:ivp-forward} and \eqref{eq:ivp-backward}.
On the other hand, for $\Pi^E$, the \mbox{SEASD-nf} and SEASD results almost always overlap each other over the entire $\pe$ range for both bidisperse and monodisperse suspensions.  \mbox{SEASD-nf} satisfactorily captures both contributions of the suspension osmotic pressures, $\Pi^B$ and $\Pi^E$.

\subsubsection{Normal stress differences}
\label{sec:norm-stress-diff}

\begin{figure}
  \centering
  \includegraphics[width=6in]{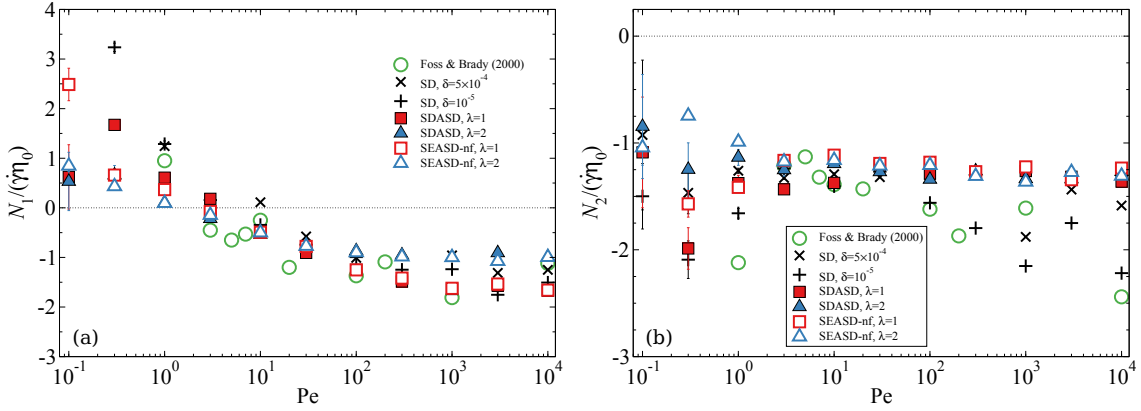}
  \caption{(Color online) The normal stress differences: (a) the first normal stress difference $N_1$ and (b) the second normal stress difference $N_2$ as functions of P\'{e}clet number $\pe$.  The volume fraction is $\phi = 0.45$ in both cases and the bidisperse composition is $\lambda = 2$ and $y_2 = 0.5$.
}
  \label{fig:N1N2}
\end{figure}

The first normal stress difference $N_1$ and the second normal stress difference $N_2$, defined as 
\begin{equation}
  \label{eq:n1n2-def}
  N_1= \langle\tens{\Sigma}\rangle_{xx} - 
\langle\tens{\Sigma}\rangle_{yy}\text{ and }
  N_2= \langle\tens{\Sigma}\rangle_{yy} - \langle\tens{\Sigma}\rangle_{zz},
\end{equation}
describe the stress anisotropy in sheared suspensions, and are important for understanding phenomena such as the shear-induced particle migrations~\cite{normal-stress-modelling-non-colloidal_morris_jor99}.
The normal stress differences $N_1$ and $N_2$ are respectively shown in Fig.~\ref{fig:N1N2}a and Fig.~\ref{fig:N1N2}b.  The monodisperse data are shown in squares and the bidisperse data in triangles, with SEASD results in filled symbols and \mbox{SEASD-nf} results in open symbols.  In addition, in Fig.~\ref{fig:N1N2}, the SD results of Foss \& Brady~\cite{sd-brownian-susp_brady_jfm2000} are presented in circles, and the SD computations at $N=30$ with $\delta = 5\times 10^{-4}$ and $10^{-5}$ are respectively shown in crosses and pluses.

In general, the first normal stress difference $N_1$ in Fig.~\ref{fig:N1N2}a changes sign from positive to negative with increasing $\pe$, and the second normal stress $N_2$ in Fig.~\ref{fig:N1N2}b remains negative for all $\pe$ studied and exhibits weak $\pe$ dependence.  The data with small systems are strongly scattered, particularly at small $\pe$.  For monodisperse suspensions, the excluded volume parameter $\delta$ has little effect on $N_1$ or $N_2$, as there lacks a qualitative difference for the SD results at $\delta = 5\times 10^{-4}$ and $10^{-5}$ in Fig.~\ref{fig:N1N2}.  These SD results in general agree with the data of Foss \& Brady~\cite{sd-brownian-susp_brady_jfm2000} when $\pe>1$.  At smaller $\pe$, the data exhibit large errors due to fluctuations in Brownian stresses, making quantitative comparisons difficult.

In Fig.~\ref{fig:N1N2} the SEASD results at $\lambda = 1$ follow the SD data with the same qualitative behaviors.  The differences at low $\pe$ is likely associated with the difficulties in measuring the fluctuating Brownian normal stresses.   In addition, the SEASD results show clearer trends at high $\pe$ thanks to larger system sizes: both $N_1$ and $N_2$ asymptote toward constant values with increasing $\pe$.  Particle size polydispersity weakens the influences of $\pe$ on the first normal stress difference $N_1$.  In Fig.~\ref{fig:N1N2}a, the bidisperse $N_1$ are less sensitive to $\pe$ compared to the monodisperse case, and as $\pe\rightarrow \infty$, the bidisperse $N_1$ asymptotes towards a negative value with a smaller magnitude.  On the other hand, the size polydispersity has little effect on the second normal stress $N_2$, as the bidisperse $N_2$ almost overlaps the monodisperse $N_2$, especially at large $\pe$.

The \mbox{SEASD-nf} and the SEASD results agree satisfactorily when $\pe\geq 10$ for both the monodisperse and bidisperse suspensions.  As expected, larger differences are found at low $\pe$, as the mean-field Brownian approximation in \mbox{SEASD-nf} explicitly removes the anisotropy in the far-field mobility tensor.  However, the \mbox{SEASD-nf} results still capture the qualitative aspect of $N_1$ and $N_2$ even in the low $\pe$ limit.

\subsubsection{Species stress distribution}

\begin{figure}
  \centering
  \includegraphics[width=6in]{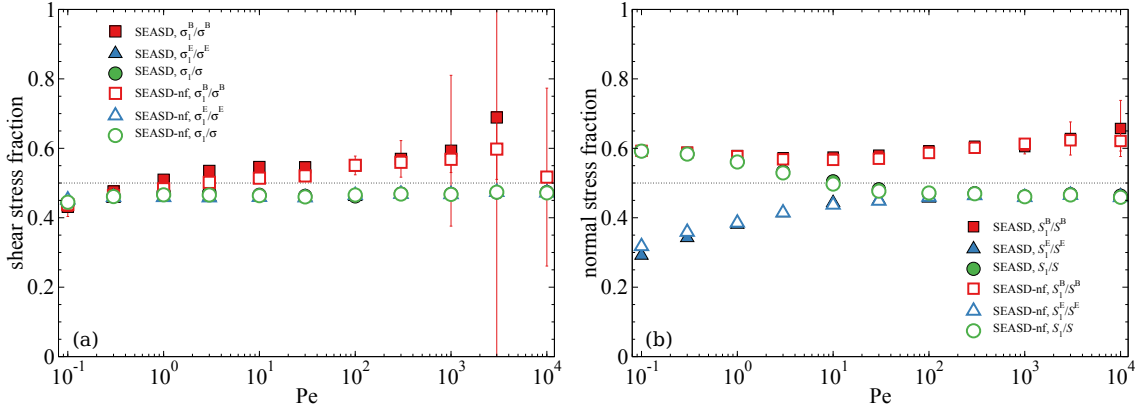}
  \caption{(Color online) The fraction of stresses taken up by the small particles (species 1) in a bidisperse suspension: (a) the fraction of the shear stress and (b) the fraction of the normal stress.  The stress fractions are shown as functions of $\pe$.  The composition of the bidisperse hard-sphere suspension is $\phi=0.45$, $\lambda = 2$, and $y_2 = 0.5$.
}
  \label{fig:frps}
\end{figure}

Stress distributions across different species are key to understand the phenomena of particle migration and segregation in polydisperse suspensions~\cite{migration-bidisperse_vdSman_fd2012}, and are presently only accessible from simulations.  Fig.~\ref{fig:frps} presents the stress distribution, expressed as the stress fraction taken up by the small particles (species 1), as functions of $\pe$ for bidisperse suspensions with $\phi=0.45$, $\lambda = 2$, and $y_2=0.5$.  Fig.~\ref{fig:frps}a shows various shear stress fractions.
In terms of the definitions in Eqs.~\eqref{eq:bulk-stress-st}--\eqref{eq:sb}, $\sigma_1/\sigma$ (circles), $\sigma^B_1/\sigma^B$ (squares) , and $\sigma^E_1/\sigma^E$ (triangles) in Fig.~\ref{fig:frps}a are
\begin{equation}
  \label{eq:frac-shear}
\sigma_1/\sigma = x_1 \langle \tens{\Sigma} \rangle_{1,xy}/ \langle \tens{\Sigma} \rangle_{xy},\;
\sigma^B_1/\sigma^B = x_1 \langle \tens{S}^{\mathrm{B}} \rangle_{1,xy}/ \langle 
\tens{S}^{\mathrm{B}} \rangle_{xy},\text{ and }
\sigma^E_1/\sigma^E = x_1 \langle \tens{S}^{\mathrm{E}} \rangle_{1,xy}/ \langle 
\tens{S}^{\mathrm{E}} \rangle_{xy},
\end{equation}
where $\langle \cdot\rangle_\alpha$ indicates averaging with respect to species $\alpha$.
Fig.~\ref{fig:frps}b presents various normal stress fractions.  The normal stress fractions $S_1/S$ (circles), $S^B_1/S^B$ (squares), and $S^E_1/S^E$ (triangles) in Fig.~\ref{fig:frps}b are similarly defined as 
\begin{equation}
  \label{eq:frac-norm}
S_1/S = x_1 (\bI:\langle  \tens{\Sigma} \rangle_{1})/ (\tI: \langle \tens{\Sigma} \rangle),\;
S^B_1/S^B = x_1 (\tI: \langle \tens{S}^{\mathrm{B}} \rangle_{1})/  
(\tI: \langle \tens{S}^{\mathrm{B}} \rangle),\text{ and }
S^E_1/S^E = x_1 (\tI: \langle \tens{S}^{\mathrm{E}} \rangle_{1})/ 
(\tI: \langle \tens{S}^{\mathrm{E}} \rangle).
\end{equation}
In both figures, the SEASD results are shown in filled symbols and the \mbox{SEASD-nf} results are shown in open symbols.

Fig.~\ref{fig:frps}a illustrates that the total shear stress is roughly equally partitioned between the two species, and the fraction $\sigma_1/\sigma$ is almost constant with respect to $\pe$.  This is largely because the flow shear stress fraction $\sigma^E_1/\sigma^E$ is insensitive to $\pe$.  The Brownian shear stress fraction  $\sigma^B_1/\sigma^B$, on the other hand, exhibits weak $\pe$ dependence: the ratio $\sigma^B_1/\sigma^B$ increases with $\pe$ from less than $0.45$ at $\pe = 0.1$ to close to $0.6$ at $\pe=100$.  At higher $\pe$, the Brownian stress fraction shows large fluctuations, also due to the difficulties associated with the anisotropic structures.  However, in this limit, the Brownian contribution to the total stress is small, and the large fluctuations in Fig.~\ref{fig:frps}a is inconsequential.  
On the other hand, the total normal stress fraction $S_1/S$ in Fig.~\ref{fig:frps}b shows stronger $\pe$ dependency, and it decreases from $0.6$ at $\pe = 0.1$ to $0.45$ at $\pe = 10^{4}$.  Contrary to shear stress distributions in Fig.~\ref{fig:frps}a, the Brownian normal stress distribution $S^B_1/S^B$ is almost constant at $0.6$, but $S^E_1/S^E$ increases from $0.3$ at $\pe = 0.1$ and asymptotes towards $0.45$ as $\pe\rightarrow \infty $.  Since the Brownian stresslet dominates at low $\pe$ and the flow stresslet dominates at high $\pe$, the normal stress distributions in Fig.~\ref{fig:frps}b are distinctively affected by both the flow and the Brownian contributions.
Fig.~\ref{fig:frps} demonstrates that both the shear and the normal stresses in bidisperse suspensions are distributed based on the species volume and the distribution weakly depends on $\pe$.  This is a useful insight for modelling polydisperse systems.

The stress distributions from \mbox{SEASD-nf} accurately capture the SEASD results except the Brownian shear stress distribution $\sigma_1^B/\sigma^B$ at high $\pe$ in Fig.~\ref{fig:frps}a, where the \mbox{SEASD-nf} results is slightly lower.  This difference, however, is expected since the mean-field Brownian approximation ignores the structural anisotropy in the suspension. Moreover, the discrepancies are only evident at P\'{e}clet numbers where the Brownian stress does not affect the overall suspension rheology.  From this perspective, the overall quality of the \mbox{SEASD-nf} approximation is deemed satisfactory.

\subsubsection{Long-time diffusion}

\begin{figure}
  \centering
  \includegraphics[width=6in]{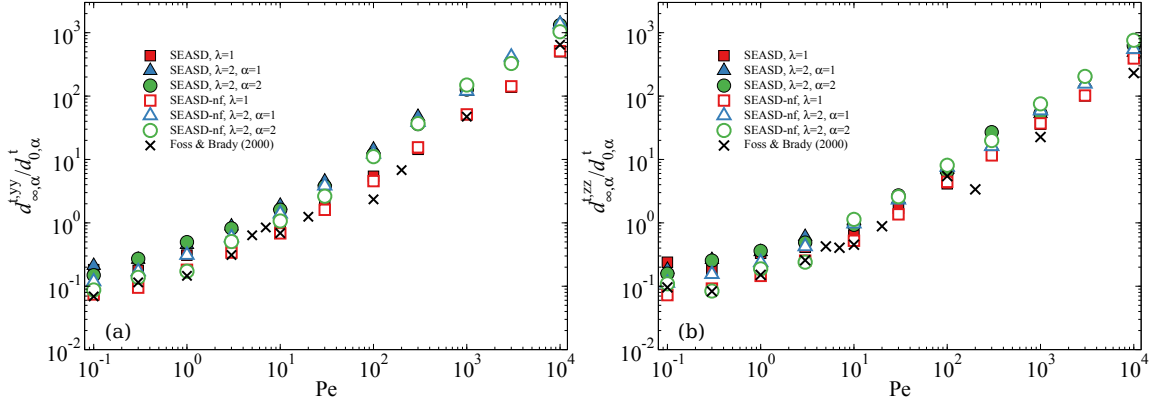}
  \caption{(Color online) The species long-time self-diffusivities: (a) the velocity gradient direction diffusivity $d^{t,yy}_{\infty,\alpha}$ and (b) the vorticity direction diffusivity $d^{t,zz}_{\infty,\alpha}$ of monodisperse and bidisperse hard-sphere suspensions as functions of $\pe$. The volume fraction is $\phi = 0.45$ for both cases, and the bidisperse composition is $\lambda = 2$ and $y_2 = 0.5$.
}
  \label{fig:dinfyz}
\end{figure}

An important characterization of the overall suspension dynamics is the translational long-time self-diffusivities.  The long-time limit refers to a time scale $t \gg \tau_D $, where, recall that, $\tau_D = 6\pi\eta_0 a_\mathrm{p}^3/\kT$ is the single particle diffusive time scale.  In this limit, the particle movement is diffusive due to extensive interactions with their neighbors.  The corresponding diffusivities are obtained from the particle mean-square displacement.  In the velocity gradient and the vorticity directions, these self-diffusivities are respectively defined as
\begin{equation}
  \label{eq:dinf-def}
  d^{t,yy}_{\infty,\alpha} = \lim_{t\rightarrow \infty} 
\tfrac{1}{2} \dd \langle (\Delta y)^2 \rangle_\alpha/\dd t
\text{ and } 
d^{t,zz}_{\infty,\alpha} = \lim_{t\rightarrow \infty} 
\tfrac{1}{2}\dd \langle (\Delta z)^2 \rangle_\alpha/ \dd t,
\end{equation}
where $\Delta y$ and $\Delta z$ are the particle trajectory fluctuations in $y$- and $z$-directions.  Fig.~\ref{fig:dinfyz}a and \ref{fig:dinfyz}b respectively present the long-time diffusivities $d^{t,yy}_{\infty,\alpha}$ and $d^{t,zz}_{\infty,\alpha}$ as functions of P\'{e}clet number.  The monodisperse results are shown in squares.  For bidisperse suspensions, the small and the large particle long-time self-diffusivities are presented in triangles and circles, respectively.  For comparison, Fig.~\ref{fig:dinfyz} also shows the results from Foss \& Brady~\cite{sd-brownian-susp_brady_jfm2000} in crosses.  Moreover, the SEASD and the \mbox{SEASD-nf} results are shown in filled and open symbols, respectively.

For monodisperse suspensions in Fig.~\ref{fig:dinfyz}, both $d^{t,yy}_{\infty}$ and  $d^{t,zz}_{\infty}$ grow with $\pe$ due to the imposed shear flow, with the velocity direction diffusivity $d^{t,yy}_{\infty}$ slightly higher.  At low $\pe$, $d^{t,yy}_{\infty}$ and $d^{t,zz}_{\infty}$ grow weakly with $\pe$, and at large $\pe$, both diffusivities are proportional to $\pe$.  The SEASD results is consistent with the SD results of Foss \& Brady~\cite{sd-brownian-susp_brady_jfm2000} at intermediate $\pe$.  The differences at large and small $\pe$ are most likely due to the system size, as in this work $N=200$ while in  Foss \& Brady~\cite{sd-brownian-susp_brady_jfm2000} $N=27$.  For bidisperse suspensions, the long-time self-diffusivities $d^{t,yy}_{\infty,\alpha}$ and $d^{t,zz}_{\infty,\alpha}$ for both species exhibit similar $\pe$ dependencies as the monodisperse case.  However, introducing a second species to the suspension apparently enhances the long-time self-diffusivities of both species, particularly at high $\pe$.  This mutual diffusivity enhancement is in contrast to the short-time diffusivities in Fig.~\ref{fig:dstr}a, where at $\phi=0.45$, the small particle diffusivity enhancement is always accompanied by the large particle diffusivity supression. Moreover, the diffusivity enhancement in $y$-direction is stronger than those in $z$-direction.

In Fig.~\ref{fig:dinfyz} the diffusivities from \mbox{SEASD-nf} in general agree with the SEASD results for both monodisperse and bidisperse suspensions.  At low $\pe$, the \mbox{SEASD-nf} diffusivity is lower, particularly for the large particles.  The agreement between SEASD and \mbox{SEASD-nf} improves with increasing $\pe$ due to the reduced influences of Brownian motion.

\subsubsection{Suspension structures}

\begin{figure}
  \centering
  \includegraphics[width=5.2in]{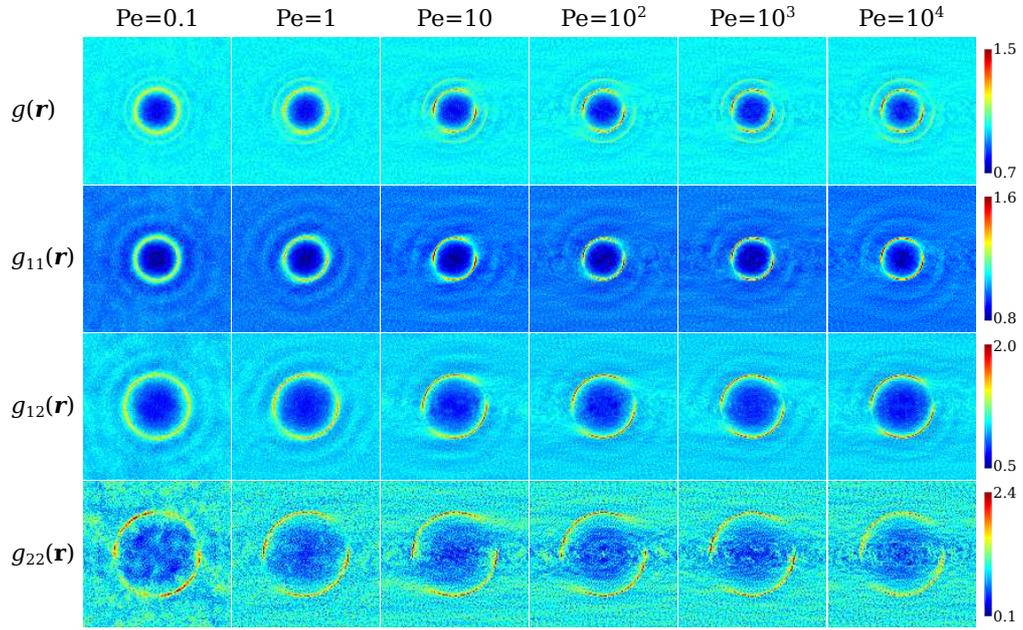}
  \caption{(Color online) The velocity-velocity gradient ($xy$-) plane projection of the pair-distribution function $g(\br)$ and the partial pair-distribution functions $g_{\alpha\beta}(\br)$ at various $\pe$ for bidisperse suspensions with $\phi = 0.45$, $\lambda = 2$, and $y_2 = 0.5$.
}
  \label{fig:g12}
\end{figure}

\begin{figure}
  \centering
  \includegraphics[width=5.2in]{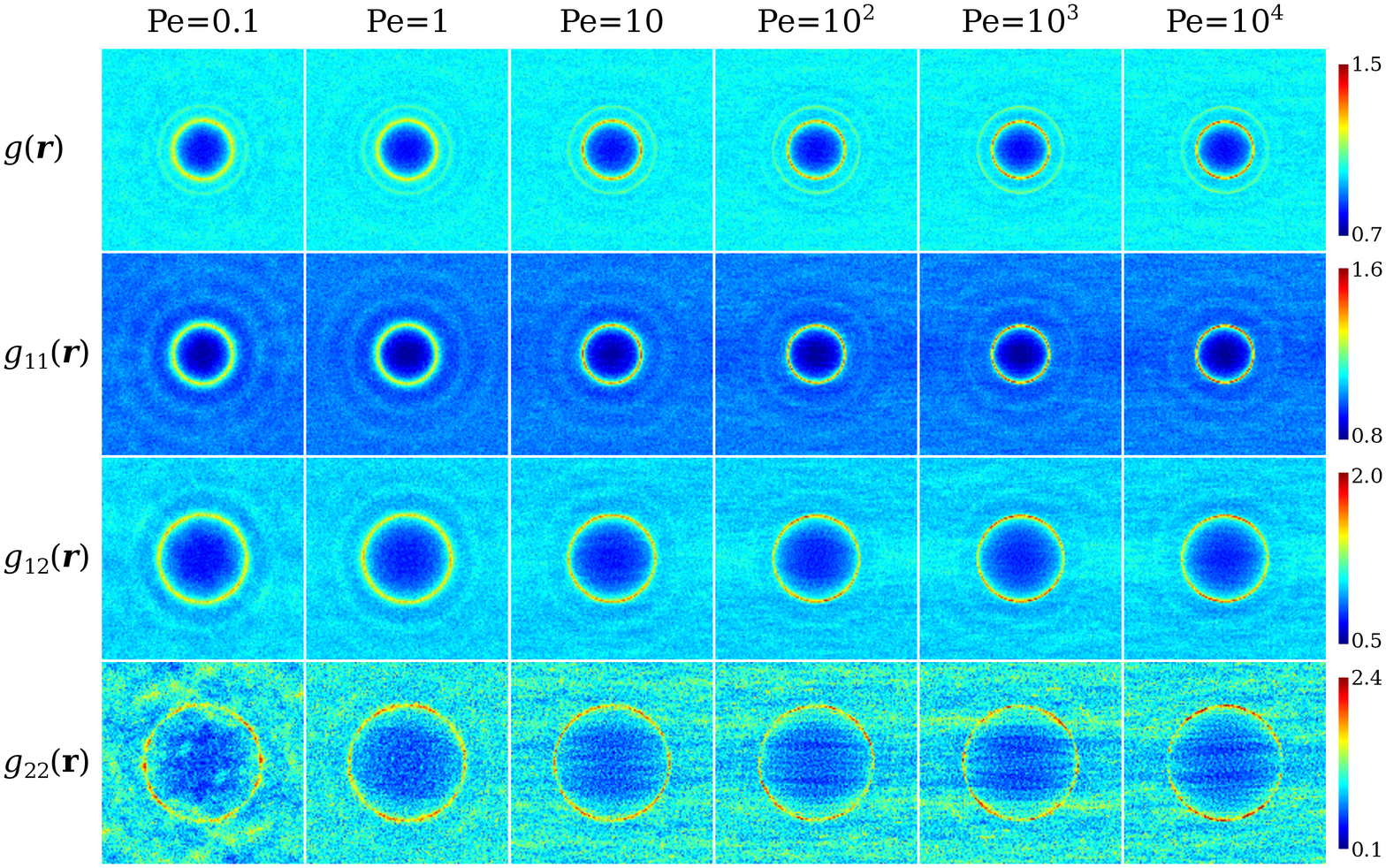}
  \caption{(Color online) The velocity-vorticity ($xz$-) plane projection of the pair-distribution function $g(\br)$ and the partial pair-distribution functions $g_{\alpha\beta}(\br)$ at various $\pe$ for bidisperse suspensions with $\phi = 0.45$, $\lambda = 2$, and $y_2 = 0.5$.
}
  \label{fig:g13}
\end{figure}

\begin{figure}
  \centering
  \includegraphics[width=5.2in]{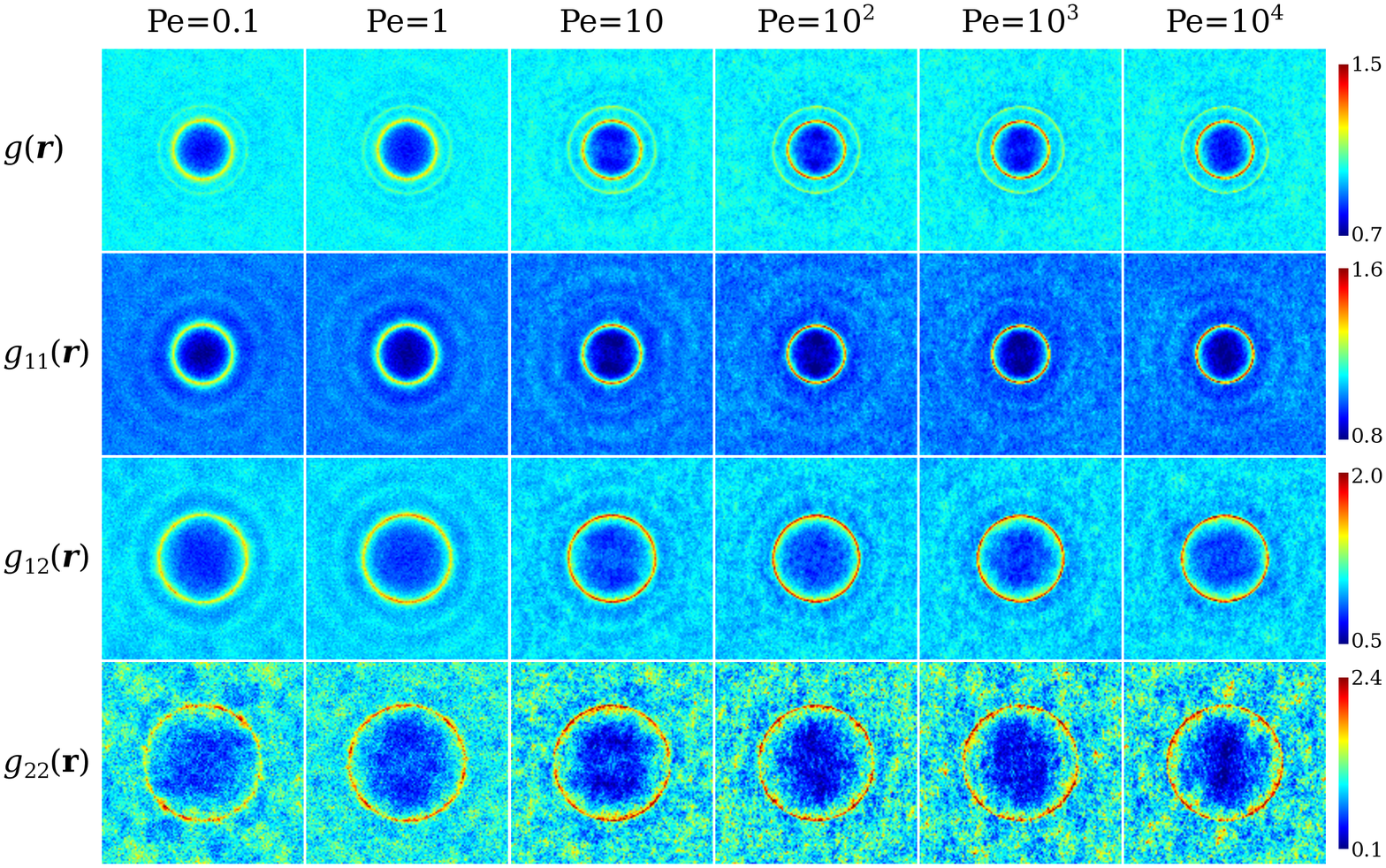}
  \caption{(Color online) The velocity gradient-vorticity ($yz$-) plane projection of the pair-distribution function $g(\br)$ and the partial pair-distribution functions $g_{\alpha\beta}(\br)$ at various $\pe$ for bidisperse suspensions with $\phi = 0.45$, $\lambda = 2$, and $y_2 = 0.5$.
}
  \label{fig:g23}
\end{figure}

Finally, we examine the structures of sheared bidisperse suspensions via the projections of the partial pair-distribution functions $g_{\alpha\beta}(\br)$, which are defined as the conditional probability of finding another particle in species $\beta$ given a particle of species $\alpha$, \ie,
 \begin{equation}
  \label{eq:gab-def}
  g_{\alpha\beta}(\br) = \frac{1}{n_\alpha n_\beta}\bigg\langle 
\sideset{}{'}\sum_{i\in\alpha, \atop j\in \beta} \frac{1}{V}\delta(\br - \br_i + 
\br_j) \bigg\rangle.
\end{equation}
They are related to the pair-distribution function $g(\br)$ through
\begin{equation}
  \label{eq:gr-def}
  g(\br) = \sum_{\alpha, \beta} x_\alpha x_\beta g_{\alpha\beta}(\br).
\end{equation}
Fig.~\ref{fig:g12}, \ref{fig:g13}, and \ref{fig:g23} present projections of $g(\br)$ and $g_{\alpha\beta}(\br)$ on the velocity-velocity gradient ($xy$-) plane, the velocity-vorticity ($xz$-) plane, and the velocity gradient-vorticity ($yz$-) plane, respectively, at selected P\'{e}clet numbers.  These figures are based on particle trajectories from SEASD simulation, and are indistinguishable from the \mbox{SEASD-nf} results.

Fig.~\ref{fig:g12} clearly displays the structural anisotropy caused by the shear flow in the $xy$-plane, characterized by the distortion of the otherwise isotropic pair-distribution rings.  With increasing $\pe$, the overall pair-distribution function $g(\br)$ shows an accumulation of neighboring particles in the compressional quadrant.  This is indicated by the brightening and thinning of the rings at $2a_1$, $a_1+a_2$, and $2a_2$, corresponding to the particle pairs of two small particles, a large and a small particle, and two large particles, respectively.  Meanwhile, the particle pairs are depleted in the extensional quadrant.

Specific changes in different types of particle pairs are revealed by examining the corresponding partial pair-distribution function $g_{\alpha\beta}(\br)$ in Fig.~\ref{fig:g12}.  The distribution of the small-small particle pairs is presented in $g_{11}(\br)$.  Similarly to $g(\br)$, $g_{11}(\br)$ is increasingly distorted and compressed in the compressional quadrant with increasing $\pe$, forming a boundary layer.  At higher $\pe$, the pair structure remain approximately unchanged.  In the extensional quadrant, the pair breakup point shifts from the extensional axis towards the velocity ($x$-) direction due to the lubrication interactions, with a clear tail of high probability outlining the trajectory of small-small pair disengagement.  The distribution of the small-large particle pairs in $g_{12}(\br)$ shows a similar structural distortion in the compressional quadrant with increasing $\pe$.  Moreover, in the extensional quadrant, the trajectory of particle disengagement is more diffusive compared to $g_{11}(\br)$ at the same $\pe$.  This suggests that particle movement in bidisperse suspensions are facilitated by the breakup of small-large particle pairs, and partially explains the mutual enhancement of long-time self-diffusivity in Fig.~\ref{fig:dinfyz}.  For the distribution of large-large particle pairs, $g_{22}(\br)$ also exhibits anisotropy with increasing $\pe$ in Fig.~\ref{fig:g12}.  However, due to the limited particle number, information beyond the first coordinate shell is difficult to analyze.

Fig.~\ref{fig:g13} displays the total and partial pair-distribution function projections in the $xz$-plane.  Unlike the $xy$-plane projections in Fig.~\ref{fig:g12} which exhibits strong anisotropy, the suspension structures here are less sensitive to $\pe$.  With increasing $\pe$, the particles are compressed towards each other, which is evidenced by the thinning and brightening of the first coordinate shells.  More interestingly, at higher $\pe\geq 100$, $g_{12}(\br)$ shows a belt of particle enrichment along the flow direction, while $g_{11}(\br)$ and $g_{22}(\br)$ exhibit a corresponding particle depletion.  This indicates that the small-large pairs are preferred in the $xz$-plane, and that the shear flow promotes species mixing in the flow direction.

Fig.~\ref{fig:g23} shows the projection of $g(\br)$ and $g_{\alpha\beta}(\br)$ in the $yz$-plane.  With increasing $\pe$, the shear flow also compresses the particle pairs in this plane without apparent anisotropy.  Note that even at $\pe = 10^4$, the suspension does not exhibit string ordering~\cite{fossandbrady2000} due to the HIs.  The lack of structural formation is also confirmed by the continuous increase of the long-time self-diffusivities $d_{\infty,\alpha}^{t,yy}$ and $d_{\infty,\alpha}^{t,yy}$ with $\pe$ in Fig.~\ref{fig:dinfyz}.

\section{Conclusions}
\label{sec:conclusions}

In this work we presented the Spectral Ewald Accelerated Stokesian Dynamics (SEASD) for dynamic simulations of polydisperse colloidal suspensions.  Using the framework of Stokesian Dynamics (SD), SEASD can accurately and rapidly compute HIs in dense polydisperse suspensions.  Other features of SEASD include (\emph{i}) direct inclusion of the solvent compressibility and pressure evaluations; (\emph{ii}) the use of the Spectral Ewald (SE) method for accurate mobility computation with flexible error control; (\emph{iii}) a far-field preconditioner to accelerate the convergence of the nested iterative scheme; (\emph{iv}) GPGPU accelerated mobility evaluation for almost an order of magnitude speed improvement; and (\emph{v}) the incorporation of \mbox{SEASD-nf}, an extension of the mean-field Brownian approximation of Banchio \& Brady~\cite{asd-brownian_banchio_jcp2003} to polydisperse suspensions.

We extensively discussed the accuracy of mobility computation using the SE method, established the baseline for parameter selection, and demonstrated the adequate accuracy in the GPU single precision (SP) mobility computation.  We found that compared to the full SEASD computations, \mbox{SEASD-nf} can achieve significant speedup without substantially sacrificing accuracy.  Indeed, for all the dynamic simulations in this work, the SEASD and \mbox{SEASD-nf} results agree satisfactorily.  In addition, we verified the $\bigO(N\log N)$ computational scaling of SEASD and \mbox{SEASD-nf} in dynamic simulations.

We rigorously validated SEASD and \mbox{SEASD-nf} for monodisperse and bidisperse colloidal suspensions via: (\emph{i}) the short-time transport properties, (\emph{ii}) the equilibrium osmotic pressure and viscoelastic moduli, and (\emph{iii}) the steady Brownian shear rheology at $\phi=0.45$.
For (\emph{i}), the SEASD diffusivities and shear viscosity agree with the conventional SD calculations.  The SEASD sedimentation velocity differ qualitatively from the SD results due to the absence of a mean-field quadrupole term in the mobility computation.  For the bulk viscosity computation, different procedures to eliminate the spurious HIs lead to slight differences in the SEASD and the SD results.
In (\emph{ii}), SEASD and \mbox{SEASD-nf} reproduced the equilibrium suspension osmotic pressure for monodisperse and bidisperse suspensions within the error tolerance, with the SEASD data higher.
For the steady shear rheology in (\emph{iii}),  the agreement between \mbox{SEASD-nf} and SEASD is satisfactory in the suspension mechanics, dynamics, and structures.  Moreover, we found that the particle size polydispersity reduces the suspension viscosity and osmotic pressure, and enhances the long-time translational self-diffusivities of both species.  Our rheological simulations also improve our understanding on the structure, dynamics, and rheology of polydisperse suspensions.

The SEASD and \mbox{SEASD-nf} developed in this work are important tools for studying dynamics of dense, polydisperse colloidal suspensions, and have significantly extended the parameter space accessible to computational studies.  
For example, they can provide otherwise inaccessible details on a wide range of experimental observations including the yielding phenomena in glass rheology and the continuous and discontinuous shear-thickening.

Finally, through SEASD and \mbox{SEASD-nf} we demonstrated the generality and versatility of the SD framework, particularly the splitting of the far- and near-field interactions: with a suitable far-field computation, the lubrication interactions can be added pairwise for free.  We believe that many far-field HI computational methods can and should be used with the SD framework to expand their accessible parameter range, particularly for dense systems.

\section*{Acknowledgements}
We thank Wen Yan for helpful discussions on GPGPU programming and particle mesh techniques.  M.W. gratefully acknowledges supports from the Natural Sciences and Engineering Research Council of Canada (NSERC) by a Postgraduate Scholarship (PGS), and the National Science Foundation (NSF) grant CBET-1337097.

\section*{References}
\bibliographystyle{elsarticle-num} 
\bibliography{ref-seasd}

\end{document}